\documentclass[12pt,a4paper]{article}

\usepackage{jheppub,amsmath,amssymb,color,graphicx}

\usepackage[utf8]{inputenc}
\usepackage{tikz}
\usepackage{tikz-feynman}
\usepackage{subcaption}

\newcommand{\ptjv}{p_{T,\mathrm{veto}}}
\newcommand{\ETmiss}{{E\!\!\!/}_{\mathrm T}}

\title{Anomalous triple gauge couplings in the light of dimension-8 operators in $W^+W^-$}
\date{\today}

\author[a]{Daniel Gillies,}
\author[a]{Andrea Banfi,}
\author[b]{Adam Martin,}

\affiliation[a]{Department of Physics and Astronomy, University of Sussex, Brighton BN1 9QH, U.K.}
\affiliation[b]{Department of Physics, University of Notre Dame, Notre Dame, IN 46556, USA}

\abstract{We compare the size of dimension-8 effects on $W^+W^-$
  production at the LHC arising from the $q\bar q$ and $\gamma\gamma$
  initial states.  In particular, we consider bosonic operators, which
  contribute to the anomalous triple and quadruple gauge
  couplings. The relevant dimension-6 and dimension-8 operators are
  matched to the anomalous triple gauge couplings that contribute to
  the $q\bar q$ channel, allowing for the resummation of large
  logarithmic contributions arising in the presence of a jet-veto
  through the program MCFM-RE.  For the $\gamma\gamma$ channel, which
  receives contributions from both triple and quadruple gauge
  couplings, such resummation can be performed through the program
  MadGraph, by simply setting the factorisation scale to the jet-veto
  scale.  We find that, when neglecting fermionic dimension-8
  operators, the $\gamma\gamma$ channel has a dominant bosonic
  dimension-8 contribution and this channel can be used to understand
  the range of validity of the Effective Field Theory. With this in
  mind, and carefully considering theoretical and experimental
  uncertainties, we provide constraints on higher dimensional
  operators using current and future data.

 } 
\begin{document}

\maketitle

\newpage

\newpage

\section{Introduction}
\label{sec:intro}

The production of W boson pairs ($W^+ W^-$) in high-energy
proton-proton collisions provides one of the most sensitive testbeds
for electroweak (EW) precision observables. As a purely EW process
that has been precisely calculated up to NNLO in
QCD~\cite{Grazzini:2016ctr}, $W^+W^-$ production probes the gauge
structure of the SM including non-Abelian interactions of the $W$, $Z$
and $\gamma$ bosons. These have been extensively tested at the LHC and
previously at LEP, looking for anomalous triple gauge couplings
(aTGCs)~\cite{ALEPH:2013dgf, Corbett:2012ja, ALEPH:2010aa, CDF:2007aqs, D0:2006eed, Villa:2004dx,OPAL:2000rnf}. 
There has also been work
searching for the presence of neutral triple gauge couplings in the
$ZZ$ channel~\cite{Ellis:2025jgt, Mimasu:2025ulk, Guo:2024qyx,
  Ellis:2025ghl, Rahaman:2016pqj}, while $W^+W^-$ production provides
strong constraints on charged triple gauge couplings.  However, the
language of aTGCs has been largely replaced by more sophisticated
analyses using Effective Field Theory (EFT) methods. Instead of
looking for modifications of the triple gauge couplings, searches for
the presence of EFT operators have been
performed~\cite{ElFaham:2025fow, ElFaham:2024uop, Butter:2016cvz,
  Green:2016trm, Zhang:2016zsp, Baglio:2017bfe, Ellis:2018gqa,
  Baglio:2018bkm, Baglio:2019uty, Bellan:2021dcy, Almeida:2021asy,
  Degrande:2012wf, Falkowski:2016cxu, Aoude:2019cmc, DiNoi:2025uhu,
  ElFaham:2024egs}.  These are generally more controlled approaches
which maintain the gauge invariance of $W$ and $Z$ bosons and can be
matched to frameworks like the Standard Model Effective Field Theory
(SMEFT). The SMEFT look for signs of new physics in a
model-independent fashion assuming only that new physics exists at a
decoupled higher energy scale ($\Lambda$) and that the symmetries of
the Standard Model (SM) continue to hold up to those energies.

In $W^+ W^-$ ($WW$) production, the leading Standard Model
contribution is given by $q\bar q$ initiated processes.  Bosonic
operators, which we define as operators containing only bosonic
fields, corresponding to aTGCs can enter into this process by
modifying the $Z W^+ W^-$ and $\gamma W^+ W^-$ vertices which appear
in the $q\bar q$ $s$-channel.  For a study of non-bosonic ($qqWW$)
dimension-8 operators for this channel, we refer the reader to a
recent study~\cite{ElFaham:2025fow}.  Studying bosonic operators alone
is not generally valid in the SMEFT since, using the SM equations of
motion, higher-derivative bosonic operators may be traded for
fermionic operators. This analysis should therefore be considered to
apply specifically to universal EFTs~\cite{Wells:2015uba} which are
defined as those theories which can be mapped entirely onto bosonic
higher dimensional operators. Universal Theories are basis dependent~\cite{Wells:2015uba}. 
We are working within a Warsaw-like basis, where 
as many derivatives as possible are removed by the use 
of integration by parts and/or the equations of motion.
Previously, it has been found that, in this
channel, the Standard Model $\gamma\gamma$ contribution is not
negligible.  In fact at higher energies, the $\gamma\gamma$
contribution becomes much greater than the $gg$ contribution which is
usually considered to be the second-largest contribution to this
channel~\cite{Gillies:2024mqp}. All the dimension-6 operators which
modify the $q\bar q$ process via the $Z W^+ W^-$ and $\gamma W^+ W^-$
vertices also modify the $\gamma\gamma$ contribution. Some of these
operators also induce a photon-photon-Higgs ($\gamma\gamma h$) contact interaction. This
motivates the inclusion of the subleading $\gamma\gamma$ contributions
into fits of these operators and the study of their impact at
dimension-6 and dimension-8.

When performing an EFT analysis of a high energy observable, special
attention must be paid to ensuring EFT validity.  In the EFT validity
region, operators are expanded in powers of $(\sqrt{\hat s}/\Lambda)^N$,
where $N$ is set by the dimension of the operator $d$ as $N=d-4$ and
$\sqrt{\hat s}$ is the partonic centre-of-mass energy of the process
being probed. When $\sqrt s \ll \Lambda$, one can truncate the
expansion to leading order by considering only the dimension-6
operators.  In this channel, $\sqrt s = M_{WW}$, the invariant mass of
the $WW$ pair. When both $W$ decay leptonically, the case studied in
this work, this quantity is not known due to the missing neutrino
energy.  In a previous letter~\cite{Gillies:2026tpc}, we showed that
enforcing a cut of $M_{WW} < \Lambda$ only on the EFT simulation is
not sufficient to ensure that the dimension-8 contribution is less
than the dimension-6 contribution. This is a possible indication of
EFT breakdown due to the fact that the hierarchy of EFT operators is
no longer respected. By looking at all dimension-8 squared
contributions, including those of subleading channels such as $gg$ and
$\gamma\gamma$, direct probes of the EFT validity can be obtained by
comparison to their dimension-6 counterparts.

However, even if EFT validity has been ensured, there is one caveat,
which is that the true leading contribution of the dimension-6
operators is their interference with the Standard Model. When this
interference is small, constraints are often placed using the
dimension-6 squared contribution, which is formally next to leading
order in the EFT expansion. The dimension-8 interference and
dimension-6 squared contributions are both of order $1/\Lambda^4$ and
so due care must be taken to account for the size of dimension-8
interference pieces when deriving constraints on dimension-6 operators
with their squared contributions.  If dimension-8 interference terms
are also found to be small, then they can be neglected.

Fortunately for this channel, dimension-8 bosonic operators that
contribute to the $q\bar q$ channel have been found to have a small
interference with the Standard Model~\cite{Corbett:2023qtg}, thus
motivating the inclusion of dimension-6 squared
contributions. However, the dimension-8 $q\bar q$ bosonic operators
only modify the triple gauge interaction and are therefore limited in
the extent to which they can grow with energy.

There are also bosonic dimension-8 operators which modify the
$\gamma\gamma WW$ vertex which grow much more rapidly with energy and
could therefore overtake the dimension-8 bosonic $q\bar q$
contributions.  The size of these operators should be profiled to
ensure that both the dimension-8 interference and squared terms in the
$\gamma\gamma$ channel are also small. Furthermore, the size of these
contributions allows us to understand better what the region of EFT
validity is for this channel in the absence of direct measurements of
$M_{WW}$.

Analysis of this channel is made more complex by the inclusion of a
jet-veto or $b$-jet-veto.  This is used to suppress large backgrounds
from $t\bar t$ production, usually considered a separate process
despite the decay into two $W$ bosons. In the case of a full jet-veto,
the $gg$ and $qq$ channels must be resummed to avoid the breakdown of
the peturbative series in $\alpha_s$. This is due to large logarithms
appearing alongside powers of $\alpha_s$ in some diagrams. This
jet-veto also has implications for the photon channel, despite the
fact it itself does not radiate gluons, the photon channel is still
affected by the jet-veto through the photon PDFs. Recently a new
measurement of $WW$ production been released by ATLAS with the
replacement of the full jet-veto with a $b$-jet-veto via
$b$-tagging~\cite{ATLAS:2025dhf}. Currently, this is modelled by using
fixed order calculations with no jet-veto. We study the effect of the
inclusion or exclusion of this jet-veto on the quality of the
constraints that can be obtained from this channel at the HL-LHC. The
recent analysis also provides constraints for some available bosonic
EFT operators using an alternative method of ensuring EFT
validity~\cite{ATLAS:2025dhf, Brivio:2022pyi} - we discuss these
constraints and compare them to previous and future data.

In the following sections we present the EFT operators considered for
this analysis at dimension-6 and their effects on the quark and photon
channels for $WW$ production. We then present the full list of bosonic
dimension-8 operators for both quark and photon channels. For those
which contribute to $q\bar q$, we provide a dictionary to convert the
EFT operators into corresponding aTGC values which can then be
inputted into the Monte Carlo event generator MCFM-RE. This allows for
the first the resummation of large logarithmic contributions due to a
jet-veto, for these operators, at the next-to-leading logarihtmic
(NLL) accuracy. We also present updated SM predictions alongside some
discussion of sources of error for this channel. The effect of the
jet-veto on processes induced by gluon, quark, and photon fusion. In
particular the method for obtaining resummed predictions for the
$\gamma \gamma$ channel is also discussed.

The relative size of photon and quark contributions is then presented
at dimension-6 and dimension-8 for the various operators which
contribute to this channel. This indicates which operators are best
placed to be constrained using this channel. The relative sizes of
dimension-6 and dimension-8 contributions is also used to understand
better the EFT valid regime for the observable we consider, the
distribution in $M_{e\mu}$, the invariant mass of an electron and a
muon.

Finally, constraints are presented for this channel using 2019 ATLAS measurement ~\cite{ATLAS:2019rob} and sensitivity studies are 
performed for HL-LHC. Constraints are given at multiple values of $\Lambda$ allowing for multivariate information
between Wilson coefficients and the high energy scale to be extracted whilst always ensuring EFT validity holds.
The effect of squared contributions and the jet-veto on the quality of the constraints is also discussed.

\section{EFT Analysis of Bosonic Operators}
\label{sec:EFT}

In this section, we present the CP-even dimension-6 and dimension-8
operators which contribute to $WW$ production. 
We restrict our analysis to CP-even operators, assuming CP-conservation in the underlying new physics.
We consider only leading order
tree-level BSM interactions.
Also, we focus on higher dimensional operators without any fermion field,
which we refer to as bosonic operators. In particular, we study those
which contribute to the $q\bar q$ and $\gamma \gamma$ induced
processes. At dimension-6, there are four bosonic operators which
contribute to $q\bar q$ or $\gamma \gamma$ channels at leading order.
They are presented in table~\ref{table:d_6operators}, in the Warsaw
basis~\cite{Grzadkowski:2010es}.  To determine which operators
contribute to the process at hand, we used the
\texttt{FeynRules}~\cite{Alloul:2013bka} package to generate the
Feynman rules for the complete set of dimension-6 operators which was
given by the SMEFTsim package~\cite{Brivio:2020onw, Brivio:2017btx}.
The operator $\mathcal{O}_{HB}$ does not contribute to the $q\bar q$
channel as it does not modify any of the anomalous triple gauge
couplings. Instead it couples the photons directly to the Higgs boson
through a contact interaction which grows with energy. The operators
$\mathcal{O}_{HW}$, $\mathcal{O}_{HWB}$ and $\mathcal{O}_{WWW}$ all
contribute to both $q\bar q$ and $\gamma \gamma$ channels.
\begin{table}
\renewcommand{\arraystretch}{1.5}
\begin{center}
\begin{tabular}{ |c|c|c| } 
 \hline
  & Dimension-6 Operator & Processes Contributing to \\ 
 \hline
 $\mathcal{O}_{HW}$ & $\phi^\dagger \phi\, W^{I,\mu\nu} W^{I}_{\mu\nu}$ & $q\bar q$ and $\gamma\gamma$\\
 \hline
 $\mathcal{O}_{HB}$ & $\phi^\dagger \phi\, B^{\mu\nu} B_{\mu\nu}$ & $\gamma\gamma$ only\\
 \hline
 $\mathcal{O}_{HWB}$ & $\phi^\dagger  \sigma^I\phi\, W^I_{\mu\nu}B^{\mu\nu}$ & $q\bar q$ and $\gamma\gamma$\\ 
 \hline
 $\mathcal {O}_{WWW}$ & $\epsilon_{IJK} W^{I\, \nu}_{\mu}W^{J\,\rho}_{\nu}W^{K\, \mu}_{\rho}$ & $q\bar q$ and $\gamma\gamma$\\ 
 \hline
\end{tabular}
\caption{List of the bosonic dimension-6 SMEFT operators which affect $q \bar q$- and $\gamma\gamma$-induced
  $WW$ production, in the Warsaw basis~\cite{Grzadkowski:2010es}.
Here $\phi$ is the Higgs field, $W^{I}_{\mu\nu}$ and $B_{\mu\nu}$ the field strenghts corresponding to the $W^I_{\mu}$ and $B_\mu$ gauge fields, before spontaneous symmetry breaing.
\label{table:d_6operators}}
\end{center}
\end{table}

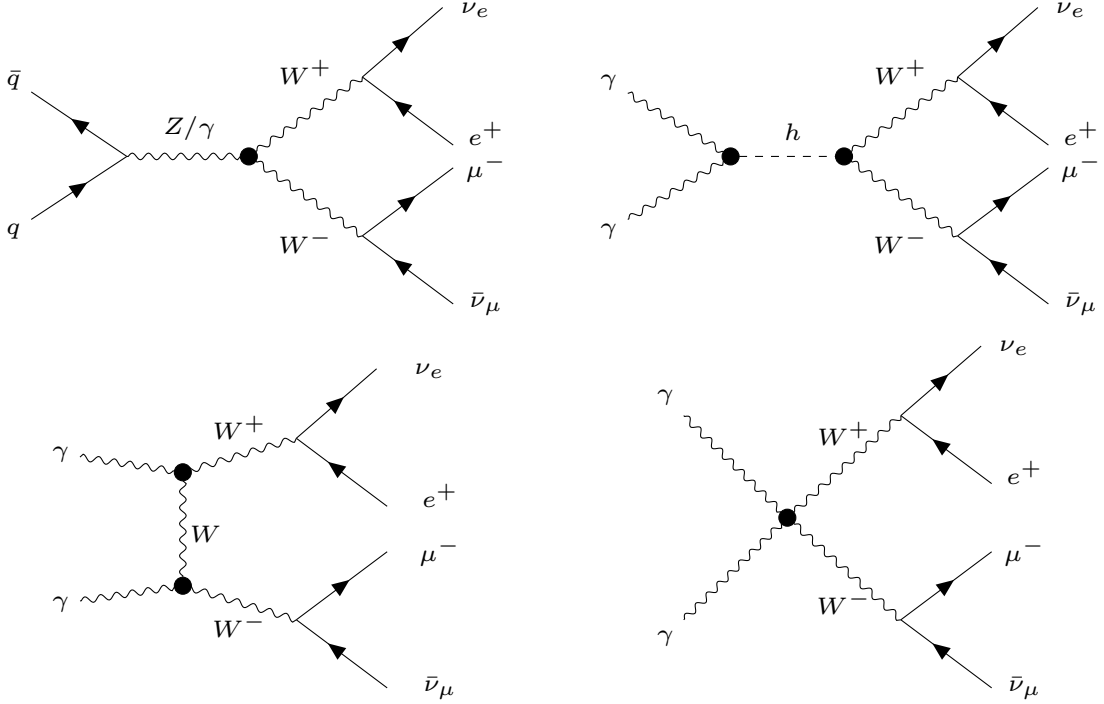
\begin{figure}[htbp]
  \begin{minipage}[t]{0.48\textwidth}
      \centering
\begin{tikzpicture}[scale=1.5, transform shape]
      \begin{feynman}
        \vertex (a); 
        \vertex [right = 1cm  of a] (v0);
        \vertex [right = 0.75cm  of a] (q0);
        \vertex [right = 1cm  of v0, dot] (v1) {};
        \vertex [right = 1cm  of v1] (c);
        \vertex [right = 0.55cm  of v0] (labelhelp);
        \vertex [node font=\tiny, above = 0.05cm  of labelhelp](label){\(Z/\gamma\)};
        \vertex [right = 0.5cm  of v1] (aux);
        \vertex [node font=\tiny, above = 0.55cm  of aux] (label1){\(W^+\)};
        \vertex [node font=\tiny, below = 0.55cm  of aux] (label2){\(W^-\)};
        \vertex [node font=\tiny, above = 0.5cm of a] (a1) {\(\bar q\)};
        \vertex [node font=\tiny, below = 0.5cm of a] (a2) {\(q\)};
        \vertex [node font=\tiny, above = 0.5cm of q0] (q1);
        \vertex [node font=\tiny, below = 0.5cm of q0] (q2);
        \vertex [above = 0.7cm of c] (v2);
        \vertex [below = 0.7cm of c] (v3);
        \vertex [right = 0.8cm  of v2] (d);
         \vertex [right = 0.8cm  of v3] (e);
         \vertex [node font=\tiny, above = 0.6cm of d] (l1) {};
         \vertex [node font=\tiny, right = 0.2cm of l1] (l12) {};
         \vertex [node font=\tiny, below = 0.1cm of l12] (l13){\(\nu_{e}\)};
        \vertex [node font=\tiny, below = 0.6cm of d] (l2);
        \vertex [node font=\tiny, right = 0.2cm of l2] (l22) {};
        \vertex [node font=\tiny, above = 0.1cm of l22] (l23){\(e^+\)};
        \vertex [node font=\tiny, above = 0.6cm of e] (l3);
        \vertex [node font=\tiny, right = 0.2cm of l3] (l32) {};
        \vertex [node font=\tiny, above = 0cm of l32] (l33){\(\mu^{-}\)};
        \vertex [node font=\tiny, below = 0.6cm of e] (l4);
        \vertex [node font=\tiny, right = 0.2cm of l4] (l42) {};
        \vertex [node font=\tiny, below = 0cm of l42] (l43){\(\bar\nu_{\mu}\)};
        \diagram* {
          (v0) -- [fermion] (a1), 
          (a2) -- [fermion] (v0),
          (v0) -- [boson] (v1), 
          (v1) -- [boson] (v2), 
          (v1) -- [boson] (v3), 
          (v2) -- [fermion] (l1), 
          (v2) -- [anti fermion] (l2), 
          (v3) -- [fermion] (l3), 
          (v3) -- [anti fermion] (l4),
        };
      \end{feynman}
       \end{tikzpicture}
      \end{minipage}
      \hfill
\begin{minipage}[t]{0.48\textwidth}
      \centering
  \begin{tikzpicture}[scale=1.5, transform shape]
      \begin{feynman}
        \vertex (a); 
        \vertex [right = 1cm  of a, dot] (v0) {};
        \vertex [right = 0.75cm  of a] (q0);
        \vertex [right = 1cm  of v0, dot] (v1) {};
        \vertex [right = 1cm  of v1] (c);
        \vertex [right = 0.55cm  of v0] (labelhelp);
        \vertex [node font=\tiny, above = 0.05cm  of labelhelp](label){\(h\)};
        \vertex [right = 0.5cm  of v1] (aux);
        \vertex [node font=\tiny, above = 0.55cm  of aux] (label1){\(W^+\)};
        \vertex [node font=\tiny, below = 0.55cm  of aux] (label2){\(W^-\)};
        \vertex [node font=\tiny, above = 0.5cm of a] (a1) {\(\gamma\)};
        \vertex [node font=\tiny, below = 0.5cm of a] (a2) {\(\gamma\)};
        \vertex [node font=\tiny, above = 0.5cm of q0] (q1);
        \vertex [node font=\tiny, below = 0.5cm of q0] (q2);
        \vertex [above = 0.7cm of c] (v2);
        \vertex [below = 0.7cm of c] (v3);
        \vertex [right = 0.8cm  of v2] (d);
         \vertex [right = 0.8cm  of v3] (e);
         \vertex [node font=\tiny, above = 0.6cm of d] (l1) {};
         \vertex [node font=\tiny, right = 0.2cm of l1] (l12) {};
         \vertex [node font=\tiny, below = 0.1cm of l12] (l13){\(\nu_{e}\)};
        \vertex [node font=\tiny, below = 0.6cm of d] (l2);
        \vertex [node font=\tiny, right = 0.2cm of l2] (l22) {};
        \vertex [node font=\tiny, above = 0.1cm of l22] (l23){\(e^+\)};
        \vertex [node font=\tiny, above = 0.6cm of e] (l3);
        \vertex [node font=\tiny, right = 0.2cm of l3] (l32) {};
        \vertex [node font=\tiny, above = 0cm of l32] (l33){\(\mu^{-}\)};
        \vertex [node font=\tiny, below = 0.6cm of e] (l4);
        \vertex [node font=\tiny, right = 0.2cm of l4] (l42) {};
        \vertex [node font=\tiny, below = 0cm of l42] (l43){\(\bar\nu_{\mu}\)};
        \diagram* {
          (v0) -- [boson] (a1), 
          (a2) -- [boson] (v0),
          (v0) -- [scalar] (v1), 
          (v1) -- [boson] (v2), 
          (v1) -- [boson] (v3), 
          (v2) -- [fermion] (l1), 
          (v2) -- [anti fermion] (l2), 
          (v3) -- [fermion] (l3), 
          (v3) -- [anti fermion] (l4),
        };
      \end{feynman}
       \end{tikzpicture}
      \end{minipage}
  \vspace{0.8em}
    \begin{minipage}[t]{0.48\textwidth}
      \centering
    \begin{tikzpicture}[scale=1.5, transform shape]
      \begin{feynman}
        \vertex (a); 
        \vertex [right = 1cm  of a, dot] (v0){};
        \vertex [right = 0cm  of v0] (v1);
        \vertex [below = 1cm  of v0, dot] (v4){};
         \vertex [below = 1cm  of v1] (v5);
        \vertex [right = 1cm  of v1] (c);
        \vertex [right = 0.5cm  of v1] (aux);
        \vertex [node font=\tiny, right = 0.2cm  of v1] (aux2);
        \vertex [node font=\tiny, above = 0.2cm  of aux] (label1){\(W^+\)};
        \vertex [node font=\tiny, below = 1.15cm  of aux] (label2){\(W^-\)};
        \vertex [node font=\tiny, below = 0.38cm  of aux2] (label3){\(W\)};
        \vertex [node font=\tiny, above = 0cm of a] (a1) {\(\gamma\)};
        \vertex [node font=\tiny, below = 1cm of a] (a2) {\(\gamma\)};
        \vertex [above = 0.3cm of c] (v2);
        \vertex [below = 1.3cm of c] (v3);
        \vertex [right = 0.8cm  of v2] (d);
         \vertex [right = 0.8cm  of v3] (e);
         \vertex [node font=\tiny, above = 0.6cm of d] (l1) {};
         \vertex [node font=\tiny, right = 0.2cm of l1] (l12) {};
         \vertex [node font=\tiny, below = 0.1cm of l12] (l13){\(\ \ \ \ \nu_{e}\)};
        \vertex [node font=\tiny, below = 0.6cm of d] (l2);
        \vertex [node font=\tiny, right = 0.2cm of l2] (l22) {};
        \vertex [node font=\tiny, above = 0.1cm of l22] (l23){\(\ \ \ \ e^+\)};
        \vertex [node font=\tiny, above = 0.6cm of e] (l3);
        \vertex [node font=\tiny, right = 0.2cm of l3] (l32) {};
        \vertex [node font=\tiny, above = 0cm of l32] (l33){\(\ \ \ \ \mu^{-}\)};
        \vertex [node font=\tiny, below = 0.6cm of e] (l4);
        \vertex [node font=\tiny, right = 0.2cm of l4] (l42) {};
        \vertex [node font=\tiny, below = 0cm of l42] (l43){\(\ \ \ \ \bar\nu_{\mu}\)};
        \diagram* {
          (v0) -- [boson] (a1), 
          (a2) -- [boson] (v4),
          (v4) -- [boson] (v0), 
          (v1) -- [boson] (v2), 
          (v5) -- [boson] (v3), 
          (v2) -- [fermion] (l1), 
          (v2) -- [anti fermion] (l2), 
          (v3) -- [fermion] (l3), 
          (v3) -- [anti fermion] (l4),
        };
      \end{feynman}
       \end{tikzpicture}
      \end{minipage}
    \hfill
    \begin{minipage}[t]{0.48\textwidth}
      \centering
    \begin{tikzpicture}[scale=1.5, transform shape]
      \begin{feynman}
        \vertex (a); 
        \vertex [right = 1cm  of a, dot] (v1) {};
        \vertex [right = 1cm  of v1] (c);
        \vertex [right = 0.5cm  of v1] (aux);
        \vertex [node font=\tiny, above = 0.55cm  of aux] (label1){\(W^+\)};
        \vertex [node font=\tiny, below = 0.55cm  of aux] (label2){\(W^-\)};
        \vertex [node font=\tiny, above = 0.9cm of a] (a1) {\(\gamma\)};
        \vertex [node font=\tiny, below = 0.9cm of a] (a2) {\(\gamma\)};
        \vertex [above = 0.9cm of c] (v2);
        \vertex [below = 0.9cm of c] (v3);
        \vertex [right = 0.8cm  of v2] (d);
       	\vertex [right = 0.8cm  of v3] (e);
       	\vertex [node font=\tiny, above = 0.6cm of d] (l1) {};
       	\vertex [node font=\tiny, right = 0.2cm of l1] (l12) {};
       	\vertex [node font=\tiny, below = 0.1cm of l12] (l13){\(\nu_{e}\)};
        \vertex [node font=\tiny, below = 0.6cm of d] (l2);
        \vertex [node font=\tiny, right = 0.2cm of l2] (l22) {};
        \vertex [node font=\tiny, above = 0.1cm of l22] (l23){\(e^+\)};
        \vertex [node font=\tiny, above = 0.6cm of e] (l3);
        \vertex [node font=\tiny, right = 0.2cm of l3] (l32) {};
        \vertex [node font=\tiny, above = 0cm of l32] (l33){\(\mu^{-}\)};
        \vertex [node font=\tiny, below = 0.6cm of e] (l4);
        \vertex [node font=\tiny, right = 0.2cm of l4] (l42) {};
        \vertex [node font=\tiny, below = 0cm of l42] (l43){\(\bar\nu_{\mu}\)};
        \diagram* {
          (a1) -- [boson] (v1), 
          (a2) -- [boson] (v1), 
          (v1) -- [boson] (v2), 
          (v1) -- [boson] (v3), 
          (v2) -- [fermion] (l1), 
          (v2) -- [anti fermion] (l2), 
          (v3) -- [fermion] (l3), 
          (v3) -- [anti fermion] (l4),
        };
      \end{feynman}
       \end{tikzpicture}
  \end{minipage}
  
      \caption{Feynman diagrams for $W^+W^-$ production via EFT operators at dimension-6 and dimension-8. 
      Diagrams for the following channels are shown:
      $q\>\bar q\! \to \! W^+\>W^-$ with modifications to the $Z\, W^+W^-$ or $\gamma\, W^+W^-$ triple gauge couplings (top-left),
      as well as $\gamma\>\gamma\! \to \! W^+\>W^-$ with an induced $\gamma\gamma h$ coupling or modified $W^+W^- h$ coupling (top-right),
       with modifications to the $\gamma\,W^+W^-$ triple gauge coupling (bottom-left), or with modifications to the $\gamma\,\gamma\,W^+W^-$ quadruple gauge coupling (bottom-right).
  \label{fig:FeynmanDiagramqq}
  }
\end{figure}

At dimension-8, we have identified $20$ purely bosonic operators which
could potentially contribute to either the $q\bar q$ or $\gamma\gamma$
channels. These can be identified from previous work~\cite{Murphy:2020rsh, Corbett:2024yoy, Li:2020gnx},
which was informed by Hilbert series methods~\cite{Lehman:2015via, Henning:2015daa, Lehman:2015coa, Henning:2015alf}.
These are presented in table~\ref{table:d_8operators}. The
first five operators of table~\ref{table:d_8operators} generate triple
gauge couplings and therefore contribute to the $q \bar q$ channel.
The operator $\mathcal{O}_{8HDHW2}$ only generates couplings via the
covariant derivative. Since the Higgs does not couple to the
photon, this operator introduces only a $Z W^+ W^-$ anomalous
coupling. This therefore only modifies the $q \bar q$-induced
contribution. However, the final fifteen operators in
table~\ref{table:d_8operators} induce direct $\gamma \gamma W^+ W^-$
contact interactions which grow with energy. They therefore can only
be seen when considering the $\gamma \gamma$ channel for $W^+W^-$
production. The operator $\mathcal {O}_{8D2H2X24}$ was found not to
generate this quadruple gauge interaction and therefore has no
contribution to this channel at LO.
\begin{table}
\renewcommand{\arraystretch}{1.5}
\begin{center}
  \footnotesize
\begin{tabular}{ |c|c|c| } 
 \hline
  & Dimension-8 Operator & Processes Contributing to \\ 
 \hline
  $\mathcal {O}_{8HDHB}$ & $i\phi^\dagger \phi\, (D^{\mu}\phi)^\dagger D^{\nu}\phi\, B_{\mu\nu}$ & $q\bar q$ and $\gamma\gamma$\\ 
 \hline
  $\mathcal {O}_{8HDHW}$ & $i\phi^\dagger \phi\, (D^{\mu}\phi)^\dagger \sigma^I D^{\nu}\phi\, W^I_{\mu\nu}$ & $q\bar q$ and $\gamma\gamma$\\ 
 \hline
  $\mathcal {O}_{8HDHW2}$ & $i \epsilon^{IJK} \phi^\dagger \sigma^I \phi\, (D^{\mu}\phi)^\dagger \sigma^J D^{\nu}\phi\, W^K_{\mu\nu}$ & $q\bar q$ only\\ 
 \hline
  $\mathcal {O}_{8W}$ & $\epsilon^{IJK} \phi^\dagger \phi\, W^{I\,\nu}_{\mu} W^{J\,\rho}_{\nu} W^{K\,\mu}_{\rho}$ & $q\bar q$ and $\gamma\gamma$\\ 
 \hline
  $\mathcal {O}_{8W2}$ & $\epsilon_{IJK}\phi^\dagger \sigma^I \phi\, W^{J\,\nu}_{\mu} W^{K\,\rho}_{\nu} B^{\ \mu}_{\rho}$ & $q\bar q$ and $\gamma\gamma$\\ 
 \hline
  $\mathcal {O}_{8HW}$ & $(\phi^\dagger \phi)^2\, W^{I,\mu\nu} W^{I}_{\mu\nu}$ & $q\bar q$ and $\gamma\gamma$\\ 
 \hline
  $\mathcal {O}_{8HW2}$ & $(\phi^\dagger  \sigma^I\phi)\,(\phi^\dagger  \sigma^J\phi)\, W^{I,\mu\nu} W^{J}_{\mu\nu}$ & $\gamma\gamma$ only\\ 
 \hline
  $\mathcal {O}_{8HB}$ & $ (\phi^\dagger \phi)^2\, B^{\mu\nu} B_{\mu\nu}$ & $\gamma\gamma$ only\\ 
 \hline
  $\mathcal {O}_{8HWB}$ & $ (\phi^\dagger \phi)\phi^\dagger  \sigma^I\phi\, W^I_{\mu\nu}B^{\mu\nu}$ & $q\bar q$ and $\gamma\gamma$\\ 
  \hline
  $\mathcal {O}_{8WWWW1}$ & $W^I_{\mu\nu} W^{I\, \rho\sigma} W^{J\, \mu\nu} W^{J}_{\rho\sigma}$ & $\gamma\gamma$ only\\ 
 \hline
  $\mathcal {O}_{8WWWW2}$ & $W^I_{\mu\nu} W^{I\, \mu\nu} W^J_{\rho\sigma} W^{J\, \rho\sigma}$ & $\gamma\gamma$ only\\ 
 \hline
  $\mathcal {O}_{8WWWW3}$ & $W^I_{\mu\nu} \tilde W^{I\, \mu\nu} W^J_{\rho\sigma} \tilde W^{J\, \rho\sigma}$ & $\gamma\gamma$ only\\ 
 \hline
  $\mathcal {O}_{8WWWW4}$ & $W^I_{\mu\nu} W^{I\, \rho\sigma} \tilde W^{J\, \mu\nu} \tilde W^{J}_{\rho\sigma}$ & $\gamma\gamma$ only\\ 
 \hline
  $\mathcal {O}_{8BBWW1}$ & $B_{\mu\nu} B^{\rho\sigma} W^{J\, \mu\nu} W^{J}_{\rho\sigma}$ & $\gamma\gamma$ only\\ 
 \hline
  $\mathcal {O}_{8BBWW2}$ & $B_{\mu\nu} B^{\mu\nu} W^J_{\rho\sigma} W^{J\, \rho\sigma}$ & $\gamma\gamma$ only\\ 
 \hline
  $\mathcal {O}_{8BBWW3}$ & $B_{\mu\nu} \tilde B^{\mu\nu} W^J_{\rho\sigma} \tilde W^{J\, \rho\sigma}$ & $\gamma\gamma$ only\\ 
 \hline
  $\mathcal {O}_{8BBWW4}$ & $B_{\mu\nu} B^{\rho\sigma} \tilde W^{J\, \mu\nu} \tilde W^{J}_{\rho\sigma}$ & $\gamma\gamma$ only\\ 
 \hline
  $\mathcal {O}_{8D2H2X21}$ & $(D^{\mu}\phi)^\dagger D_{\mu}\phi\, W^I_{\nu\rho}W^{I\,\nu\rho}$ & $\gamma\gamma$ only\\ 
 \hline
  $\mathcal {O}_{8D2H2X22}$ & $(D^{\mu}\phi)^\dagger D_{\nu}\phi\, W^{I}_{\mu\rho}W^{I,\,\nu\rho}$ & $\gamma\gamma$ only\\ 
 \hline
 $\mathcal {O}_{8D2H2X23}$ & $(D^{\mu}\phi)^\dagger \sigma^{I} D_{\mu}\phi\, B_{\nu\rho}W^{I\,\nu\rho}$ & $\gamma\gamma$ only\\ 
 \hline
 $\mathcal {O}_{8D2H2X24}$ & $(D^{\mu}\phi)^\dagger \sigma^{I}D_{\nu}\phi\, (B_{\mu\rho}W^{I,\,\nu\rho} - B^{\nu\rho}W^{I}_{\mu\rho})$ & No contribution\\ 
 \hline
 $\mathcal {O}_{8D2H2X25}$ & $(D^{\mu}\phi)^\dagger \sigma^{I}D_{\nu}\phi\, (B_{\mu\rho}W^{I,\,\nu\rho} + B^{\nu\rho}W^{I}_{\mu\rho})$ & $\gamma\gamma$ only\\ 
 \hline
 $\mathcal {O}_{8D2H2X26}$ & $(D^{\mu}\phi)^\dagger D_{\mu}\phi\, B_{\nu\rho}B^{\ \nu\rho}$ & $\gamma\gamma$ only\\ 
 \hline
  $\mathcal {O}_{8D2H2X27}$ & $(D^{\mu}\phi)^\dagger D_{\nu}\phi\, B_{\mu\rho}B^{\nu\rho}$ & $\gamma\gamma$ only\\ 
 \hline
  $\mathcal {O}_{8D2H2X28}$ & $i\epsilon_{IJK}\,(D^{\mu}\phi)^\dagger \sigma^{I} D_{\mu}\phi\, W^J_{\nu\rho}W^{K\,\nu\rho}$ & No contribution\\ 
 \hline
\end{tabular}
\caption{List of the bosonic dimension-8 EFT operators which affect $q \bar q$ and $\gamma\gamma$
production for this channel.\label{table:d_8operators}}
\end{center}
\end{table}
\newpage
This channel can be studied in the context of anomalous triple gauge
couplings (aTGCs). In the EFT framework, these correspond to operators
which can modify the Standard Model $s$-channel contribution to
$q\bar q \to W^+W^-$ by modifying the $ZW^+W^-$ and $\gamma W^+W^-$ vertices
as shown in figure~\ref{fig:FeynmanDiagramqq} (top left). When adding
the EFT operators which generate aTGC-like modifications to the
$q\bar q$ channel, we note that these operators can also induce
changes to other vertices.  In the case of $\gamma \gamma \to W^+W^-$,
the same EFT operators considered above can modify both the $t$ and
$u$-channel diagrams through aTGCs (figure~\ref{fig:FeynmanDiagramqq},
bottom left) but can also induce a $\gamma\gamma h$ (or modified $hW^+W^-$) coupling which
allows for an $s$-channel Higgs-mediated contribution
(figure~\ref{fig:FeynmanDiagramqq}, top right). Finally, the new
operators can also modify the $\gamma\gamma W^+W^-$ quadruple gauge
coupling (figure~\ref{fig:FeynmanDiagramqq}, bottom-right). These may
have non-negligible impact on the size of both the dimension-6 contribution of
these operators and of linear or quadratic dimension-8
contributions. This is investigated in section~\ref{sec:Numerics}.

In order to generate resummed predictions for the operators'
contribution to the $q\bar q$ channel at NLL accuracy (required in the
presence of a jet-veto), these operators were matched to the
Lagrangian for anomalous triple gauge couplings which is quoted in the
MCFM documentation~\cite{mcfm_re_d8, Campbell:2015qma} as:
\begin{multline}
\mathcal{L_{\mathrm{anom}}}=-ig_{WW(Z/A)}\big[\Delta g_1^{Z/\gamma}\left(W^+_{\mu\nu}W^{-\,\mu}(Z/A)^{\nu} - W^-_{\mu\nu}W^{+\,\mu}(Z/A)^{\nu} \right)\\
 + \Delta \kappa_1^{Z/\gamma}W^+_{\mu}W^-_{\nu}(Z/A)^{\mu\nu} + \frac{\lambda^{Z/\gamma}}{M^2_{W}}W_\mu^{+\,\nu} W_\nu^{-\,\rho} (Z/A)_\rho^{\ \mu}\big]\,.
\end{multline}
We found that dimension-6 and dimension-8 operators can be mapped to
the contributions described above in all but one case, namely the
dimension-8 operator $\mathcal {O}_{8HDHW2}$. This generates a new
hermitian triple gauge coupling, which does not contain a factor of
the imaginary unit $i$ and is instead proportional to:
\begin{equation}
\left(W^+_{\mu\nu}W^{-\,\mu}(Z/A)^{\nu} + W^-_{\mu\nu}W^{+\,\mu}(Z/A)^{\nu} \right).
\end{equation}
This operator, as well as all CP-odd dimension-6 contributions, are not currently
 implemented in MCFM.  However, at LO, its contribution was
found to be smaller than that of $\mathcal {O}_{8HDHW}$.  We therefore
leave study or implementation of this operator alongside CP-odd
dimension-6 operators to future work.  The conversion of the other
operators into anomalous triple gauge couplings is found in
Appendix~\ref{sec:AppendixMatch}.  In equations~\eqref{eq:dictionary}, we present the dictionary
relating the triple gauge couplings as implemented in MCFM to the
dimension-6 and dimension-8 operators. Note that we keep factors of
$M_W$ in the conversions for $\lambda^{Z/A}$ to match the MCFM
conventions:
\begin{subequations} 
\label{eq:dictionary}
\begin{align}
\Delta g_1^Z &= -c_{HW}\frac{v^2}{\Lambda^2} - c_{8HW}\frac{v^4}{2\Lambda^4} 
+ c_{8HDHW}\frac{e\, v^4}{ 8\, \cos^2{\theta_W}\, \sin\theta_W\, \Lambda^4} \\[1em]
\Delta g_1^\gamma &=  -c_{HW}\frac{v^2}{\Lambda^2} - c_{8HW}\frac{v^4}{2\Lambda^4} \\[1em]
\Delta \kappa^Z &= -c_{HW}\frac{v^2}{\Lambda^2} 
- c_{HWB}\frac{\sin\theta_W\, v^2}{\cos\theta_W \Lambda^2} - c_{8HW}\frac{v^4}{2\Lambda^4} \\[1em]
& - c_{HWB}\frac{\sin\theta_W\, v^4}{2\cos\theta_W \Lambda^4}
+ c_{8HDHB}\frac{e\, v^4}{8\, \cos\theta_W\, \Lambda^4}
- c_{8HDHW}\frac{e\, v^4}{8\, \sin\theta_W\, \Lambda^4}
\\[1em]
\Delta \kappa^\gamma &= -c_{HW}\frac{v^2}{\Lambda^2} 
+ c_{HWB}\frac{\sin\theta_W\, v^2}{\cos\theta_W \Lambda^2} - c_{8HW}\frac{v^4}{2\Lambda^4}  \\[1em]
& - c_{HWB}\frac{\sin\theta_W\, v^4}{2\cos\theta_W \Lambda^4}
- c_{8HDHB}\frac{e\, v^4\, \cos\theta_W}{8\, \sin^2\theta_W\, \Lambda^4}
- c_{8HDHW}\frac{e\, v^4}{8\, \sin\theta_W\, \Lambda^4}
\\[1em]
\lambda^Z &= c_{WWW}\frac{6\, \sin\theta_W\, M_{W}^2}{e\, \Lambda^2} 
+ c_{8W}\frac{3\, \sin\theta_W\, M_{W}^2 v^2}{e\, \Lambda^4} 
+ c_{8W2}\frac{\sin^2\theta_W\, M_{W}^2\, v^2}{e\, \cos\theta_W\, \Lambda^4} 
\\[1em]
\lambda^\gamma &= c_{WWW}\frac{6\, \sin\theta_W\, M_{W}^2}{e\, \Lambda^2} 
+ c_{8W}\frac{3\, \sin\theta_W\, M_{W}^2 v^2}{e\, \Lambda^4} 
- c_{8W2}\frac{\cos\theta_W\, M_{W}^2 v^2}{e\, \Lambda^4} 
\end{align}
\end{subequations}

\section{Numerical Predictions}
\label{sec:Numerics}

In this section, we study SM and BSM predictions for $WW$ production
at the LHC, with fully leptonic decays of different flavours (electron
and muon). SM predictions for this process have been presented in our
previous work~\cite{Gillies:2024mqp}. Following from that work, in
this section we present results with a centre-of-mass energy
$\sqrt s=14\,$TeV, with jets reconstructed according to the anti-$k_t$
algorithm~\cite{Cacciari:2008gp} with a jet radius $R = 0.4$.  We
adopt the fiducial cuts on leptons and jets detailed in
table~\ref{tab:fiducial-cuts}. These are the cuts of the experimental
analysis performed by the ATLAS collaboration in~\cite{ATLAS:2019rob},
which we assume to also be similar for studies of this channel at the
HL-LHC.
\begin{table}[!htbp]
\begin{center}
\begin{tabular}{c|c}
  \hline\hline
  Fiducial selection requirement  & Cut value \\
  \hline\hline
  $p_T^{\ell}$ & $>27\,\mathrm{GeV}$ \\
  $|y_{\ell}|$ & $<2.5$ \\
  $M_{e\mu}$ & $>55\,\mathrm{GeV}$ \\
  $|\vec{p}_T^{\ e}+\vec{p}_T^{\ \mu}|$ & $>30\,\mathrm{GeV}$ \\
  Number of jets with $p_T> 35\,\mathrm{GeV}$ & 0 \\
  $\ETmiss$ & $>20\,\mathrm{GeV}$ \\
  \hline\hline
\end{tabular}
 \end{center}
 \caption{
   Definition of the $WW\rightarrow e\mu$ fiducial phase space, where
   $\vec{p}_T^{\ \ell},y_\ell$ are the transverse momentum and rapidity of either an
   electron or a muon, $M_{e\mu}$ is the invariant mass of the electron-muon
   pair, and $\ETmiss$ is the missing transverse energy.
 }
   \label{tab:fiducial-cuts}
\end{table}
For the following results, we set electroweak constants using the $G_\mu$ scheme. 
We use input parameters as given in table~\ref{tab:input-params}.
\begin{table}[!htbp]
\begin{center}
\begin{tabular}{c|c}
  \hline\hline
   Input Parameter & Value \\
  \hline\hline
  $G_\mu$ & $1.16637\times 10^{-5}\,\mathrm{GeV}^{-2}$ \\
  $M_W$ & $80.385\,\mathrm{GeV}$ \\
  $M_Z$ & $91.1876\,\mathrm{GeV}$ \\
  $m_t$ & $173\,\mathrm{GeV}$ \\
  $m_b$ & $4.66\,\mathrm{GeV}$ \\
  $M_H$ & $125\,\mathrm{GeV}$ \\
  $\Gamma_W$ & $2.093\,\mathrm{GeV}$ \\
  $\Gamma_Z$ & $2.4952\,\mathrm{GeV}$ \\
  $\Gamma_t$ & $1.4777\,\mathrm{GeV}$ \\
  $\Gamma_H$ & $4.07\times 10^{-3}\,\mathrm{GeV}$ \\
  \hline\hline
\end{tabular}
 \end{center}
 \caption{Input parameters used for the numerical results presented.}
   \label{tab:input-params}
\end{table}
In all predictions, care must be exercised in handling the interference with top production. We neglect
it in the present study by utilising a four-flavour scheme for parton
distribution functions, the
NNPDF31$\_$nnlo$\_$as$\_$0118$\_$luxqed$\_$nf$\_$4 PDF
set~\cite{Bertone:2017bme}.

We notice that the the cuts outlined in Tab.~\ref{tab:fiducial-cuts}
include the jet-veto condition that all jets must have a transverse
momentum smaller than $\ptjv=35\,$GeV. Whenever such a
condition is imposed, large logaritms $\ln(M_{WW}/\ptjv)$
arise at all-orders in QCD. We can resum those logarithms at the
next-to-NLL (NNLL) accuracy for the SM background and at NLL accuracy
for all BSM contributions.

\subsection{SM Predictions, including EW corrections}

\begin{figure}[htbp]
  \centering
  \includegraphics[width=.7\textwidth]{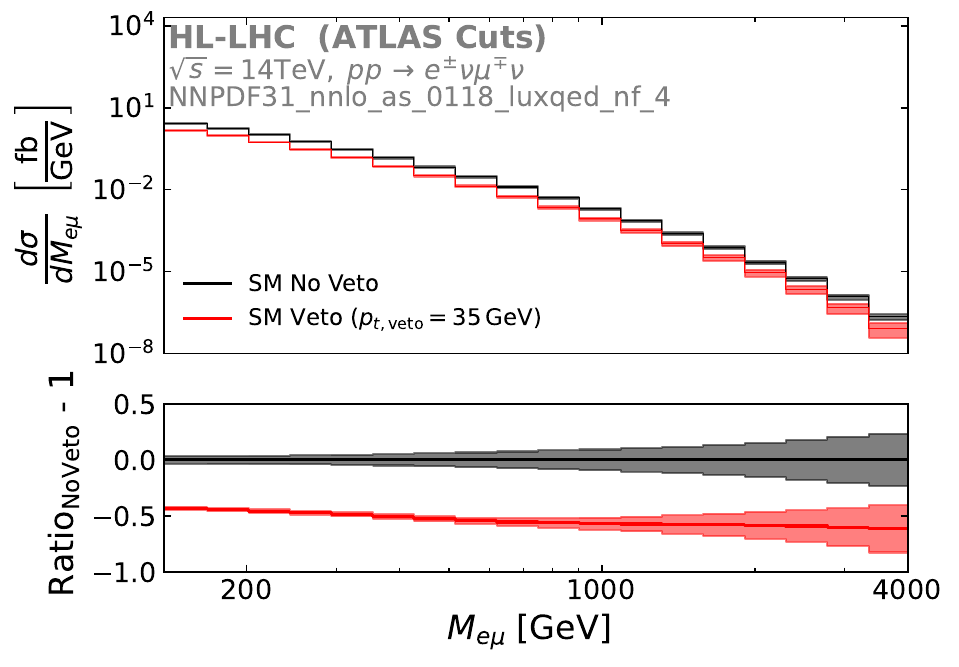} 
  \caption{SM predictions for $W^+W^-$ production at the LHC at $\sqrt{s}=14\,$TeV. 
  This is shown both with a jet-veto (red) at $\ptjv=35\,$GeV and without a jet-veto
  (black).}
  \label{fig:EW_Effects_total}
\end{figure}

It is known that, in the high-energy tails of diboson distributions,
EW effects become large and negative due to the presence of Sudakov
logarithms~\cite{Grazzini:2019jkl,Bierweiler:2013dja,Bierweiler:2012kw,Accomando:2004de,Denner:2000jv}.
In our previous work~\cite{Gillies:2024mqp} we included EW corrections
and found them to be large, but did not include an estimate of the
uncertainty from higher order combined QCD-EW corrections.  For this
study, we more carefully analyse the size of electroweak corrections
and include their uncertainties in our analysis.  As our
reference prediction, we consider the multiplicative scheme, defined in~\cite{Grazzini:2019jkl}
as:
\begin{equation}
	\label{eq:multiplicative}
	\mathrm{d}\sigma_{\mathrm{NNLL+NNLO\,\,QCD}\times \mathrm{EW}_{q\bar{q}}}=\mathrm{d}\sigma^{q\bar{q}}_{\mathrm{NNLL+NNLO\,\,QCD}}\big(1+\delta^{q\bar{q}}_{\mathrm{EW}}\big)+\mathrm{d}\sigma^{\gamma\gamma}_{\mathrm{NLO}}+\mathrm{d}\sigma^{gg}_{\mathrm{NLL}}\,,
\end{equation}
where $\delta^{q\bar{q}}_{\mathrm{EW}}$ are the NLO EW corrections to
the LO quark induced process. In order to ascertain the error arising
from such a scheme, this can be compared with the additive scheme,
also defined in~\cite{Grazzini:2019jkl}. In general, uncertainties
associated with missing EW corrections are demonstrated to be one of
the largest sources of errors for this channel, likely requiring
EW-resummation in the future. These errors are shown and compared with
other sources of error in Appendix~\ref{sec:ERRs}. The final SM
prediction, including EW theoretical errors, can be found in
figure~\ref{fig:EW_Effects_total}.

\subsection{Jet-veto effects on the $\gamma\gamma$ channel}
\label{sec:jetvetogammagamma}

\begin{figure}[htbp]
\centering
  \includegraphics[width=.7\textwidth]{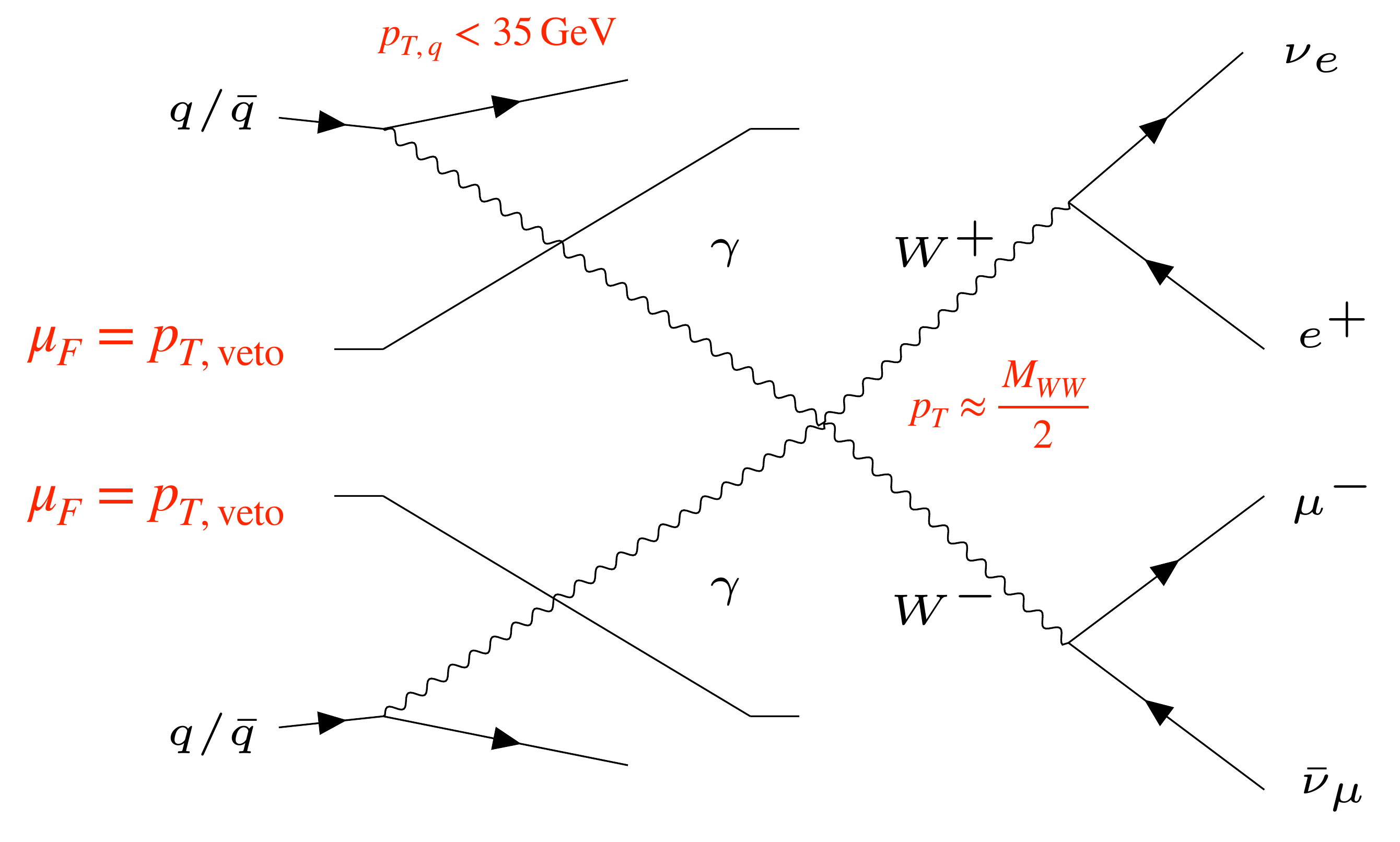}
  \caption{Pictorial representation of the factorisation of the PDFs from the hard process
  used for producing predictions with Monte Carlo event generators. The 
  hard process has energy equal to $M_{WW}/2$ which is usually the scale chosen 
  for the factorisation scale $\mu_F$. The production of a photon from a proton occurs
  via radiation from a quark. In the presence of a jet-veto we do not want any of the 
  partons to remain with a transverse mass greater than $\ptjv$. Therefore choosing
  a factorisation scale $\mu_F = \ptjv$ ensures that all radiation from the 
  proton remains below the jet-veto scale.}
  \label{fig:photonresummation}
\end{figure}

The $gg$ and $q\bar q$ initial states can radiate gluons and therefore
require a full resummation in the presence of a jet-veto. In the case
of the $\gamma\gamma$ contribution, the colourless nature of the
photons means that the NLL resummation can be performed using only the
PDFs. Initial state jets can only arise from the $\gamma\gamma$
contribution in the case that an incoming quark radiates a photon and
then goes on to form a jet. Such contributions contain collinear
logarithms that are resummed in the PDFs. In particular, all collinear
emissions with a transverse momentum smaller than the factorisation
scale are automatically included in the PDFs.  This is shown
pictorially in figure~\ref{fig:photonresummation}. As there are no
emissions with transverse momentum larger than $\ptjv$, we set the
factorisation scale equal to the jet-veto scale.

\begin{figure}[htbp]
\centering
  \includegraphics[width=.7\textwidth]{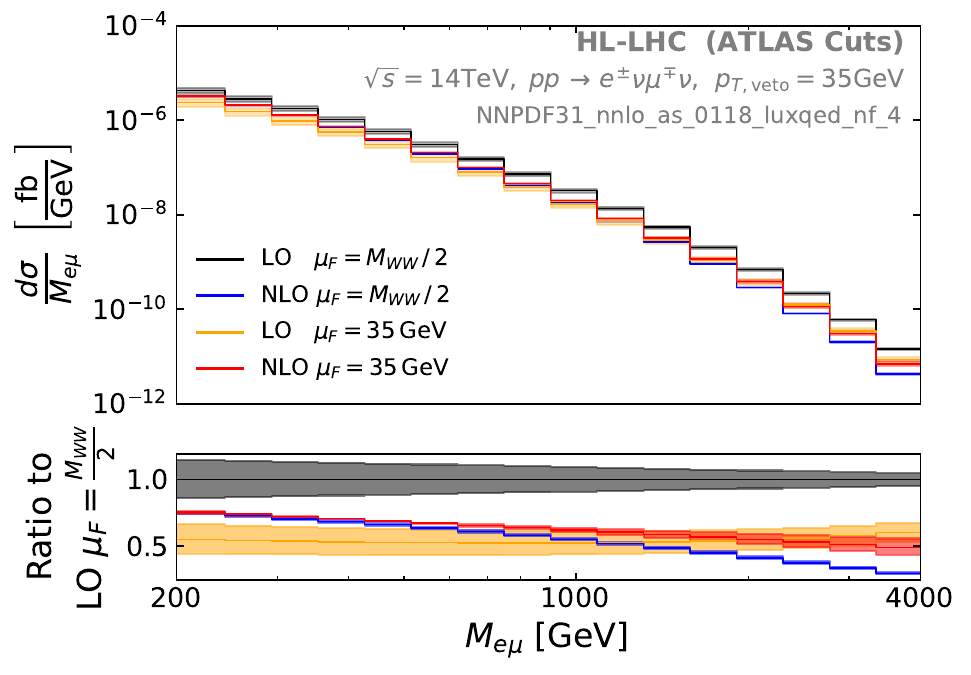}
  \caption{Comparison of the convergence between LO and NLO for the
  Standard Model photon induced contribution at HL-LHC ($14\,$TeV)
  with ATLAS cuts. This is shown with a jet-veto at $p_{T\,,\mathrm{veto}} = 35\,$GeV.
  The factorisation scale is either chosen to be dynamic with $\mu_F = M_{WW}/2$ or fixed at the
  jet-veto scale. With the dynamic scale, the convergence between LO (black) and NLO (blue)
  is poor. With the fixed scale, which automatically resums the jet-veto for the photon contribution,
  the NLO (red) and LO (yellow) agree within errors.
    }
  \label{fig:gammagammaconvergence}
\end{figure}

At NLO the picture becomes more complex, as we must apply the jet-veto
also to the real emission contributions. This gives rise to a term
$\ln(\mu_F/\ptjv)$, which vanishes if we set
$\mu_F = \ptjv$. This corresponds to reabsorbing all
logarithmically enhanced terms $\ln(\mu_F/\ptjv)$ into the PDFs. In
figure~\ref{fig:gammagammaconvergence}, the convergence of LO and NLO
EW is studied. There, we show the predictions in the presence of a
jet-veto with factorisation scale $\mu_F = \ptjv$, as
well as with the usual $\mu_F = \frac{M_{WW}}{2}$. It can be seen that
the convergence is much better for the $\mu_F = \ptjv$
case. After setting the factorisation scale to the jet-veto scale, we
can see the suppression that the jet-veto should cause for the photon
contribution already at LO. This is because, although photons are not
colour charged and are not affected by Sudakov suppression, they are
still affected by the veto through collinear logarithms. At NLO,
similar predictions are obtained, with better high-energy convergence
for the $\mu_F = p_{T, \mathrm{veto}}$ case.

\begin{figure}[htbp]
\centering
  \includegraphics[width=.7\textwidth]{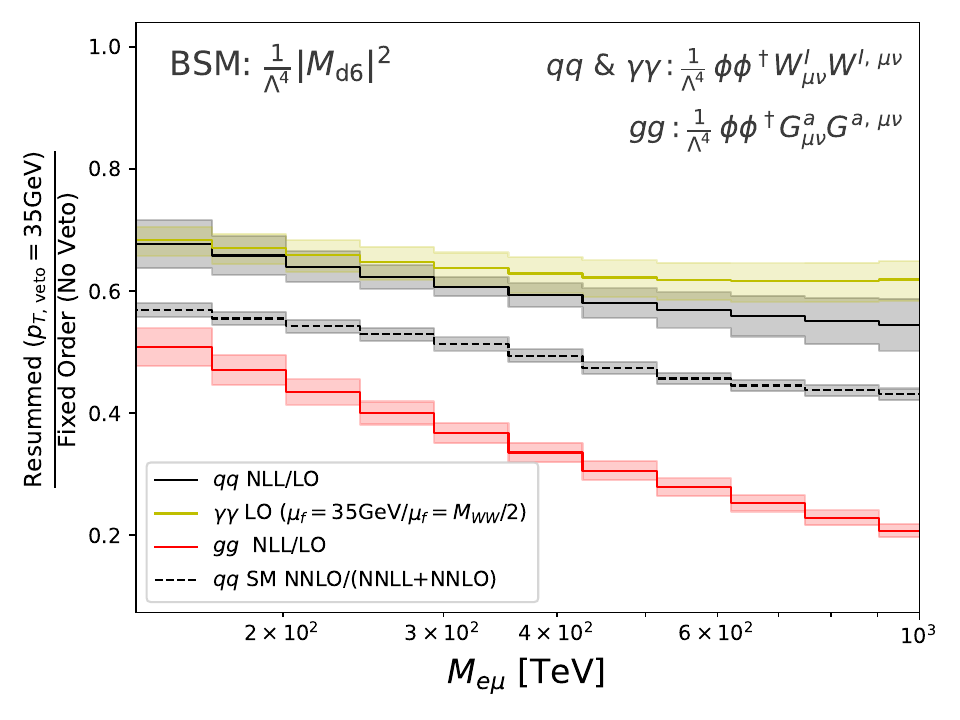}
  \caption{Comparison of the effect of the jet veto on different channels: quark induced (black),
  gluon induced (red) and photon induced (yellow). This is shown for the operators $\mathcal{O}_{GH}$
  and $\mathcal{O}_{HW}$. The ratio 
  of the jet-veto to no jet-veto case is also shown for the Standard Model $q\bar q$ channel for comparison
  (black dashed).
  }
  \label{fig:vetoeffectratios}
\end{figure}

Finally, we compare the effect of the jet veto on the $\gamma \gamma$
channel relative to the $q \bar q$ and $gg$ channels.  This is shown
in figure~\ref{fig:vetoeffectratios}. In order to attempt a ``fair''
comparison between the channels, we use two operators with the same
form:
($\mathcal{O}_{GH} = \phi^\dagger \phi\,G^a_{\ \mu\nu}G^{a\,\mu\nu}$
and
$\mathcal{O}_{HW} = \phi^\dagger \phi\,W^I_{\
  \mu\nu}W^{I\,\mu\nu}$). These generate $hWW$, $ggh$, and
$\gamma\gamma h$ contact interactions with the same Lorentz structure
as the $q\bar q$, $gg$, and $\gamma\gamma$-contributions
respectively. Typically, the jet-veto prohibits coloured particles
radiating jets, which they do proportionally to their colour
charge. Gluons have the largest colour charge and so they are affected
the most by the veto, as can be seen in
figure~\ref{fig:vetoeffectratios}. Meanwhile, although the photons are
not colour charged, missing real emissions with transverse momentum
larger than $\ptjv$ lead to a suppression with respect to the fully
inclusive case. This is particularly evident when setting the
factorisation scale to $\mu_F = \ptjv$, which resums the effects of
these emissions to all orders. This results in the fact that the BSM
$\gamma\gamma$ channel does not have a significant relative
enhancement with respect to the $q\bar q$ BSM channel in the presence
of a jet-veto. In fact, the quark channel suppression is milder that
that of the gluon channel, as quarks have a smaller colour charge. The
SM suppression is also shown in the figure as a reference.  Since in
the inclusive case the NNLO contribution gives a large K-factor, the
jet-veto suppression on the SM $q\bar q$ contribution appears much
larger than the suppression on the BSM $q\bar q$ contribution. This
has implications for the ability of uncertainties based on variation
of scales to profile the true uncertainty on the $q\bar q$ BSM
fixed-order contributions.

\subsection{BSM Predictions with Dimension-6 Operators}
\label{sec:bsm_preds}

In this section, we present predictions for the $M_{e\mu}$
distributions with dimension-6 operators at both SM-interference and
squared amplitude level. We show predictions at a reference value of
$\Lambda=2\,$TeV and $c_i = 1$ , for the ATLAS cuts from
table~\ref{tab:fiducial-cuts}, and at a centre-of-mass energy of
$\sqrt s=14\,$TeV. For the $q\bar q$ predictions, we obtain NLL
accuracy from MCFM by matching the operators to the anomalous triple
gauge couplings as shown in equation~\eqref{eq:dictionary}.  For the
$\gamma\gamma$ case, we use the LO prediction from
MadGraph~\cite{Alwall:2014hca} using a UFO~\cite{Darme:2023jdn} file,
ensuring to set $\mu_F = p_{T, \mathrm{veto}} = 35\,$GeV in order to
correctly model the effect of the jet veto on this channel. For the
dimension-6 $q \bar q$ contributions, we also show the $gg$
contribution arising from the dimension-6 operator
$\mathcal{O}_{GH} = \phi^\dagger\phi\ G^a_{\ \mu\nu}G^{a\,\mu\nu}$,
with $G^a_{\mu\nu}$ the field strength corresponding to the gluon
field.  This is to compare the size of the $gg$ channel's dimension-6
squared and interference contribution to that of $q\bar q$ and
$\gamma\gamma$.
  
\begin{figure}[htbp]
  \centering
  \includegraphics[width=.49\textwidth]{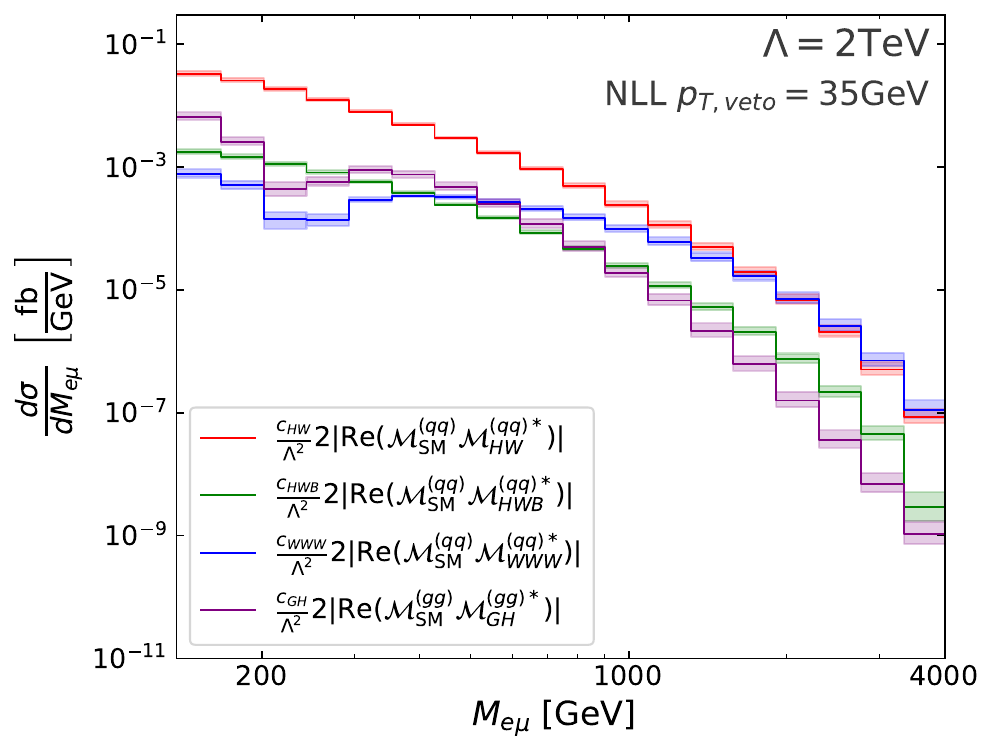}
  \includegraphics[width=.49\textwidth]{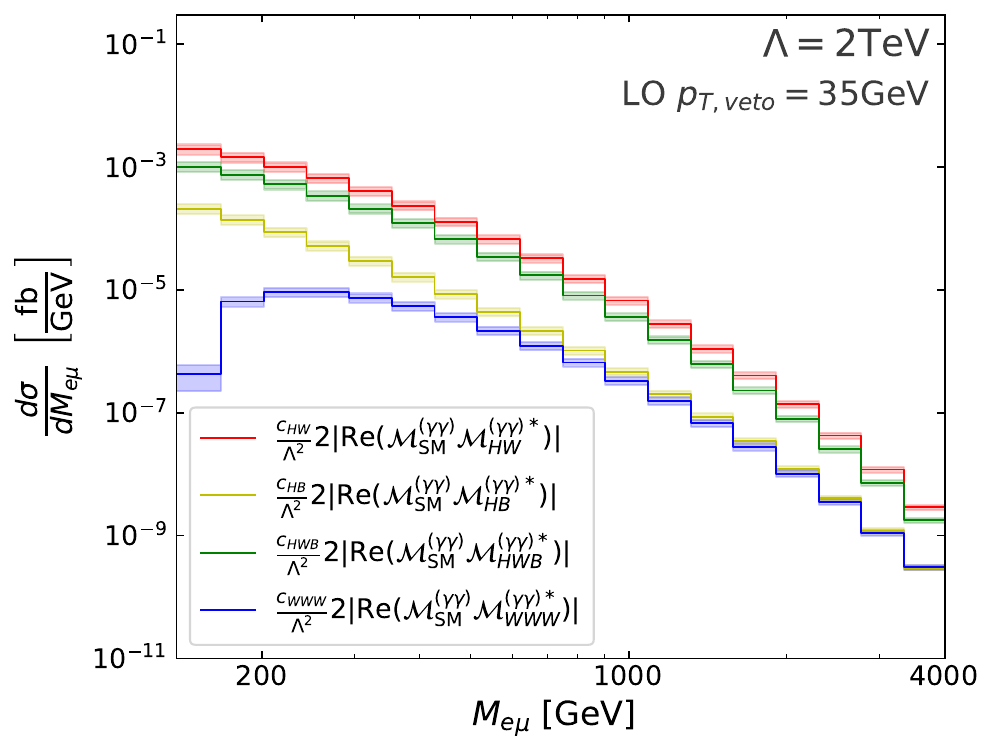}  
  \caption{Comparison of dimension-6 interference of the $q \bar q$ (left) and $\gamma\gamma$ (right) contributions at EFT mass scale $\Lambda=2\,$TeV. The $q \bar q$ contributions
    are shown at (NLL) accuracy with a jet-veto resummation, whilst the $\gamma\gamma$ channel is shown at LO with $\mu_F = p_{T,\mathrm{veto}}$. Operator $\mathcal{O}_{HB}$ only contributes to the $\gamma\gamma$ channel.}
  \label{fig:BSMOperatorPrediction_d6int}
\end{figure}

First, in figure~\ref{fig:BSMOperatorPrediction_d6int} we show the
interference between dimension-6 and SM for the $q\bar q$ and
$\gamma\gamma$ channels.\footnote{Note that interference contributions
  can be negative.  Since we want to plot them in logarithmic scale,
  we have decided to plot their absolute value.  These leads to
  apparent discontinuities in
  figures~\ref{fig:BSMOperatorPrediction_d6int}
  and~\ref{fig:BSMOperatorPrediction_d8vvvint}} To compare the size of
the photon channel relative to the quark channel, we treat each
operator in turn. For operator $\mathcal{O}_{HW}$, the
$\gamma \gamma$-induced interference term, at low energies, is around
$6\%$ of the corresponding one for the $q\bar q$ channel. This drops
to $2-3\%$ at higher energies. Operator $\mathcal{O}_{HB}$ does not
contribute to the $q\bar q$ channel.  Its $\gamma \gamma$ contribution
is an order of magnitude smaller than that of the $\mathcal{O}_{HWB}$
$q\bar q$ contribution. The $\gamma \gamma$-induced interference is
most prominent for the operator $\mathcal{O}_{HWB}$, being between
15\% and $60\%$ of the $q\bar q$ contribution across the range of
dilepton masses considered. For operator $\mathcal{O}_{WWW}$, the
$\gamma\gamma$ contribution is mostly less than $1\%$ of the $q\bar q$
contribution, except at energies of $200-400\,$GeV, where it is around
$5\%$ of that contribution. Overall, the $\gamma\gamma$ contribution
cannot be neglected if the desired accuracy is less than a few percent.

\begin{figure}[htbp]
  \centering
  \includegraphics[width=.49\textwidth]{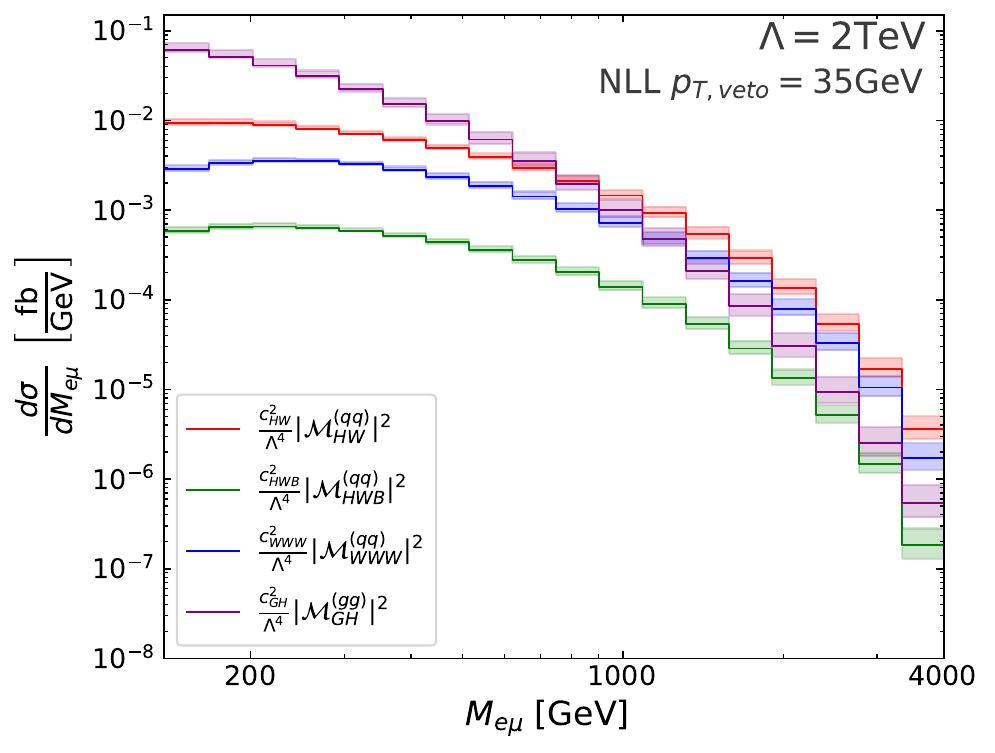}
  \includegraphics[width=.49\textwidth]{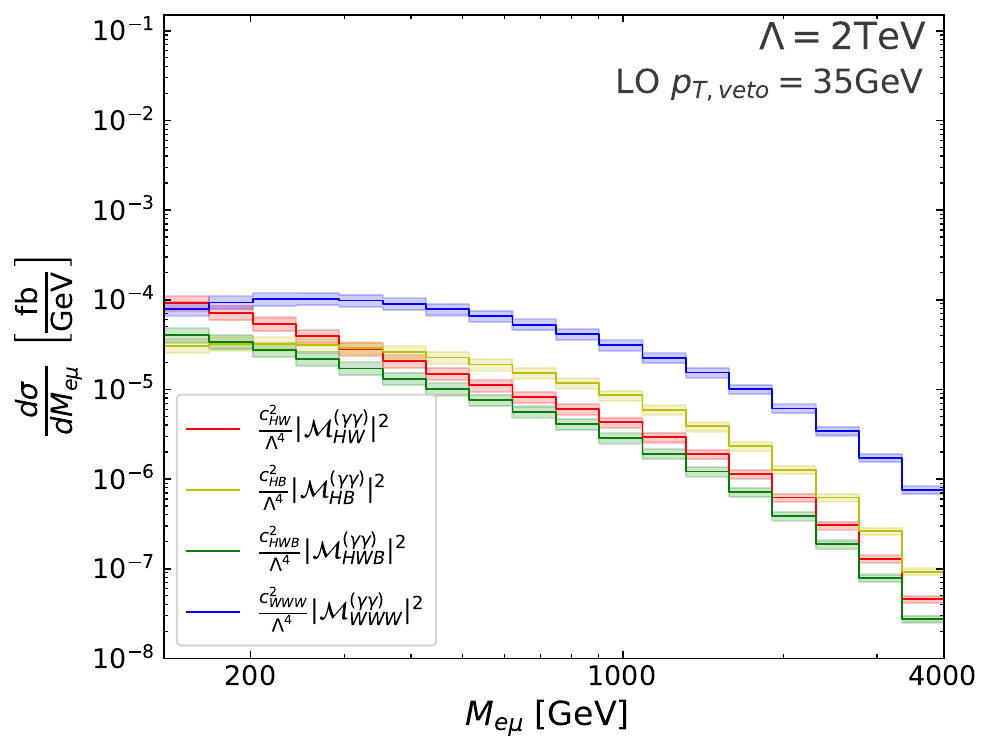}  
  \caption{Comparison of dimension-6 squared contribution for the $q \bar q$, $gg$, (left) and $\gamma\gamma$ (right) channels at EFT mass scale $\Lambda=2\,$TeV. The $q \bar q$ contributions
    are shown at (NLL) accuracy with a jet-veto resummation, whilst the $\gamma\gamma$ channel is shown at LO with $\mu_F = p_{T,\mathrm{veto}}$. Operator $\mathcal{O}_{HB}$ only contributes to the $\gamma\gamma$ channel.}
  \label{fig:BSMOperatorPrediction_d6sq}
\end{figure}

The $\gamma\gamma$ SM contribution is much smaller that of the SM
$q\bar q$ channel. For that reason, in
figure~\ref{fig:BSMOperatorPrediction_d6sq} we also consider the
squared dimension-6 contributions.  For operator $\mathcal{O}_{HW}$,
the $\gamma\gamma$ contribution is less than $1\%$ of the
$q\bar q$ contribution. For operator $\mathcal{O}_{HWB}$, the
$\gamma\gamma$ contribution is between $5-10\%$ of the $q\bar q$
channel across the dilepton mass range. Similarly to its interference term, 
the $\mathcal{O}_{HB}$ squared
contribution is an order of magnitude smaller than that of the
$\mathcal{O}_{HWB}$ quadratic $q\bar q$ contribution. For the
$\mathcal{O}_{WWW}$ operator, the $\gamma\gamma$ contribution starts
at $3\%$ for low energies, rising up to $40\%$ in the final bin.

\newpage
\subsection{BSM Predictions with Dimension-8 Operators}
\label{sec:bsm_preds_d8}
\begin{figure}[htbp]
  \centering
  \includegraphics[width=.49\textwidth]{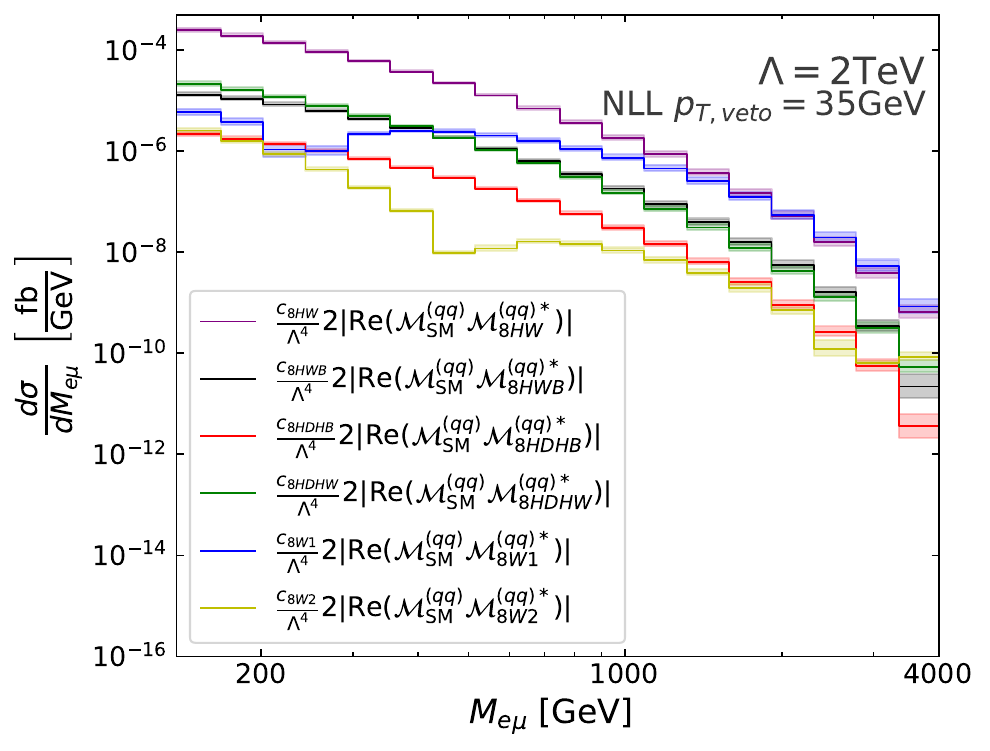}
  \includegraphics[width=.49\textwidth]{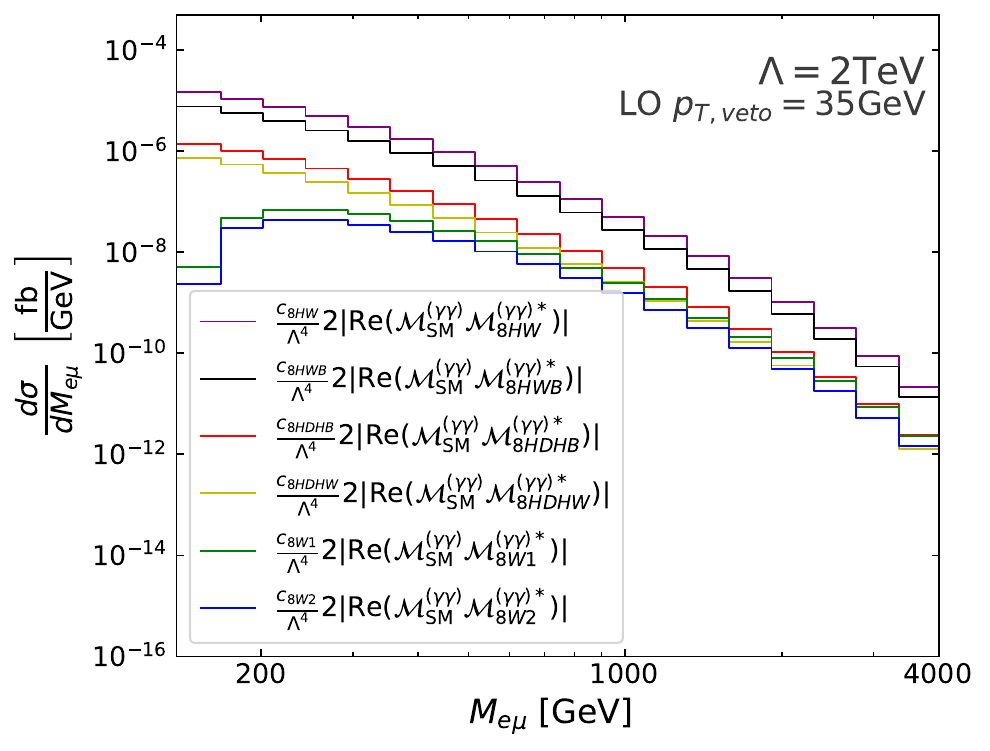}  
  \caption{SM interference contribution of the bosonic dimension-8 operators 
   for the $q\bar q$ (left) and $\gamma\gamma$ (right) channels, at the EFT mass scale $\Lambda=2\,$TeV.
    We only show dimension-8 operators which contribute to both $q\bar q$ and $\gamma\gamma$ channels.}
  \label{fig:BSMOperatorPrediction_d8vvvint}
\end{figure}

In this section, we present predictions for the $M_{e\mu}$ distribution with dimension-8
operators, under the same conditions as for the dimension-6
operators. At dimension-8, there are four operators which generate
$(\gamma/Z)\,W^+W^-$ triple gauge couplings. They can be directly
compared between the $q \bar q$ and $\gamma \gamma$ channels. The
comparison of the four operators' ($\mathcal {O}_{8HDHB}$,
$\mathcal {O}_{8HDHW}$, $\mathcal {O}_{8W1}$, and
$\mathcal {O}_{8W2}$) interference terms is shown in
figure~\ref{fig:BSMOperatorPrediction_d8vvvint}. 
We note that $\mathcal{O}_{8W1}$ has the same structure as $\mathcal{O}_{8WWW}$ for this channel, and the two 
cannot be distinguished at leading order.
It can be seen that, for
operators $\mathcal {O}_{8HDHB}$ and $\mathcal {O}_{8W2}$, the photon
and quark channels are of similar orders of magnitude. Meanwhile,
$\mathcal{O}_{8W1}$ and $\mathcal{O}_{8HDHW}$ have a negligible
$\gamma\gamma$ contribution.

\begin{figure}[htbp]
  \centering
  \includegraphics[width=.49\textwidth]{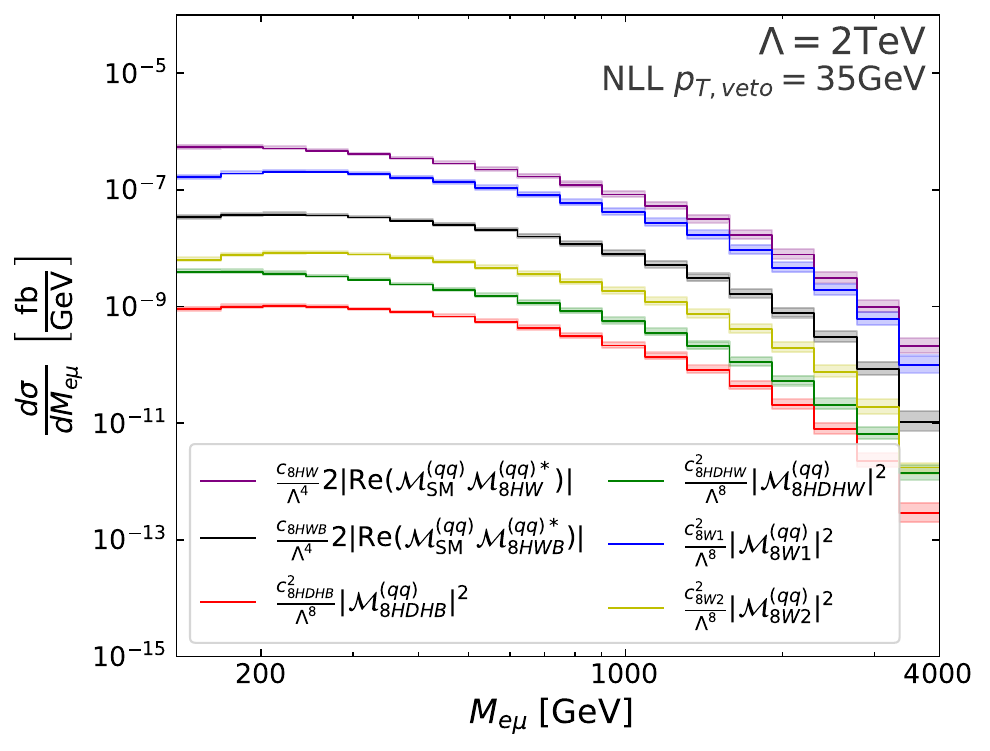}
  \includegraphics[width=.49\textwidth]{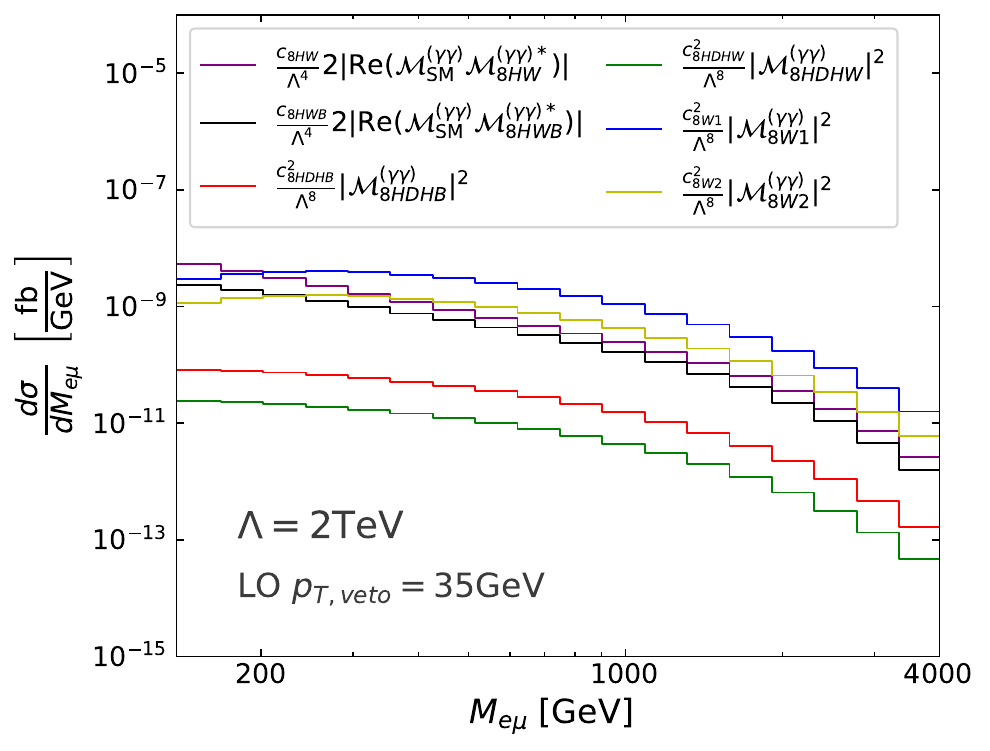}  
  \caption{Squared contribution of the bosonic dimension-8 operators 
   for the $q\bar q$ (left) and $\gamma\gamma$ (right) channels, at the EFT mass scale $\Lambda=2\,$TeV.
    We only show dimension-8 operators which contribute to both $q\bar q$ and $\gamma\gamma$ channels.
  }
  \label{fig:BSMOperatorPrediction_d8vvvsq}
\end{figure}

We also show the squared contributions from these operators in figure~\ref{fig:BSMOperatorPrediction_d8vvvsq}. 
The dimension-8 operators $\mathcal {O}_{8HDHB}$ and 
$\mathcal {O}_{8W2}$ have a larger relative photon contribution, and the $\mathcal{O}_{8W1}$ has a 
relevant photon contribution as well. The $\mathcal{O}_{8HDHW}$ operator instead continues to be dominated by the $q\bar q$ 
channel.

In the quark channel, the largest dimension-8 squared contribution is a factor of $1000$ smaller than the 
largest dimension-6 squared contribution, and does not grow any faster with energy. 
This is because the dimension-8 operators can all be mapped to the anomalous triple gauge couplings - 
albeit with additional pre-factors proportional to $\frac{v^4}{\Lambda^4} \approx \frac{1}{4000}$ 
($\frac{v^2}{\Lambda^2} \approx \frac{1}{64}$) for
the squared (interference) terms at an EFT scale of $\Lambda = 2\,$TeV. 

\begin{figure}[htbp]
  \centering
  \includegraphics[width=.49\textwidth]{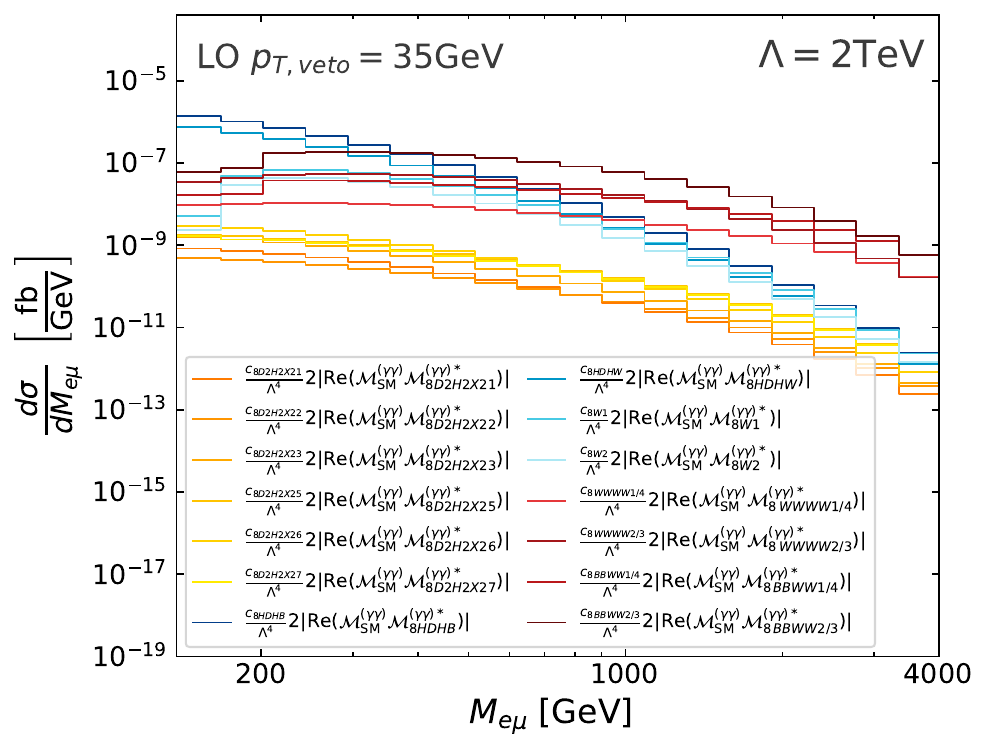}
  \includegraphics[width=.49\textwidth]{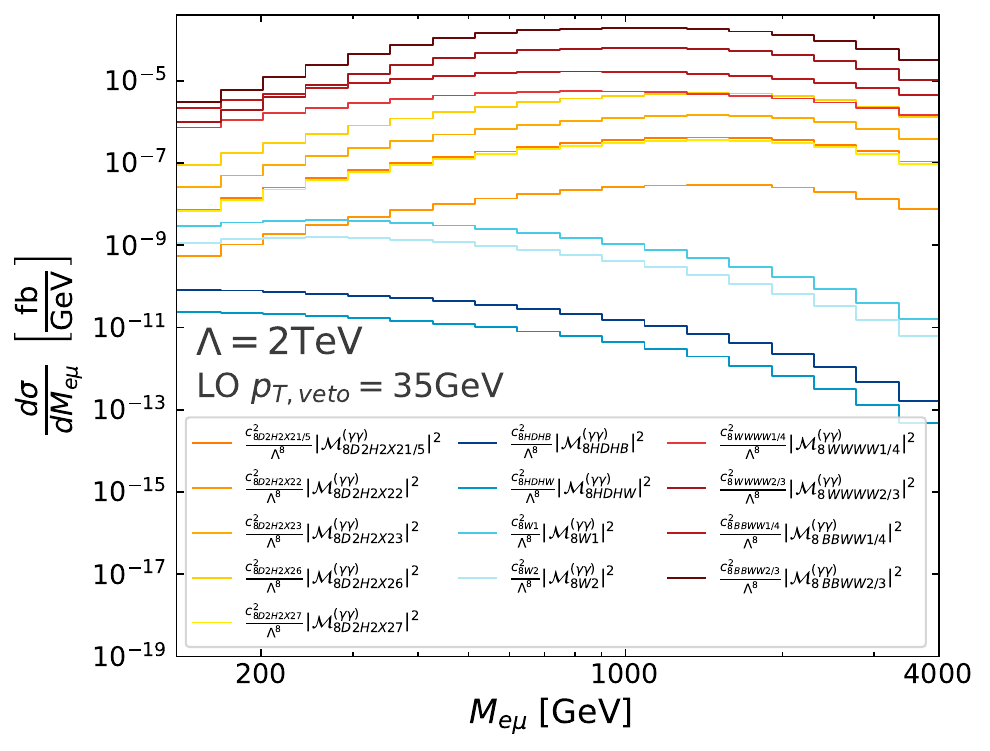}  
  \caption{Contribution of all the bosonic dimension-8 operators 
   for the $\gamma\gamma$ channel, at the EFT mass scale $\Lambda=2\,$TeV.
    On the left we show interference, and on the right we show squared contributions.
    We colour-code groups of operators according to their Lorentz structure.}
  \label{fig:BSMOperatorPrediction_d8gammagamma}
\end{figure}

Although the dimension-8 operators presented in
figure~\ref{fig:BSMOperatorPrediction_d8vvvsq} are negligible relative
to dimension-6 contributions, and do not have dominant $\gamma \gamma$
contributions, many more of the dimension-8 bosonic operators
contribute to the $\gamma \gamma$ channel than to the $q \bar q$
channel (see table~\ref{table:d_8operators}). Furthermore, those which
grow most rapidly with energy, and therefore have the biggest impact in
the tails of distributions, are the $\gamma\,\gamma\,WW$ four-point contact
operators. These have four powers of momentum in the coupling and
therefore grow rapidly with energy. In
figure~\ref{fig:BSMOperatorPrediction_d8gammagamma}, we compare the
interference of the different dimension-8 operators contributing to
the $\gamma \gamma$ channel. We have colour coded the operators into
different classes: blue for those which give triple gauge couplings only, 
red for operators with the form $VVVV$, and yellow for operators 
with the form $D\phi D\phi VV$.
In particular, we find that the four vertex generating ($VVVV$ and $D\phi D\phi VV$) bosonic
operators give the largest contribution (up to more than $1000$ times larger
than the triple-vertex operators). These could conceivably compete
with the dimension-6 contributions, and so should be studied further.

\newpage

\section{Constraining Dimension-6 Operators}
\label{sec:sensitivity}

In this section, we present constraints on the dimension-6 operators
in table~\ref{table:d_6operators} using current data, as well as
projections for future data. Unlike in the previous work~\cite{Gillies:2024mqp},
we choose not to constrain dimension-8 operators in this work. This is for 
two reasons, firstly the bosonic dimension-8 contributions from $q\bar q$ 
and $\gamma\gamma$ channels are not as large as for the dimension-8 $gg$ contributions.
The second reason is that unlike in the $gg$ case, where dimension-6 contributions
were already very well constrained from Higgs data~\cite{Gillies:2024mqp, ATLAS:2019nkf}, 
there is no competing process which 
allows us to set the dimension-6 Wilson coefficients to be negligible.

\subsection{Discussion of EFT Validity}

We first begin with a discussion of the EFT validity considerations
for this channel. This follows from the discussion in our previous
paper~\cite{Gillies:2024mqp}. The method we use involves the
comparison of dimension-6 and dimension-8 squared contributions in
order to establish the regime of EFT stability. This is done on a bin
by bin basis with each bin requiring a $\Lambda_{\min}$ given by:

\begin{equation}
   \label{eq:empirical_lambdamin}
   \Lambda_{\min} = (2\,\mathrm{TeV})\left(2\times\frac{\sigma^{(8)}_{\Lambda=2\,\mathrm{TeV}}}{\sigma^{(6)}_{\Lambda=2\,\mathrm{TeV}}}\right)^\frac{1}{4}. 
\end{equation}

For the above equation, we choose the largest squared dimension-6 and dimension-8 for a
given channel (e.g. $\gamma\gamma$ or $gg$). In the $\gamma\gamma$ case, we have found those to be
${O}_{WWW}$ and ${O}_{8BBWW3}$ respectively.

\begin{figure}[htbp]
\centering
  \includegraphics[width=.9\textwidth]{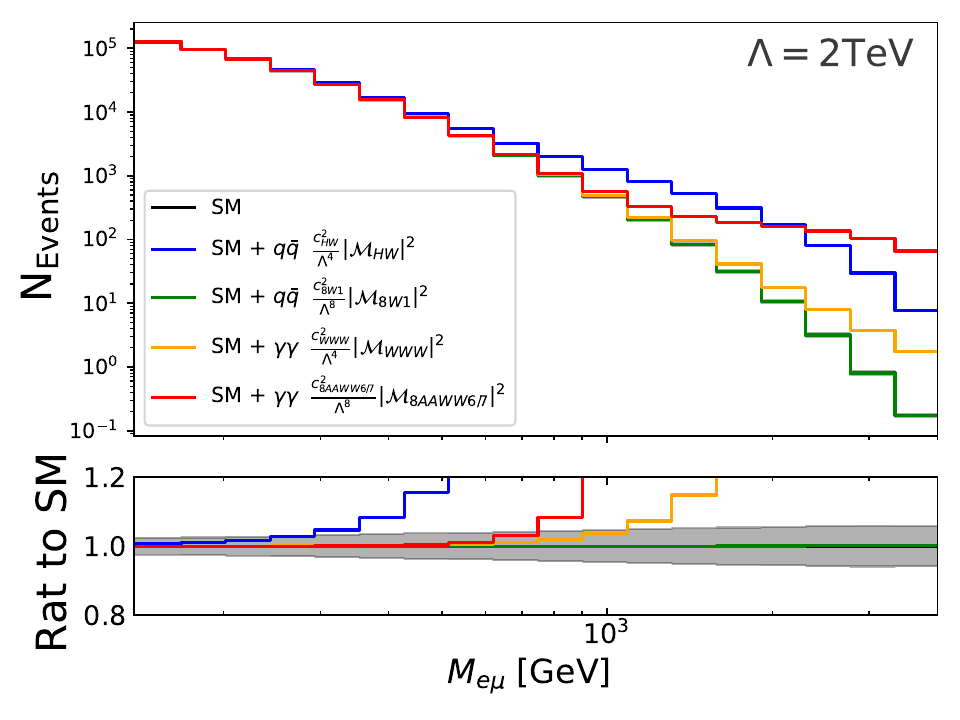}
  \caption{Comparison of bosonic dimension-8 squared contributions arising from the $q\bar q$ channel (green)
  and $\gamma\gamma$ channel (red). The SM (black) and dimension-6 squared contributions are shown 
  for both these channels (blue and yellow respectively) for comparison. The dimension-8 $q\bar q$ channel 
  does not grow rapidly with energy, whereas the dimension-8 $\gamma\gamma$ grows much more rapidly with energy
  and competes with the dimension-6 $q\bar q$ contribution at $M_{e\mu} = 2\,$TeV.}
  \label{fig:dim6dim8eftcomparison}
\end{figure}

In figure~\ref{fig:dim6dim8eftcomparison}
we compare the sizes of the dimension-6 and dimension-8 contributions for both the $q\bar q$ and $\gamma\gamma$ channel.
We set $\Lambda = 2\,$TeV as a representative benchmark scale for potential fits. 

\newpage
As was noted earlier, the bosonic dimension-8 $q\bar q$ contribution
does not grow rapidly with energy and comes with a variety of
suppressive factors arising from Higgs insertions and the EFT
mass-scale suppression.  We argue that by considering only the
$q\bar q$-initiated bosonic effects on this channel, dimension-8
contributions are vastly underestimated.

\begin{figure}[htbp]
\centering
  \includegraphics[width=.9\textwidth]{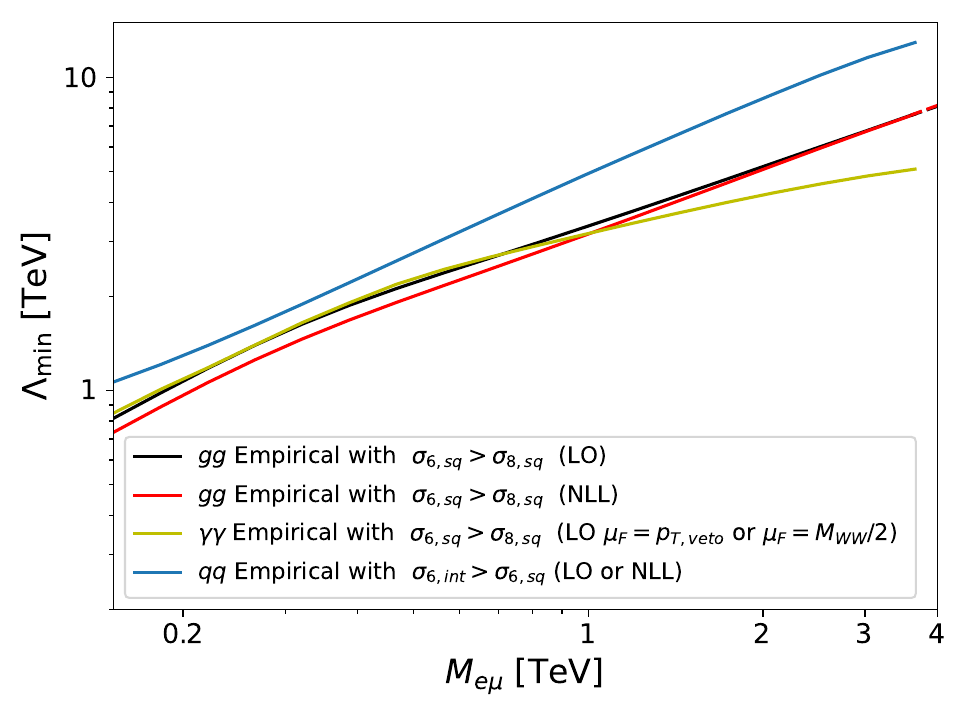}
  \caption{Minimum EFT validity scale ($\Lambda_{\min}$) required for
    a given $M_{e\mu}$ bin as required by
    equation~\eqref{eq:empirical_lambdamin}. This is calculated using
    the $gg$ or $\gamma\gamma$ dimension-6 and dimension-8 squared
    contributions. For $gg$ this is shown for NLL with a jet-veto and
    at LO without a jet veto.  We also show $\Lambda_{\min}$ obtained
    from the $q\bar q$ channel, by comparing interference and squared
    contributions.}
  \label{fig:min_lambda_with_mll}
\end{figure}

Instead, we can estimate the scale of EFT breakdown using the
$\gamma \gamma$ channel.  The dimension-8 fermionic operators which
induce $qqWW$ contact interactions do likely grow with
energy. However, an exploration of the size of these operators is
beyond the scope of this work. A viable alternative would be to compare the
$q\bar q$ dimension-6 interference and squared contributions, but this
approach is dependent on how well the dimension-6 operator interferes
with the SM amplitude. We instead compare the scale of EFT breakdown
for each bin arising from the $\gamma \gamma$ and $gg$ channels in
figure~\ref{fig:min_lambda_with_mll}. We also show the constraint
obtained from bosonic operators in the $q\bar q$ channel for
reference. The scale given by the $\gamma \gamma$ channel follows
closely that of the $gg$ channel at LO up to around $1\,$TeV. 
At higher energies, the difference in the ratio of dimension-6 and 
dimension-8 could arise from differences in the PDFs of the 
two channels. It could also arise from the fact that the dimension-6 operators used for the
$\gamma \gamma$ channel is the larger $\mathcal{O}_{WWW}$
operator. This has a different Lorentz structure with respect to the
$\mathcal{O}_{GH}$ operator, the only one that gives the $gg$ channel
contribution (via the exchange of a virtual Higgs). 
\newpage


Using the minimum EFT scale from the $\gamma\gamma$ channel, we can constrain dimension-6 operators
confidently using only the bins which satisfy EFT validity for a given
$\Lambda$. We then take the best SM prediction found from our previous
paper~\cite{Gillies:2024mqp} (now setting $\mu_F = 35\,$GeV for the
NLO $\gamma\gamma$ contribution in the jet-veto case). For a set of
$c_i$ (which we take to mean $c_i/\Lambda^N$ any time the inverse
powers of $\Lambda$ are suppressed), a prediction at either the LHC or
HL-LHC which we call $\{m_j\left(c_i\right)\}$ can be made. We can then compare
this to data points $\{n_j\}$. For the LHC, we take this data from
ATLAS~\cite{ATLAS:2019rob}. However for the HL-LHC sensitivity studies
$\{n_j\}$ are obtained from the best current SM predictions.

For the generation of exclusion plots and sensitivity studies we then use a delta chi-squared test statistic defined as:
\begin{equation}
  \label{eq:deltachisq}
    \Delta \chi^2\left(c_i\right) \equiv \chi^2\left(c_i\right) - \chi^2\left(\hat c_i\right)\,,
\end{equation}
where $\chi^2\left(c_i\right)$ is defined as:
\begin{equation}
  \label{eq:chisq}
    \chi^2\left(c_i\right) \equiv \sum_{j=1}^N \frac{\left(n_j - m_j\left(c_i\right)\right)^2}{(\Delta m_j)^2}\,,
\end{equation}
and $\hat c_i$ are values of the considered $c_i$ which
minimise $\chi^2\left(c_i\right)$.

At dimension-6 there are four operators to be considered. In order to
explore their four-dimensional parameter space and understand their
correlation (generated by interference), we use a Markov chain Monte-Carlo (MCMC)
approach, implemented via the python package~\texttt{emcee}~\cite{Foreman_Mackey_2013}. The
algorithm accepts or rejects new positions in the parameter space
generated by $c_i$ based on the probability distribution $p(\{c_i\})$.
  When performing fits with
interference terms only, we do not allow the interference to become larger
than the squared terms to avoid negative predictions for differential
cross sections.
To facilitate reproducibility of our results, we give some technical details of the procedure.
As a starting point, we generate random values of $c_i$ using a uniform distribution over the range
$|c_i| < 4\pi$, so that Wilson coefficients never grow larger
than the strongly coupled limit. Then, the probability distribution is
updated according to the $\chi^2\left(\{c_i\}\right)$ of the point
sampled, as follows
 \begin{equation}
 	\ln\left(p_{\mathrm{Next}}\right) = \ln\left(p_{\mathrm{Prev}}\right) - 0.5\,\chi^2\left(\{c_{i,\ \mathrm{Prev}}\}\right) -\sum_j \ln\left[\Delta m_j \left(\{c_{i,\ \mathrm{Prev}}\}\right)\right]\, ,
 \end{equation} 
 where $p_{\mathrm{Next}}$ is the updated probability distribution and
 $ p_{\mathrm{Prev}}$ is the previous probability distribution used to
 accept or reject the point $\{c_{i,\ \mathrm{Prev}}\}$.  Assuming the
 underlying probability is approximately gaussian, this correction
 will produce a realistic estimate for the value of
 $p_{\mathrm{Next}}$.  This process is repeated for $64$ sets of
 initial conditions (walkers). For each walker, we generate $200000$
 points and discard the first $50000$ as the burn-in phase. From these
 points, we produce corner plots with $2$-dimensional contour plots,
 with contours placed at $1\sigma, 2\sigma$ and $3\sigma$. We also
 show a $1$-dimensional projection for each operator with dashed lines
 showing the $2\sigma$ rejection point. This gives an individual
 constraint for each operator, as well as an estimate of the
 correlations for each operator.  Note that there could be situations
 where all four operators are large simultaneously but cancel due to
 interference effects; these will not appear in the 2-dimensional
 contour plots.

\subsection{Constraints from Current Data}
\label{sec:constraints_atlas}

In this section, we place constraints on dimension-6 operators using
2019 ATLAS data~\cite{ATLAS:2019rob} at a luminosity of
$36.1\,$fb$^{-1}$.  Recently, ATLAS has published results for $WW$
production at a luminosity of
$140\,$fb$^{-1}$~\cite{ATLAS:2025dhf}. This work employs $b$-quark
tagging in order to separate the $t\bar t$ background instead of using
a jet-veto. Fixed order results are then used for both signal and
background.  However, under this procedure the efficiency of the
$b$-tagging must be taken into account as well as cancellations
between real and virtual contributions. These are spoiled in the case
that an initial state gluon emission from a signal process decays into
a $b\bar b$ pair. We do not consider these effects here and leave a
more detailed study to future work.

This latest ATLAS analysis~\cite{ATLAS:2025dhf} also includes an analysis of a
subset of relevant SMEFT operators. In particular, the $c_{HWB}$ and
$c_{WWW}$ operators are constrained.  To deal with the breakdown of
the EFT method, ATLAS imposes a varying cut-off on the $M_{WW}$
distribution when generating the dimension-6 squared contributions
used in the fit.  Their results assume $\Lambda = 1\,$TeV, and
consider
$M_{WW} < M_{\mathrm{cut}} = \{0.75,\,1,\,1.5,\,2,\,2.5,\,3,\,\infty\}\,$TeV. 
The case where $M_{WW}$ is unbounded implies no EFT
validity considerations in the analysis. In order to check consistency
between our results and theirs, we also present results with no EFT
validity considerations and at a luminosity of
$140\,$fb$^{-1}$. However, to ensure consistency with our theoretical
modelling of the process, we use the older $36.1\,$fb$^{-1}$ data
set~\cite{ATLAS:2019rob} which uses a jet-veto, and rescale the number
of events so that the same data correspond to a luminosity of
$140\,$fb$^{-1}$. This procedure correctly accounts for the
statistical uncertainty, and is conservative for the
systematic uncertainty. The full results for no EFT validity are
presented in the appendix~\ref{sec:Appendix140}. We find that, using
the old ATLAS data with a luminosity of $140\,$fb$^{-1}$, we obtain a
constraint of $|c_{WWW}| < 0.4$ and $|c_{HWB}| < 1.2$ which is
comparable to the ATLAS results of around $|c_{WWW}| < 0.2$ and
$|c_{HWB}| < 2$ for $M_{\mathrm{cut}} = \infty$.  In general, we find
the hierarchy of constraints to be
$\mathcal{O}_{HW} > \mathcal{O}_{WWW} > \mathcal{O}_{HWB} >
\mathcal{O}_{HB}$. This agrees with the relative sizes of the
operators' squared contributions as found in
figure~\ref{fig:BSMOperatorPrediction_d6sq}. The fact that
$\mathcal{O}_{HB}$ is the smallest is reasonable, since it only
contributes via the subdominant $\gamma \gamma$ channel.

\begin{figure}[!htbp]
	\centering
  \includegraphics[width=.49\textwidth]{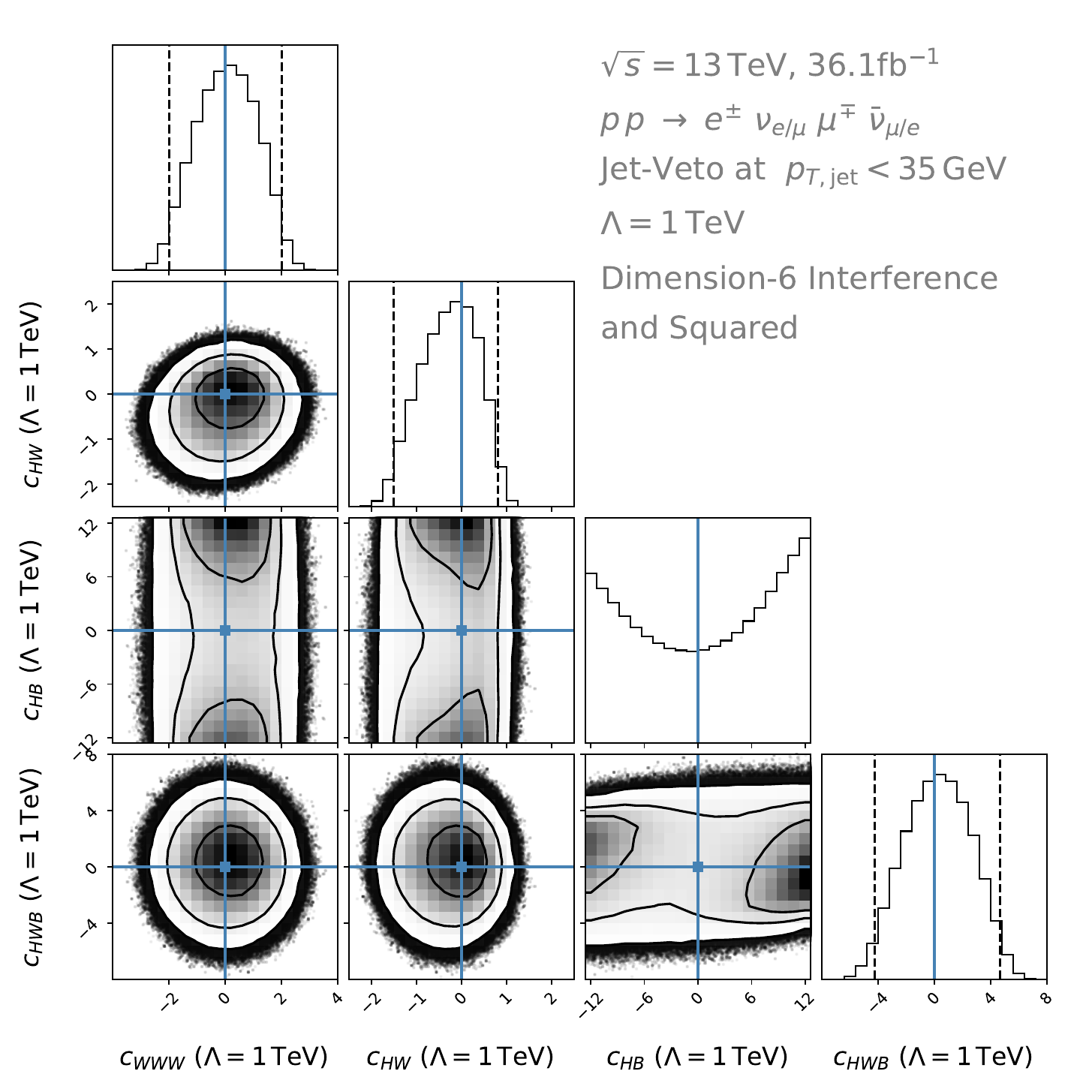}
  \includegraphics[width=.49\textwidth]{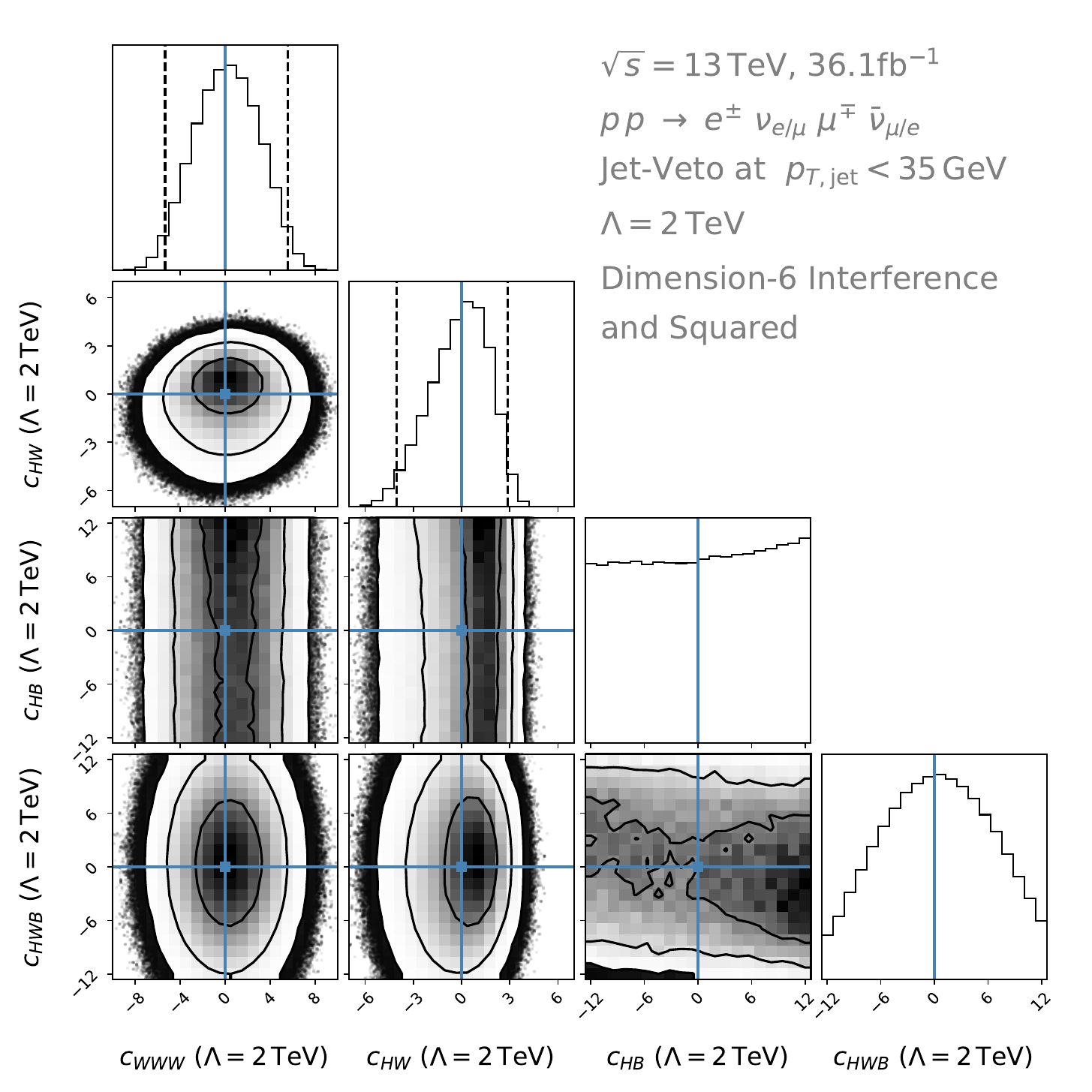}
  \caption{
 Sensitivity plots for $\{c_{HW}, c_{WWW}, c_{HWB}, c_{HB}\}$ using ($13\,$TeV, $36.1\,$fb$^{-1}$)
 ATLAS data~\cite{ATLAS:2019rob} for $WW$
  production with a jet-veto ($p_{T, \mathrm{veto}}=35\,$GeV). 
  This sensitivity study includes the interference and squared contributions of new physics
  (order $1/\Lambda^4$ in the EFT expansion).
  The exclusion contours are placed at $1$, $2$, and $3\sigma$ and the dashed lines
  correspond to univariate constraints at $2\sigma$. The value of the EFT scale is 
  given by $\Lambda=1\,$TeV (left) and $\Lambda=2\,$TeV (right)
  respectively), which changes the number of bins allowed in the constraints on the 
  Wilson coefficients. 
    }
  \label{fig:EFTvalid_ATLAS}
\end{figure}

We now present constraints with EFT validity on the 2019 ATLAS
dataset~\cite{ATLAS:2019rob} in figure~\ref{fig:EFTvalid_ATLAS}.  In
order to correctly deal with EFT validity, we take the approach that
the Wilson coefficients ($c_i$) are fit for varying values of
$\Lambda$. This allows for full information on $c_i$ and $\Lambda$ to
be given to compare with explicit models. As explained in
section~\ref{sec:sensitivity}, the value of $\Lambda$ sets the number
of bins that can be used to constrain $c_i$. We recall that, in this
and all subsequent contour plots, the coeffiecients $c_i$ are never
allowed to be larger than $4\pi$, the limit of validity of perturbation
theory. We present results at $\Lambda = \{1, 2\}\,$TeV. Due to the
fact that the SM prediction does not perfectly fit the data for all
bins, the best fit values are not usually found by turning off all of
the operators but instead in those regions of phase space where the
cross section is enhanced slightly. However, the SM prediction always
agrees with the data within $2\sigma$.

For $\Lambda=1\,$TeV, we find that it is possible to place constraints
on $c_{WWW}$, $c_{HW}$, and $c_{HWB}$.  As seen in
figure~\ref{fig:BSMOperatorPrediction_d6int}, at lower values of
$M_{e\mu}$, the operators $\mathcal{O}_{WWW}$ and $\mathcal{O}_{HWB}$
do not interfere well with the Standard Model. This leads to
approximately symmetric constraints for the first two operators at
around $|c_{WWW}| < 2$ and
$|c_{HWB}| < 4$.  In contrast, operator
$\mathcal{O}_{HW}$ interferes strongly with the SM leading to an
asymmetric constraint $-1.5 < c_{HW} < 1$.

It is not possible to constrain the operator $\mathcal{O}_{HB}$ using this 
dataset. Since the pure SM prediction does not exactly match the data
at lower energies, adding the small $c_{HB}$ squared contribution minimises the chi-squared
at $\Lambda=1\,$TeV, however the plot becomes flatter at $\Lambda=2\,$TeV as 
more bins are added to the analysis. In either case this effect is not significant
as the SM still lies within $2\,\sigma$. 

At $\Lambda=2\,$TeV, it is no longer possible to put constraints on $c_{HWB}$ within the strongly coupled limit.
As expected, constraints for $c_{WWW}$ and $c_{HW}$ also become weaker at this energy. The interference of $\mathcal{O}_{WWW}$ 
with the SM remains small at the energy ranges considered, leading to a symmetric constraint 
$|c_{WWW}| < 5$. The constraint on $c_{HW}$ becomes $-4 < c_{HW} < 3$.

\subsection{Sensitivity at the HL-LHC}

\begin{figure}[!htbp]
	\centering
  \includegraphics[width=.49\textwidth]{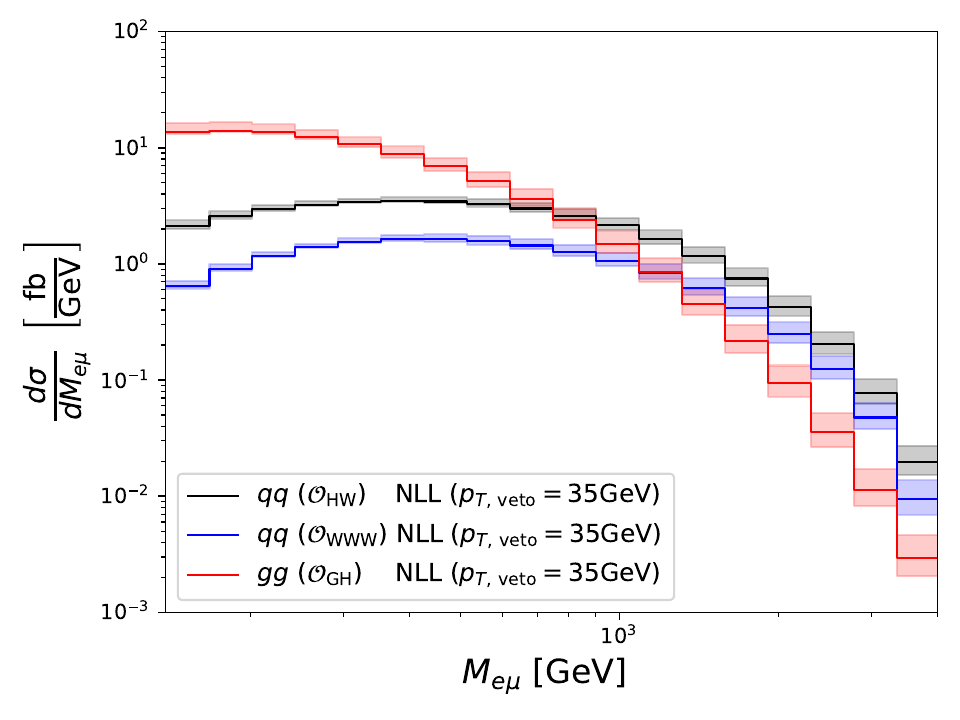}
  \includegraphics[width=.49\textwidth]{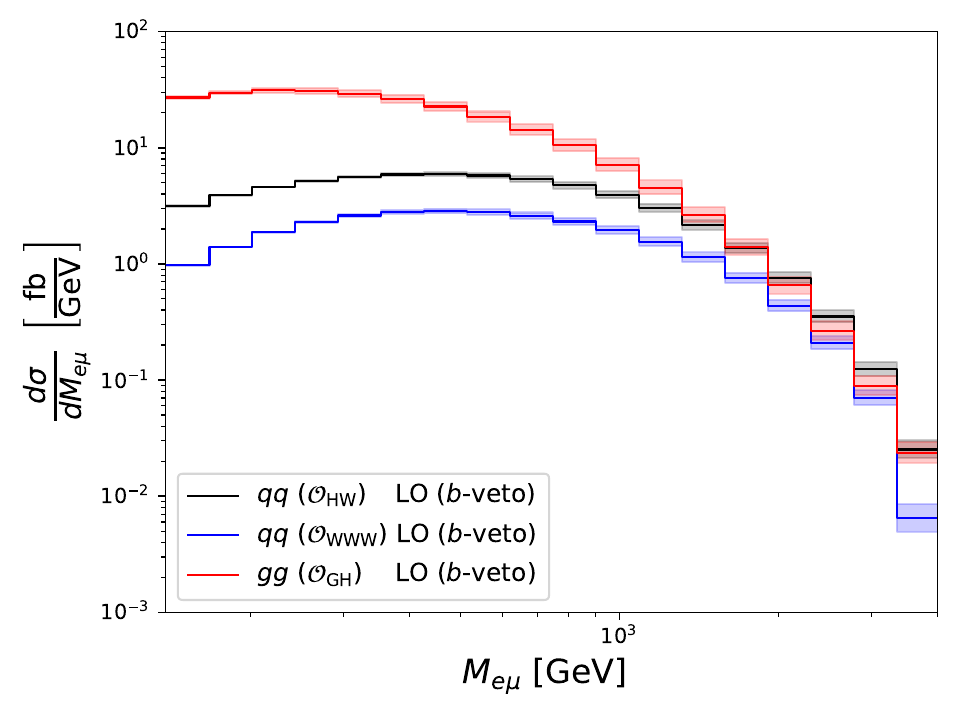}
  \caption{ Comparison of sizes of dimension-6 squared contributions
    from EFT operators which affect both the $gg$ and $q \bar q$
    channel. This is shown in the case of a jet-veto
    ($p_{T, \mathrm{veto}}=35\,$GeV) (left) and without a jet-veto
    (right).}
  \label{fig:ggqq_comparison}
\end{figure}

When properly considering EFT validity, we find that it is more
difficult than expected to put constraints on dimension-6 operators at
the HL-LHC. This is surprising given that it was possible to place
reasonably strong constraints on the subdominant (in the SM) $gg$
contribution in our previous paper~\cite{Gillies:2024mqp}. However,
when directly comparing the $q\bar q$ and $gg$ contributions at a
reference value of $1\,$TeV and $c_i = 1$ in
figure~\ref{fig:ggqq_comparison}, it can be seen that the $q \bar q$
contributions are generally smaller than the corresponding $gg$
contribution. At lower energies, where errors remain small, the $gg$
contribution is around ten times larger than the $q \bar q$
contribution both with and without a jet-veto. This is because
the $gg$ channel introduces an entirely new s-channel topology coupling
gluons directly to Higgs. Furthermore, at low energies the jet-veto 
has a smaller effect on the gluon parton luminosity which also happens to be
very high at low energies.
At higher values of $M_{e\mu}$,
the jet-veto suppresses the $gg$ contribution, which is lower than the
$q \bar q$ contribution.

\begin{figure}[!htbp]
	\centering
  \includegraphics[width=.7\textwidth]{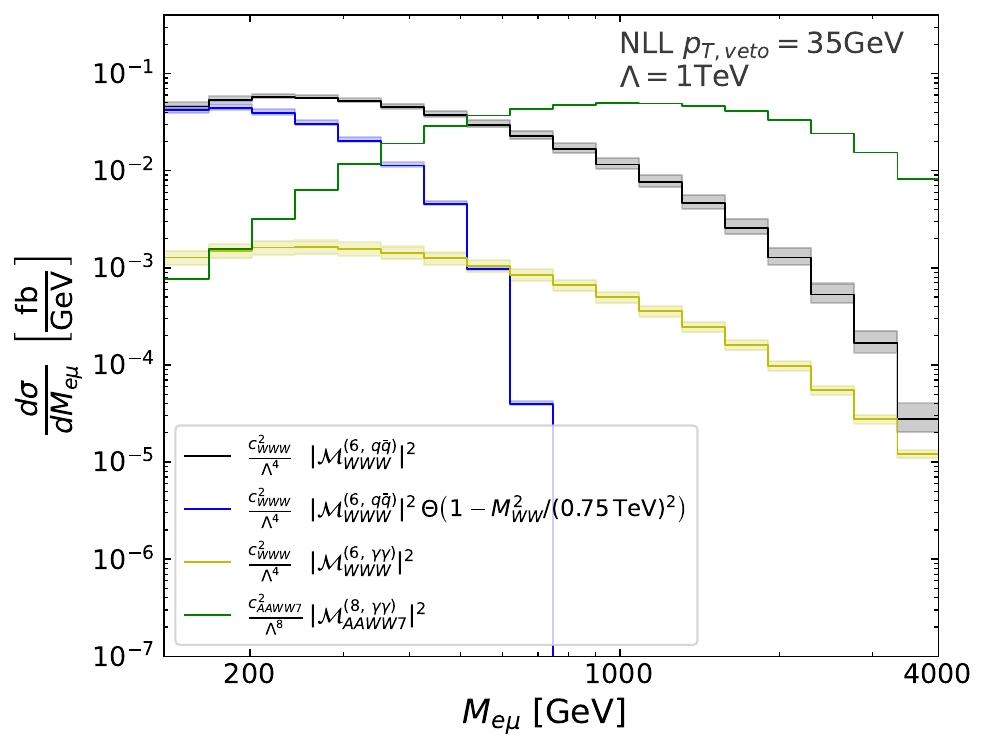}
  \caption{ Contributions of the $WWW$ dimension-6 operator to the
    $q\bar q$ and $\gamma \gamma$ channels, compared to that of the
    largest dimension-8 operator for the $\gamma \gamma$ channel. We
    also plot the same dimension-6 contribution to $q\bar q$ with a
    cut on $M_{WW}<750\,$GeV, as implemented in the simulation
    performed by ATLAS.}
  \label{fig:EFTvalid_comparison}
\end{figure}

As discussed in section~\ref{sec:constraints_atlas}, constraints have been placed on the 
operators $\mathcal{O}_{WWW}$ and $\mathcal{O}_{HWB}$ previously
by ATLAS~\cite{ATLAS:2025dhf}. However, this study did not fully consider the breakdown 
of the EFT when quoting constraints. 
They instead clipped the EFT simulation as $M_{WW} < M_{\mathrm{cut}}$ and
scanned over various values of $M_{\mathrm{cut}}$. 
However, this approach does not guarantee that the correct hierarchy
of EFT contributions.
Note that applying a cut on the EFT simulation alone is equivalent to
adding a form factor to the EFT coefficients as
$c_i \to c_i\,\Theta\left(1 - M_{WW}^2 /M_{\mathrm{cut}}^2\right)$
which makes it difficult to interpret the results of the fit in terms
of SMEFT~\cite{Gillies:2026tpc}. Furthermore, taking
$M_{\mathrm{cut}} = 0.75\,$TeV is probably the only case where the
dimension-6 contributions are likely to be larger than the dimension-8
contribution in any of the bins. This approach assumes similar
correlations between $M_{WW}$ and the observable being measured
($M_{e\mu}$) across different orders in the EFT, and this has been
shown not to hold for the $gg$ channel~\cite{Gillies:2026tpc}.  As
seen in figure~\ref{fig:EFTvalid_comparison}, with our bin-by-bin
comparision we would only use the first bin, whereas even with the
tightest cut that ATLAS, their procedure would include in the fit the
first three $M_{e\mu}$ bins, which are beyond the region of EFT
validity. The ATLAS paper~\cite{ATLAS:2025dhf} also considered bins
below $138\,$GeV, which we neglect to concentrate on the tails of the
distributions, but which could provide extra EFT information. Finally,
there are additional EW uncertainties (see appendix~\ref{sec:SMEW})
that reduce the constraining power of our analysis relative to
ATLAS's. For these reasons, our HL-LHC constraints in this section are
weaker than those quoted by ATLAS for current data at
$\Lambda=1\,$TeV, despite the fact that their data has a lower
luminosity.

\newpage

We first present the interference-only results (at order $1/\Lambda^2$
in the EFT expansion) for the HL-LHC with a jet veto
$p_{T, \mathrm{veto}} = 35\,$GeV.  We present results at
$\Lambda = 1\,$TeV and $\Lambda = 4\,$TeV and ensure the EFT expansion
is valid in all cases by restricting the bins used in the fits
appropriately.

For the $\Lambda=1\,$TeV case, we find that it is not possible to put
constraints on any coefficient other than $c_{HW}$. The operators
$\mathcal{O}_{WWW}$ and $\mathcal{O}_{HWB}$ interfere poorly at low
$M_{e\mu}$, and hence must be constrained from their squared
contributions. The constraint on $c_{HW}$ is approximately symmetric,
$|c_{HW}| < 1.2$.

At $\Lambda=4\,$TeV, higher $M_{e\mu}$ bins become available. It can be seen from figure~\ref{fig:BSMOperatorPrediction_d6int}
that, at values $M_{e\mu} > 1\,$TeV, the interference of operator $\mathcal{O}_{WWW}$ is no longer negligible. This
allows for constraints on $c_{WWW}$ at higher values of $\Lambda$. However, these are not particularly strong at 
$|c_{WWW}| < 6$. The constraint on $c_{HW}$ is also weakened slightly to 
$|c_{HW}| < 4$.
\begin{figure}[!htbp]
	\centering
  \includegraphics[width=.49\textwidth]{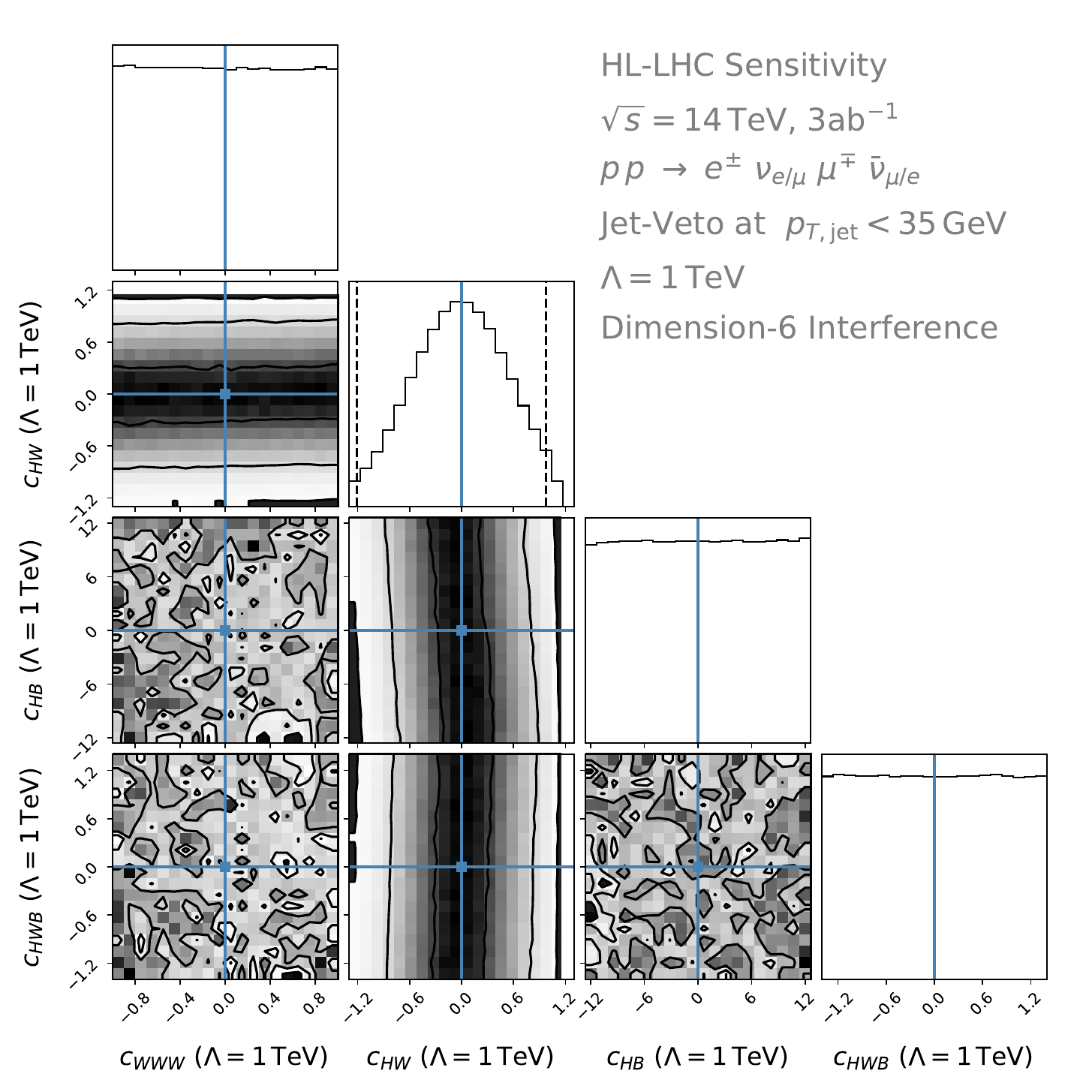}
  \includegraphics[width=.49\textwidth]{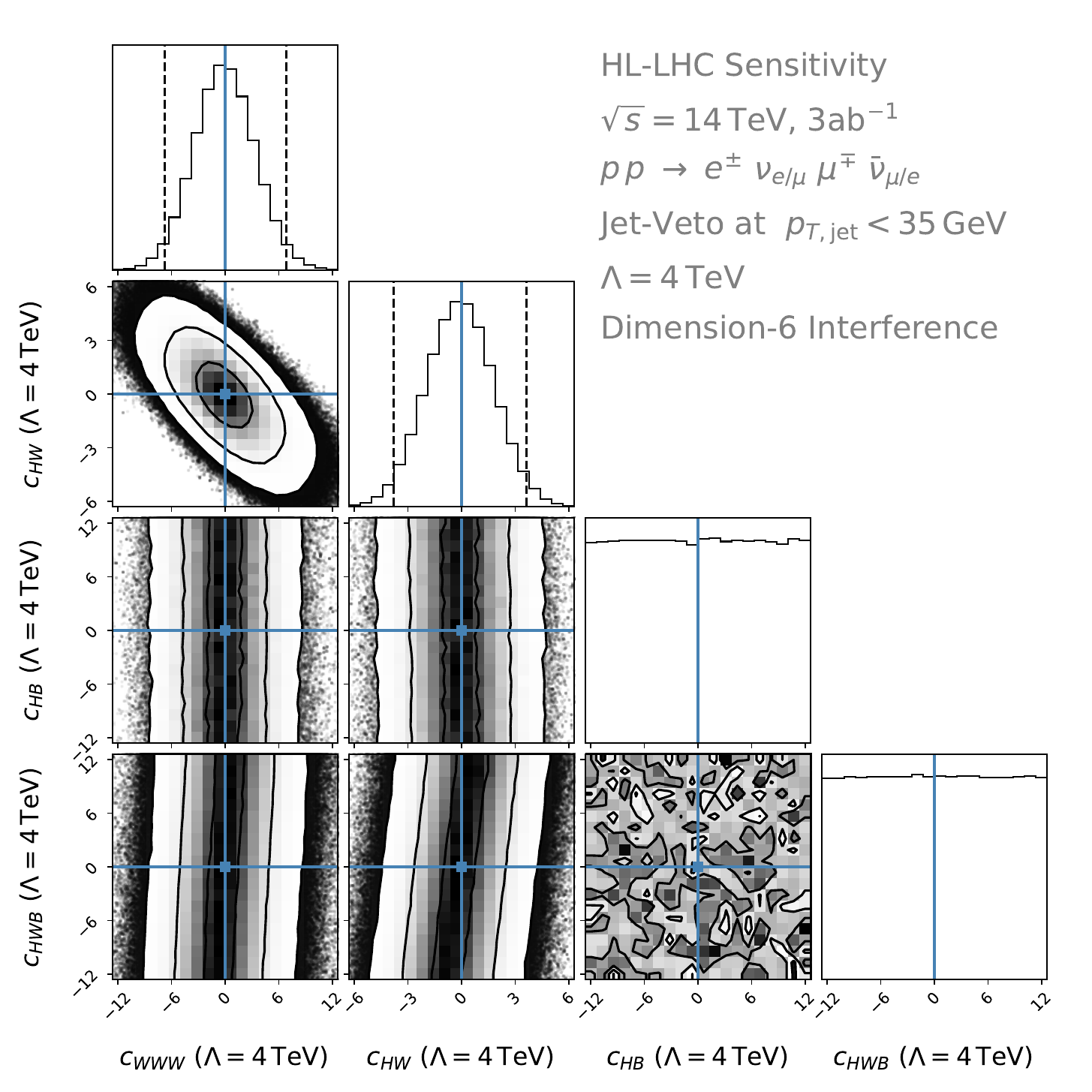}
  \caption{
 Sensitivity plots for $\{c_{HW}, c_{WWW}, c_{HWB}, c_{HB}\}$ at
   HL-LHC with ATLAS cuts ($14\,$TeV, $3\,$ab$^{-1}$) using $WW$
    production with a jet-veto ($p_{T, \mathrm{veto}}=35\,$GeV). 
    This sensitivity study only includes the interference of new physics with the 
    Standard Model (order $1/\Lambda^2$ in the EFT expansion).
    The exclusion contours are placed at $1$, $2$, and $3\sigma$ and the dashed lines
    correspond to univariate constraints at $2\sigma$. The value of $\Lambda$ is 
    changed between $\{1, 2, 4, 8\}$ (top-left, top-right, bottom-left, bottom-right 
    respectively) which changes the number of bins allowed in the constraints on the 
    Wilson coefficients. 
    }
  \label{fig:EFTvalid_HLLHC_int}
\end{figure}

\begin{figure}[!htbp]
	\centering
  \includegraphics[width=.49\textwidth]{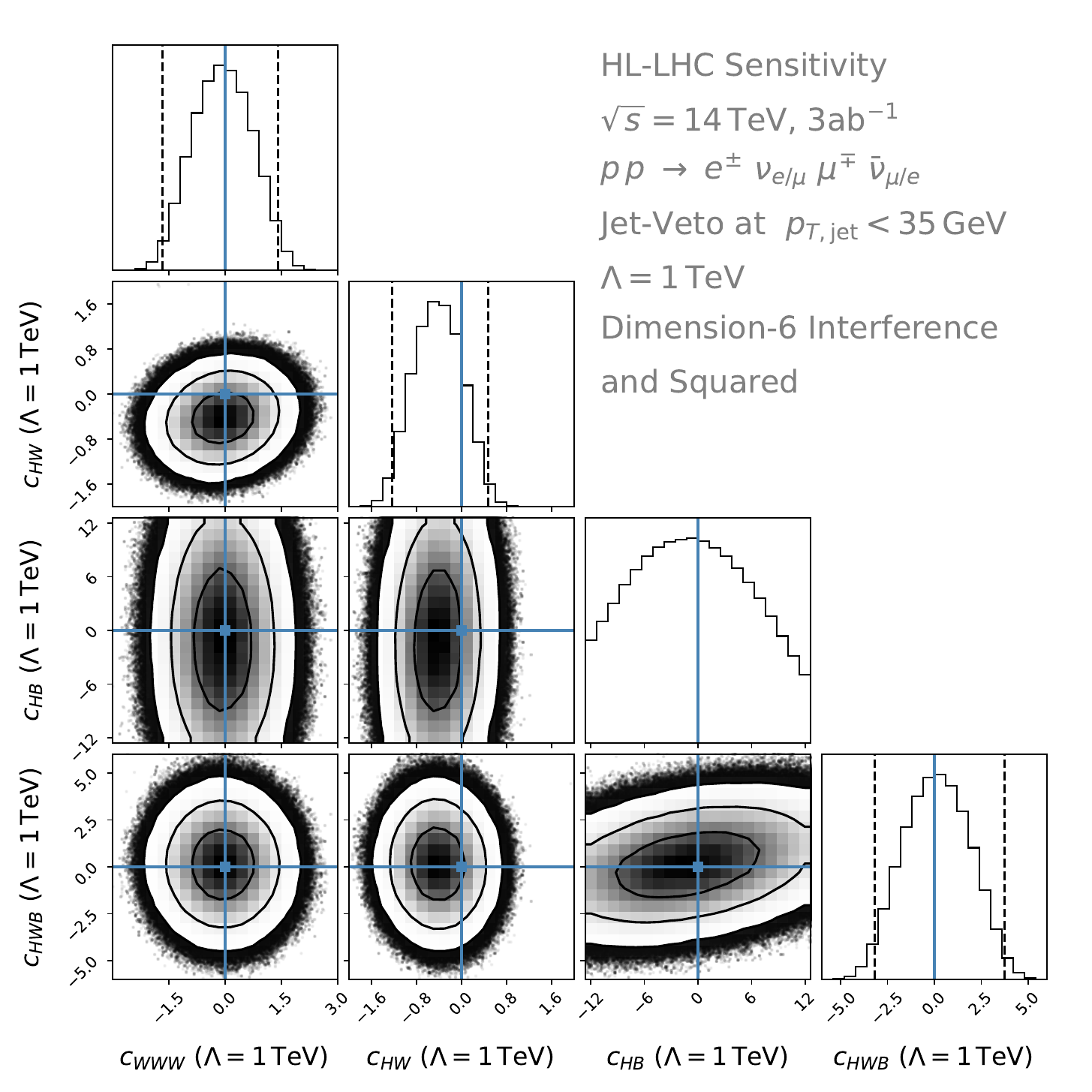}
  \includegraphics[width=.49\textwidth]{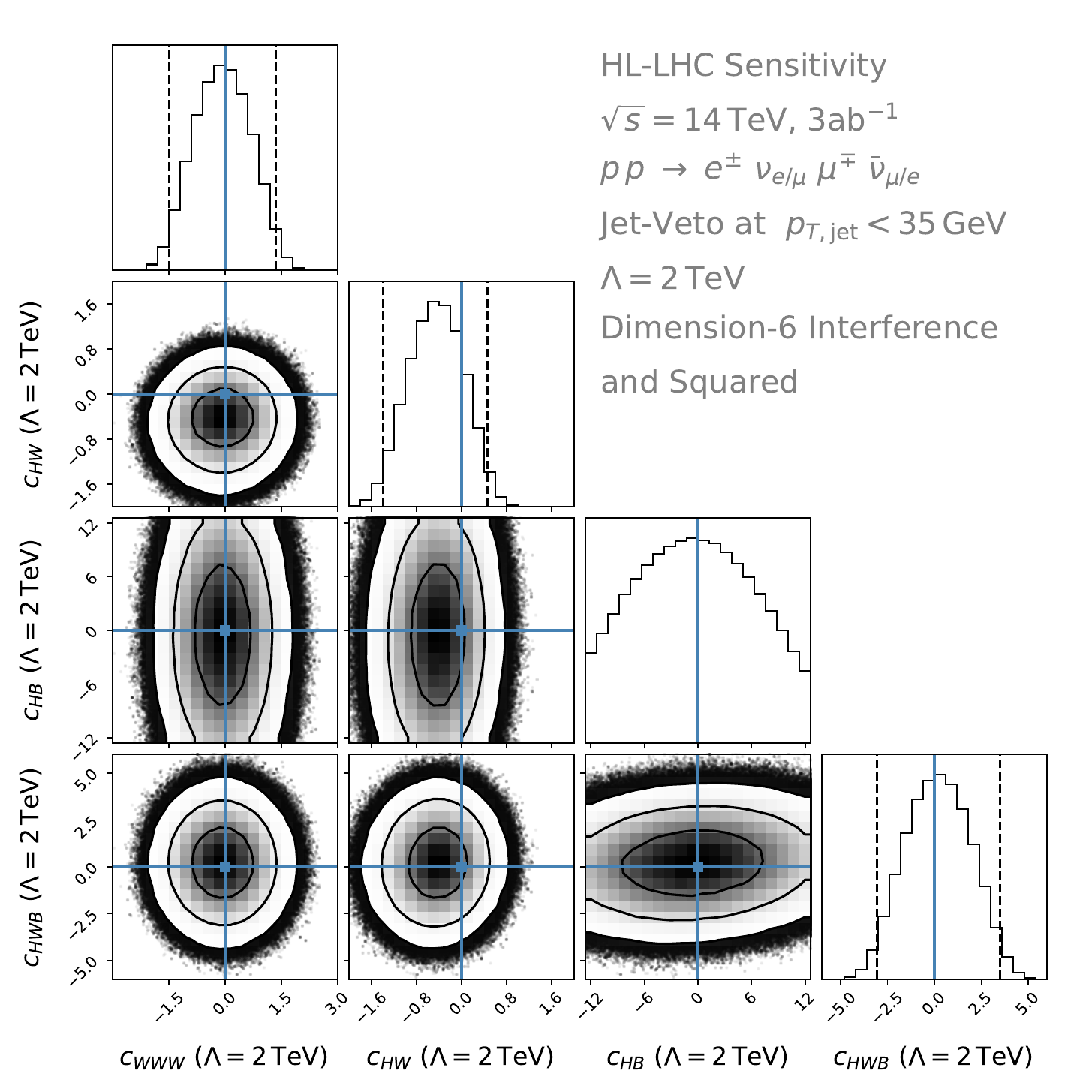}
  \includegraphics[width=.49\textwidth]{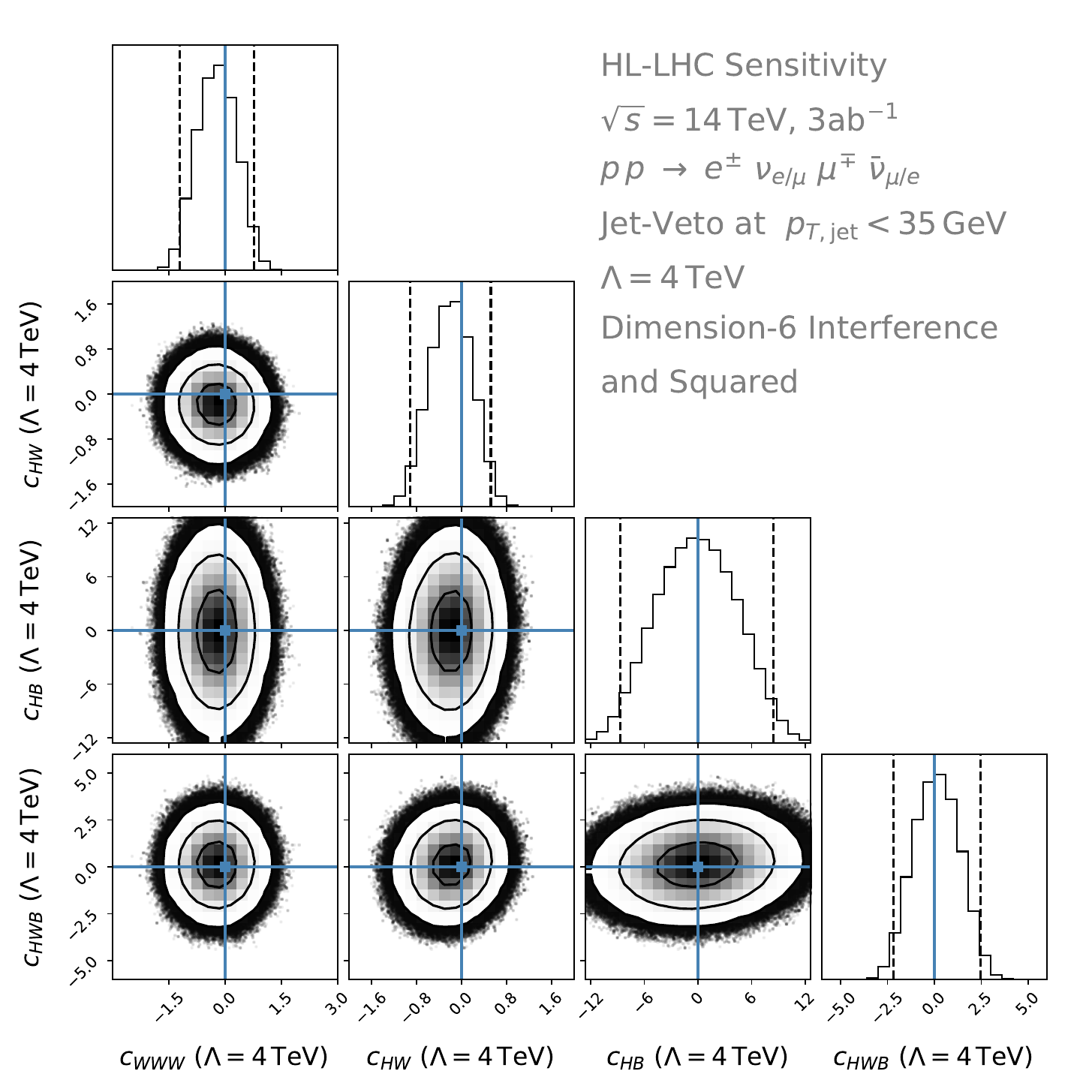}
  \includegraphics[width=.49\textwidth]{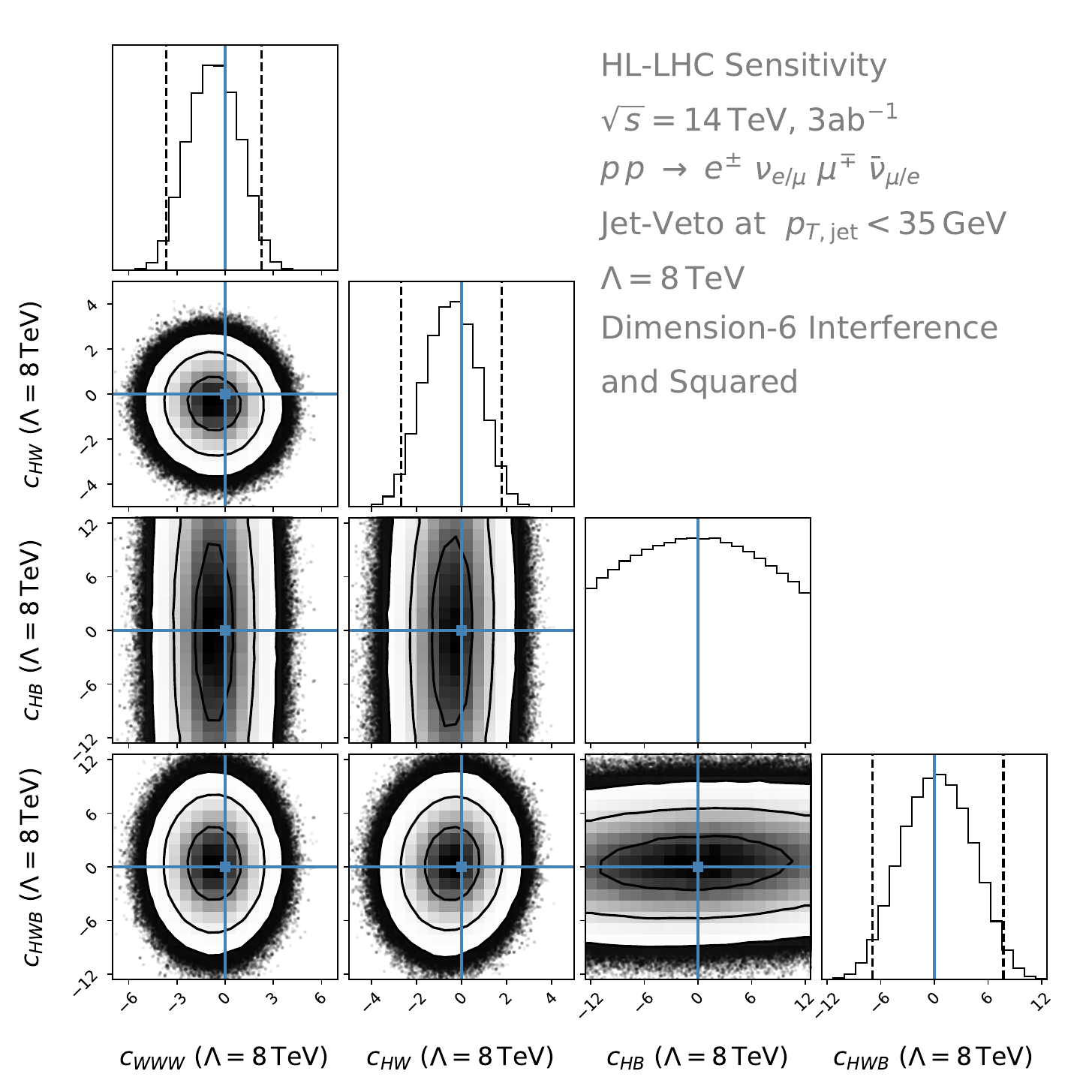}
  \caption{
 Sensitivity plots for $\{c_{HW}, c_{WWW}, c_{HWB}, c_{HB}\}$ at
   HL-LHC with ATLAS cuts ($14\,$TeV, $3\,$ab$^{-1}$) using $WW$
    production with a jet-veto ($p_{T, \mathrm{veto}}=35\,$GeV). 
    This sensitivity study includes the interference and squared contributions of new physics
    (order $1/\Lambda^4$ in the EFT expansion).
    The exclusion contours are placed at $1$, $2$, and $3\sigma$ and the dashed lines
    correspond to univariate constraints at $2\sigma$. The value of $\Lambda$ is 
    chosen at $\{1, 2, 4, 8\}$ (top-left, top-right, bottom-left, bottom-right 
    respectively), which changes the number of bins used in setting constraints on the 
    Wilson coefficients. 
    }
  \label{fig:EFTvalid_HLLHC}
\end{figure}

For the $q\bar q$ channel, when ensuring that we are in the
EFT regime, we have shown that bosonic dimension-8 interference with
the SM is small. Ensuring that the EFT regime is valid also guarantees
that the dimension-8 and higher photon squared contributions remain
negligible. We can then provide constraints using the squared
dimension-6 contributions (at order $1/\Lambda^4$ in the EFT
expansion).  These squared contributions should aid in constraining
those operators which interfere poorly with the Standard Model. The
constraints on the Wilson coefficients at
$\Lambda = \{1, 2, 4, 8\}\,$TeV are presented in
figure~\ref{fig:EFTvalid_HLLHC}.  It should be noted that for
$\Lambda > 5.1\,$TeV, all bins become available for the
fits. Therefore, the constraints from $\Lambda=8\,$TeV can be easily
mapped onto any other mass scale above $5.1\,$TeV by adjusting the value
of the Wilson coefficient appropriately.

For the $\Lambda=1\,$TeV case, we can obtain constraints on $c_{WWW}$,
$c_{HW}$, and $c_{HWB}$ with $c_{HB}$ lying just out of reach for
meaningful constraints. Once again, we obtain symmetric constraints on
$c_{WWW}$ and $c_{HWB}$ around $|c_{WWW}| < 1.5$ and $|c_{HWB}| <
3$. We can compare these to the ones quoted by
ATLAS~\cite{ATLAS:2025dhf}.  The constraints without EFT validity are
given by $|c_{WWW}|<0.17$ and $|c_{HWB}|<1.7$. As mentioned before,
these cannot be directly compared to our constraints with EFT
validity. Those obtained with the cut $M_{WW} < 0.75\,$TeV applied to
the EFT simulation are given by $|c_{WWW}|<0.5$ and $|c_{HWB}|<6$.
These are in line with our results, and show improved constraints on
$c_{HWB}$ at the HL-LHC.  However, the constraint on $c_{WWW}$ is
likely too optimistic as too many bins are being used (see
figure~\ref{fig:EFTvalid_comparison}). The strong interference of
$c_{HW}$ results in an asymmetric constraint
$-1.2 < c_{HW} < 0.4$. Note that the
cancellation between the squared and interference terms makes it
difficult to constrain in the negative $c_{HW}$ direction.

For $\Lambda=2\,$TeV, constraints on the Wilson coefficients are
similar to the $\Lambda=1\,$TeV case.  As the value of $\Lambda$
increases, more bins become available. However, the extra suppression
of the mass scale means that Wilson coefficients need to become
correspondingly larger.

The tightest constraints on the Wilson coefficients arise at
$\Lambda=4\,$TeV, where more $M_{e\mu}$ bins are available.  This is
the first case where a constraint on $c_{HB}$ can be placed --
although not a particularly strong one at $|c_{HB}| < 8$. This
symmetric constraint is consistent with the squared photon amplitude
being around an order of magnitude larger than the SM interference at
$M_{e\mu}\approx 1\,$TeV, as seen in
figures~\ref{fig:BSMOperatorPrediction_d6int}
and~\ref{fig:BSMOperatorPrediction_d6sq}.  The constraint on $c_{HWB}$
also remains roughly symmetric at $|c_{HWB}| < 2.5$.  As discussed earlier,
the operator $\mathcal{O}_{WWW}$ does interfere with the SM at higher
values of $M_{e\mu}$. This leads to asymmetric constraints for the
final two operators, namely
$-1.2 < c_{WWW} < 0.8\,(\Lambda=4\,\mathrm{TeV})$ and
$-0.9 < c_{HW} < 0.5\,(\Lambda=4\,\mathrm{TeV})$.

At $\Lambda=8\,$TeV, the constraints on the Wilson coefficients become
worse again. As the value of $\Lambda$ increases, one gets access to
more $M_{e\mu}$ bins. However, the constraining power decreases as the
SM cross section becomes very small, leading to large
statistical uncertainties with systematic uncertainties also growing. 
Furthermore, at high-$M_{e\mu}$, EW
uncertainties become huge without resummation. With the set of bins used for this analysis, the
maximum value of $M_{e\mu}$ was set to $4\,$TeV with a corresponding
minimum EFT scale of $5.1\,$TeV. Beyond this value, constraints on
Wilson coefficients necessarily become worse as $\Lambda$ increases.

These results are counter-intuitive as the inclusion of EFT validity on the bins in the distribution 
go against the na\"ive scaling with $\Lambda$, so the bounds on $c_i$ for $\Lambda=1\,$TeV and $\Lambda=2\,$TeV
are basically the same whereas a typical extrapolation from $\Lambda=1\,$TeV results would give a looser 
constraint on $c_i$ by a factor of four when considering $\Lambda=2\,$TeV. This multivariate information should
be of great interest for constraining BSM models.

We now show the effect of the jet-veto on constraints for
$\{c_{HW}, c_{WWW}, c_{HWB}, c_{HB}\}$. We compare constraints with
and without the jet-veto at $\Lambda=1\,$TeV and $\Lambda=4\,$TeV (the
scale with the most stringent constraints) in
figure~\ref{fig:vetonoveto_comparison}. For these operators, since the
SM and BSM effects are both $q\bar q$-dominated, the effect of the
jet-veto does not significantly enhance or reduce the BSM relative to
the SM case as it does when fitting operators which proceed via gluon
fusion. Therefore, the main effect the jet-veto has is to increase or
decrease errors and number of events. Overall, the introduction of the
jet-veto introduces large QCD errors which grow with energy up to
$\pm7\%$. However, without a jet-veto, NNLO$_{\mathrm{QCD}}$ and
NLO$_{\mathrm{EW}}$ contributions have a very large K-factor which
imply that the missing combined
NNLO$_{\mathrm{QCD}}\times$NLO$_{\mathrm{EW}}$ corrections, which are
estimated by varying combination scheme, are also expected to be
large. This leads to a larger EW error in the case of no jet-veto
for low $M_{e\mu}$ bins (see figure~\ref{fig:Eweffecterrors}).  Without a jet-veto, the increased number of events also
leads to reduced statistical uncertainties at higher energies.  As
shown in figure~\ref{fig:vetonoveto_comparison}, the jet-veto
constraints are tighter for $\Lambda=1\,$TeV which uses lower energy
bins and the no jet-veto constraints are tighter for $\Lambda=4\,$TeV
which makes use of all bins up to $M_{e\mu}=1\,$TeV.  In general, with
reductions in theoretical uncertainty, the no jet-veto case should
produce better constraints across most values of $\Lambda$. However,
as mentioned before, more work will be required to properly account for incomplete
cancellations that could arise due to the inclusion of the
$b$-jet-veto which is currently assumed to have negligible impact on
$WW$ production without intermediate tops.
\begin{figure}[!htbp]
	\centering
  \includegraphics[width=.49\textwidth]{figures/triangle_1_syst.pdf}
  \includegraphics[width=.49\textwidth]{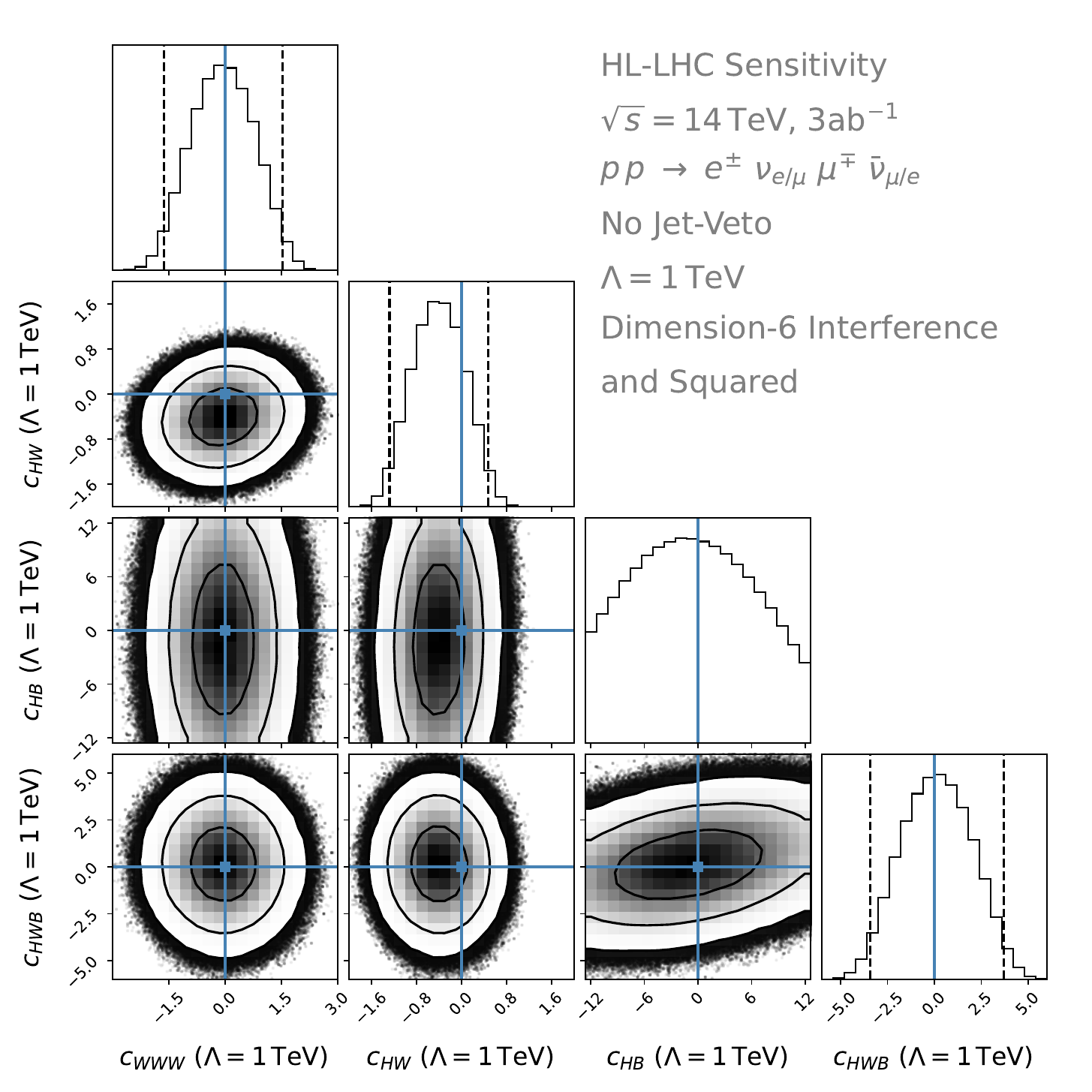}
  \includegraphics[width=.49\textwidth]{figures/triangle_4_syst.pdf}
  \includegraphics[width=.49\textwidth]{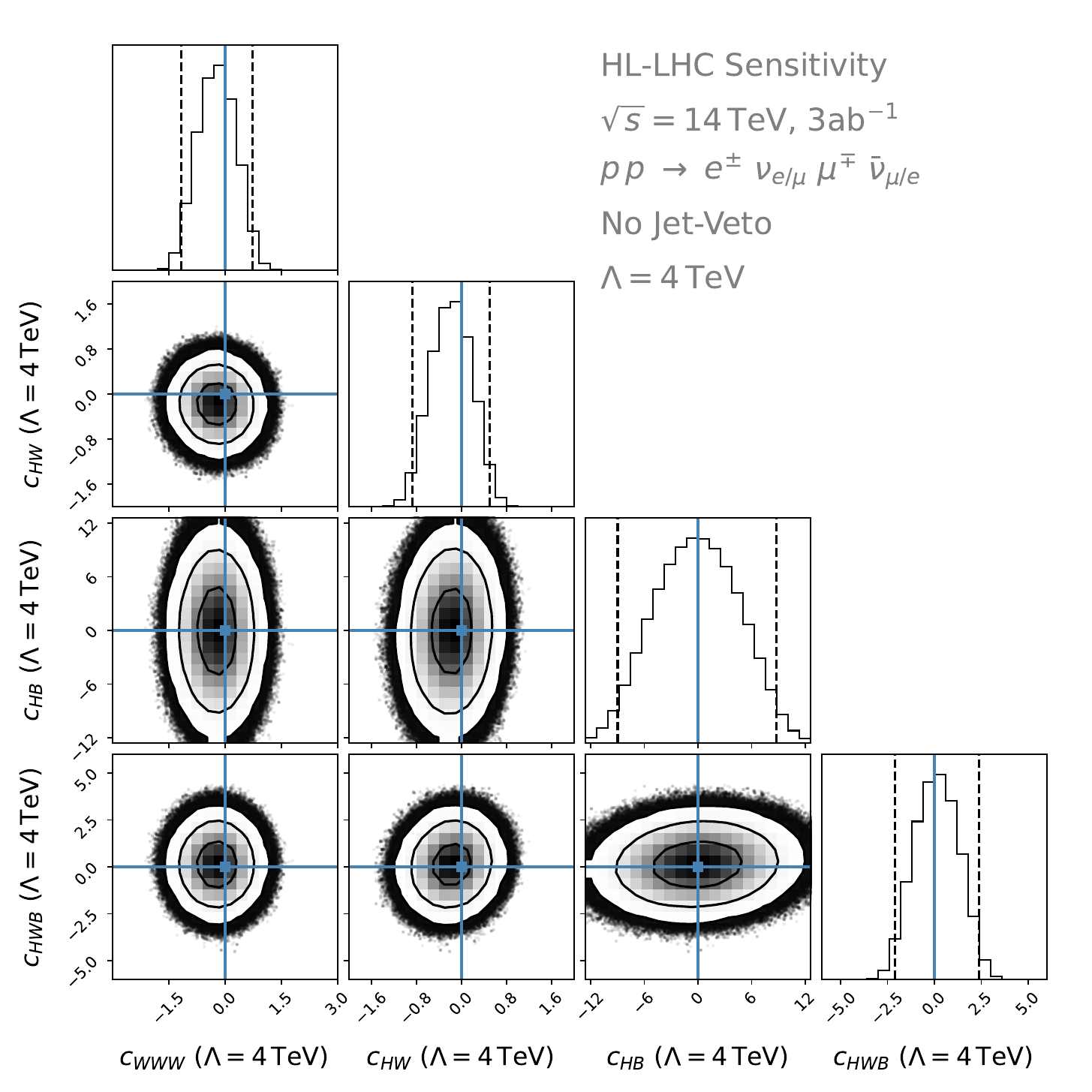}
  \caption{
 Sensitivity plots for $\{c_{HW}, c_{WWW}, c_{HWB}, c_{HB}\}$ at
   HL-LHC with ATLAS cuts ($14\,$TeV, $3\,$ab$^{-1}$) using $WW$
    production with a jet-veto ($p_{T, \mathrm{veto}}=35\,$GeV) (left)
    and without a jet-veto (or with a $b$-tagged veto) (right). 
    This sensitivity study only includes the interference of new physics with the 
    Standard Model (order $1/\Lambda^2$ in the EFT expansion).
    The exclusion contours are placed at $1$, $2$, and $3\sigma$ and the dashed lines
    correspond to univariate constraints at $2\sigma$. The constraints are given at values
    of $\Lambda=1\,$TeV (upper) and $\Lambda=4\,$TeV (lower), with only EFT valid
    bins included in the constraints.
    }
  \label{fig:vetonoveto_comparison}
\end{figure}

\newpage

We finally show the effect of EW corrections on this channel. As discussed in appendix~\ref{sec:SMEW},
missing QCD-EW uncertainties represent one of the dominant uncertainties at high energies (and consequently
at high values of $M_{e\mu}$). To assess how these uncertainties affect the sensitivity studies at the HL-LHC,
we compare how these results change under three scenarios:

\begin{itemize}
  
\item The fit used above, including the EW uncertainty estimate for 
mixed QCD-EW terms as determined by comparing multiplicative and additive schemes.

\item A fit where this uncertainty is neglected for the SM background.

\item A fit in which an additional Sudakov suppression is applied to the SMEFT 
signal. 
\end{itemize}

Constraints are achieved by following the method outlined in section~\ref{sec:sensitivity}.
After running the Markov chain we use the individual marginalised constraints for each Wilson
coefficient for values of $\Lambda = {4, 8}\,$TeV. We use larger values of $\Lambda$ as EW
effects appear in the tails of distributions which can only be used when considering appropriately large
values of $\Lambda$. These are shown in table~\ref{table:EWConstraints}. We note that to apply the Sudakov 
suppression on the SMEFT amplitudes we did not use NLO$_{\mathrm{EW}}$ constraints. Instead, we profiled
the EW Sudakov suppression with energy using the Standard Model NLO k-factor. We then found the correlation
between $M_{e\mu}$ and $M_{WW}$ at dimension-6 and dimension-8 in order to apply this suppression to the 
SMEFT contributions. This gives an estimate for the size of the Sudakov suppression but does not exactly 
model the NLO or NLL contribution for the background.

\begin{table}[!htbp]
\renewcommand{\arraystretch}{1.5}
\begin{center}
\begin{tabular}{ |c|c|c|c| } 
 \hline
 Operator & w/ EW error & no EW error & Sudakov on Signal \\ 
 \hline
 $\mathcal{O}_{HW}$ & $-0.9 < c_{HW} < 0.5$  & $-0.8 < c_{HW} < 0.4$ & $-1 < c_{HW} < 0.6$\\
 ($\Lambda = 4\,$TeV) &  &  & \\
 \hline
 $\mathcal{O}_{HWB}$ & $-2.2 < c_{HWB} < 2.5$  & $-1.9 < c_{HWB} < 2.2$ & $-2.7 < c_{HWB} < 3$\\ 
 ($\Lambda = 4\,$TeV) &  &  & \\
 \hline
 $\mathcal {O}_{WWW}$ & $-1.2 < c_{WWW} < 0.8$  & $-1.1 < c_{WWW} < 0.6$ & $-1.4 < c_{WWW} < 0.9$\\ 
 ($\Lambda = 4\,$TeV) &  &  & \\
 \hline
 $\mathcal{O}_{HW}$ & $-2.5 < c_{HW} < 1.5$  & $-2.2 < c_{HW} < 1.3$ & $-3 < c_{HW} < 2$\\
 ($\Lambda = 8\,$TeV) &  &  & \\
 \hline
 $\mathcal{O}_{HWB}$ & $-7 < c_{HWB} < 8$  & $-6 < c_{HWB} < 7$ & $-9 < c_{HWB} < 10$\\ 
 ($\Lambda = 8\,$TeV) &  &  & \\
 \hline
 $\mathcal {O}_{WWW}$ & $-3.5 < c_{WWW} < 2$  & $-3 < c_{WWW} < 1.7$ & $-4 < c_{WWW} < 2.3$\\ 
 ($\Lambda = 8\,$TeV) &  &  & \\
 \hline
\end{tabular}

\caption{Impact of EW corrections on constraints of three of the dimension-6 operators at the HL-LHC. These 
are shown for $\Lambda = 4\,$TeV and $\Lambda = 8\,$TeV. As expected the constraints are stronger when the error 
is removed and weaker when the Sudakov is applied to the signal.\label{table:EWConstraints}}
\end{center}
\end{table}

We find that the effect of EW corrections on the sensitivity to dimension-6 operators is significant.
The uncertainty of mixed EW-QCD corrections significantly impacts the ability to constrain these operators,
especially when using information from the tails of distributions such as at $\Lambda=8\,$TeV. This 
motivates reduction of these uncertainties which can be achieved through resummation of the EW contribution 
and by including or estimating the size of mixed QCD-EW contributions. We also find that the Sudakov 
suppression is enough to loosen constraints on this operator at high energies and should also be included 
in the generation of SMEFT contributions at high energies in $WW$ production.

\newpage

\section{Conclusions}
\label{sec:the-end}


In this paper, we assessed the size of bosonic dimension-6 and
dimension-8 operators corresponding to anomalous triple and quadruple
gauge couplings and contributing to $WW$ production via photon fusion,
relative to their contribution to the quark-fusion channel. We
performed this comparison both with and without the presence of a
jet-veto. These operators are typically studied at the level of
dimension-6 squared since some of them, such as $\mathcal{O}_{WWW}$,
interfere poorly with the SM at low energies.  This contribution is of
the same order as the interference of dimension-8 operators, which can
arise from either photon or quark fusion.  We found that, for
dimension-6 contributions, although the $q\bar q$ contribution is
leading, the $\gamma\gamma$ contribution can be up to $60\%$ of the
$q\bar q$ contribution at order $\frac{1}{\Lambda^2}$ and up to $40\%$
at order $\frac{1}{\Lambda^4}$. As a consequence, the $\gamma\gamma$
channel should not be neglected when performing fits on these
operators with $WW$ production and likely other diboson
processes. Furthermore, for bosonic operators, the $\gamma\gamma$
dimension-8 contribution is much larger than the corresponding
$q\bar q$ contribution. It also gives a better indication of the
validity of the EFT expansion in inverse powers of the new physics
scale $\Lambda$. Notably, we found that the $\gamma\gamma WW$ contact
interactions grow most rapidly with energy.

To model the effect of the jet-veto on photon fusion predictions, we
adjusted the PDF factorisation scale such that it was fixed at the
value of the jet-veto scale. Besides correctly resumming the
logarithms of $\ptjv/\mu_F$, this prescription gives much better
convergence between LO and NLO predictions in the presence of a
jet-veto.  We also provided a dictionary between the dimension-6 and
dimension-8 operators to the anomalous triple gauge couplings which
affect the $q\bar q$ cross section. This made it possible, for the
first time, to implement these contributions in the presence of a
jet-veto at NLL in the program MCFM-RE. We also demonstrated that
jet-veto effects the $q\bar q$ and $\gamma\gamma$ similarly.

In order to obtain reliable constraints for the considered
higher-dimensional operators, we considered various errors which will
arise at the HL-LHC. In particular, we observed that EW Sudakov
corrections are large and negative at high energy, and the missing
QCD-EW terms are an important source of error that must be included.
With a jet-veto, the QCD and EW suppressions are large, leading to a
large mixed QCD-EW error. Without a jet-veto, big QCD K-factors and EW
corrections give large uncertainties as well. We also directly 
show that the impact of these EW uncertainties is large enough to
directly affect constraints for this channel and so should be modelled
more completely for SMEFT global fits.

We demonstrated that the interference of dimension-8 operators with
the SM is generally small for both the $q\bar q$ and $\gamma\gamma$
contributions.  With this in mind, we presented constraints on the
coefficients of dimension-6 operators obtained from ATLAS 2019 data
and sensitivity studies at the HL-LHC with this channel. In order to
best account for the validity of the EFT expansion, we considered
different values of the new physics scale $\Lambda$, and fit Wilson
coefficients using only the $M_{e\mu}$ bins for which the EFT was
valid at that scale.  We found that it will be possible to constrain
$c_{HW}$ and $c_{WWW}$ at the HL-LHC, but not too much below a value
of $1$, unless theoretical and systematic uncertainties are greatly
reduced. We found $c_{HWB}$ and $c_{HB}$ harder to constrain, with the
latter only constrainable for $\Lambda = 4\,$TeV. 

This work motivates better inclusion and understanding of dimension-6
and dimension-8 photon contributions to aTGC constraints from $WW$ and
other diboson production measurements. It also motivates the
resummation of EW logarithms (such as $\ln\left(M_W/M_{WW}\right)$)
for both the SM background and BSM signal, in order to extract the
best possible constraints from future data. Remarkably, we could not
perform FCC-hh sensitivity studies because EW errors would have been
too large in the allowed ranges of $M_{e\mu}$. We also found the
photon cross section at FCC-ee not to be large enough to give
meaningful constraints.

More work must be undertaken to understand the
impact of the $b$-jet-veto on SM cross section predictions and to
ensure that using fixed order predictions is adequate, in particular
in the tails of distributions. Furthermore, increasing jet
multiplicities give rise the large K-factors. Therefore, to have
reliable predictions, it is important to fully match and merge SM and
BSM hard predictions and to breakdown distributions by jet
multiplicity. However, we remark that the jet-veto does not make a
huge difference to constraints as it does not vastly reduce errors or
increase the BSM/SM ratio. Therefore, constraints for these operators
seem to be not dramatically dependent on the fiducial cuts on the
jets.

A natural extension of this work would be the inclusion of other SMEFT
vertices in MCFM-RE.  Currently, only some of the CP-even aTGC
vertices have been included in MCFM. The vertex arising from operator
$\mathcal{O}_{8HDHW2}$ could be included, as well as the CP-odd
dimension-6 operators. Direct $qqWW$ contact interactions could also
be included. Finally, this channel should be combined with
other diboson channels to compare sensitivities, in particular to
gauge if constraints can be obtained by $ZZ$ alone or, if there are
directions in parameter space where $WW$ is complementary to other channels.

\section*{Acknowledgements}
We thank Gloria Bertolotti, Eugenia Dallari, Xavier Pritchard, Marco
Sebastianutti, Lewis Mazzei, and Claudia Muni for useful
discussions. We would also like to thank François Arleo for asking a
question during a conference which prompted the work presented
here. We thank Chloe Widdowson for running some dimension-8
$\gamma\gamma$ interference contributions on MadGraph. AB is supported
by the UK STFC under the Consolidated Grant ST/X000796/1, and thanks
Royal Holloway, University of London for hospitality while this work
was completed. The work of AM is partially supported by the National
Science Foundation under Grant Number PHY-2412701.
We acknowledge the use of computing resources made available by the Cambridge Service 
for Data Driven Discovery (CSD3), part of which is operated by the University of 
Cambridge Research Computing on behalf of the STFC DiRAC HPC Facility (www.dirac.ac.uk). 
The DiRAC component of CSD3 was funded by BEIS capital funding via STFC capital grants 
ST/P002307/1 and ST/R002452/1 and STFC operations grant ST/R00689X/1. DiRAC is part of 
the UK National e-Infrastructure.


\appendix

\section{Matching SMEFT to aTGCs}
\label{sec:AppendixMatch}

In this section, we present the matching of SMEFT to anomalous
triple-gauge couplings. Both dimension-6 and dimension-8 operators
match to SMEFT. The anomalous triple-gauge coupling Lagrangian
implemented in MCFM (and therefore in MCFM-RE) takes the following
form:
\begin{multline}
  \mathcal{L_{\mathrm{anom}}}=-ig_{WW(Z/A)}\big[\Delta g_1^{Z/A}\left(W^+_{\mu\nu}W^{-\,\mu}(Z/A)^{\nu} - W^-_{\mu\nu}W^{+\,\mu}(Z/A)^{\nu} \right)\\
  + \Delta \kappa_1^{Z/A}W^+_{\mu}W^-_{\nu}(Z/A)^{\mu\nu} +
  \frac{\lambda^{Z/A}}{M^2_{W}}W_\mu^{+\,\nu} W_\nu^{-\,\rho}
  (Z/A)_\rho^{\ \mu}\big]\,,
\end{multline}
where $g_{WWZ}=g\cos\theta_W$ and $g_{WWA}=e=g\sin\theta_W$.
We treat each of the three dimension-6 and five dimension-8 operators in turn.

\begin{itemize}
\item $\mathcal{O}_{HW}=\phi^\dagger \phi\, W^I_{\mu\nu}W^{I\,\mu\nu}$

After EW symmetry breaking:
\begin{equation}
  \begin{split}
  \phi^\dagger \phi W^I_{\mu\nu}W^{I\,\mu\nu}& \to \frac{v^2}{2} W^I_{\mu\nu}W^{I\,\mu\nu}
	\\ & =\frac{v^2}{2} \left(W^1_{\mu\nu}W^{1\,\mu\nu} +W^2_{\mu\nu}W^{2\,\mu\nu}+W^3_{\mu\nu}W^{3\,\mu\nu} \right).
  \end{split}
\end{equation}
Taking the relevant three-vertex cross terms:
\begin{equation}
	W^I_{\mu\nu}W^{I\,\mu\nu} \to 2\left(\partial_\mu W^I_\nu - \partial_\nu W^I_\mu\right)\left(-\epsilon_{IJK}gW^{J\,\mu} W^{K\,\nu}\right).
\end{equation}
%
%
Since
\begin{subequations}
  \label{eq:WAB}
  \begin{align}
  W^1_{\mu} & = \frac{1}{\sqrt{2}}\left(W^+_{\mu}+W^-_{\mu}\right)\,,\qquad
  \quad\>\> W^2_{\mu}  = \frac{i}{\sqrt{2}}\left(W^+_{\mu}-W^-_{\mu}\right)\,,\\
  W_{\mu}^3 &= \cos\theta_WZ_{\mu} +\sin\theta_WA_{\mu}\,, \qquad
  B_\mu  = \cos\theta_WA_{\mu} -\sin\theta_WZ_{\mu}\,,
  \end{align}
\end{subequations}
and defining
\begin{align}
  \label{eq:field-strengths}
  W^\pm_{\mu\nu} \equiv \partial_\mu W^\pm_\nu - \partial_\nu W^\pm_\mu\,,\quad
  Z_{\mu\nu} \equiv \partial_\mu Z_\nu - \partial_\nu Z_\mu\,,\quad
  A_{\mu\nu} \equiv \partial_\mu A_\nu - \partial_\nu A_\mu\,,
\end{align}
we generate the following couplings:
\begin{equation}
  \begin{split}
    \phi^\dagger \phi W_{\mu\nu}W^{\mu\nu}  & \to 
     		2i g_{WWZ}v^2 \left[ \left(W^+_{\mu\nu}W^{-\,\mu}-W^-_{\mu\nu}W^{+\,\mu} \right)Z^{\nu}+W^{+\,\mu} W^{-\,\nu}Z_{\mu\nu}\right] \\
& +2i g_{WWA}v^2\left[\left(W^+_{\mu\nu}W^{-\,\mu}-W^-_{\mu\nu}W^{+\,\mu} \right)A^{\nu} +W^{+\,\mu} W^{-\,\nu}A_{\mu\nu}\right]\,.
  \end{split}
\end{equation}

\item $\mathcal{O}_{HWB}=\phi^\dagger  \sigma^I\phi W^I_{\mu\nu}B^{\mu\nu}$
  
After EW symmetry breaking: 
\begin{equation}
  \begin{split}
	\phi^\dagger  \sigma^I\phi W^I_{\mu\nu}B^{\mu\nu}	\to -\frac{v^2}{2}W^3_{\mu\nu}B^{\mu\nu}
  \end{split}
\end{equation}
Selecting the relevant three-vertex terms, we obtain 
\begin{equation}
  W^3_{\mu\nu}B^{\mu\nu} \to -g\epsilon_{3JK}W^J_{\mu}W^K_{\nu}B^{\mu\nu}\,.
\end{equation}
%
%
%
Using the definitions in eqs.~(\ref{eq:WAB}) and~(\ref{eq:field-strengths}), we generate the following couplings:
\begin{equation}
  \phi \sigma^I\phi^\dagger W^I_{\mu\nu}B^{\mu\nu} \to i g v^2W^{+\,\mu} W^{-\,\nu}\left(
    \sin\theta_W Z_{\mu\nu}-\cos\theta_W A_{\mu\nu}
  \right)\,.
\end{equation}

\item $\mathcal {O}_{WWW}=\epsilon_{IJK} W^{I\, \nu}_{\mu}W^{J\,\rho}_{\nu}W^{K\, \mu}_{\rho}$


Let us consider one of the terms in the sum
\begin{equation}
  \begin{split}
  & \epsilon_{123} W^{1\, \nu}_{\mu}W^{2\,\rho}_{\nu}W^{3\, \mu}_{\rho} + 	\epsilon_{213} W^{2\, \nu}_{\mu}W^{1\,\rho}_{\nu}W^{3\, \mu}_{\rho}
	=(W^{1\, \nu}_{\mu}W^{2\,\rho}_{\nu} - W^{2\, \nu}_{\mu}W^{1\,\rho}_{\nu})W^{3\, \mu}_{\rho}\\ &
	=\frac{i}{2}((W^{+\, \nu}_{\mu}+W^{-\, \nu}_{\mu})(W^{+\,\rho}_{\nu}-W^{-\, \rho}_{\nu}) - (W^{+\, \nu}_{\mu}-W^{-\, \nu}_{\mu})(W^{+\, \nu}_{\rho}+W^{-\, \rho}_{\nu}))W^{3\, \mu}_{\rho}\,.
  \end{split}
\end{equation}
Using the definitions in eqs.~(\ref{eq:WAB})
and~(\ref{eq:field-strengths}), and selecting only the terms that
contribute to triple-gauge couplings, we obtain
\begin{equation}
(W^{1\, \nu}_{\mu}W^{2\,\rho}_{\nu} - W^{2\, \nu}_{\mu}W^{1\,\rho}_{\nu})W^{3\, \mu}_{\rho}\to -i(W^{+\, \nu}_{\mu}W^{-\, \rho}_{\nu} - W^{-\, \nu}_{\mu}W^{+\, \rho}_{\nu})W^{3\, \mu}_{\rho}\,.
\end{equation}
Similarly,
\begin{equation}
	\epsilon_{231} W^{2\, \nu}_{\mu}W^{3\,\rho}_{\nu}W^{1\, \mu}_{\rho} + 	\epsilon_{132} W^{1\, \nu}_{\mu}W^{3\,\rho}_{\nu}W^{2\, \mu}_{\rho}
%
%
%
%
	\to -iW^{3\, \nu}_{\mu}(W^{+\, \rho}_{\nu}W^{-\, \mu}_{\rho} - W^{-\, \rho}_{\nu}W^{+\, \mu}_{\rho})\,.
\end{equation}
Using the cyclic property of the trace, we obtain
\begin{equation}
	\epsilon_{IJK} W^{I\, \nu}_{\mu}W^{J\,\rho}_{\nu}W^{K\, \mu}_{\rho} \to  - 6iW^{+\, \nu}_{\mu}W^{-\, \rho}_{\nu}W^{3\, \mu}_{\rho}
\end{equation}
Using the definitions in eqs.~(\ref{eq:WAB}) and~(\ref{eq:field-strengths}), we then generate the following couplings:
\begin{equation}
  \epsilon_{IJK} W^{I\, \nu}_{\mu}W^{J\,\rho}_{\nu}W^{K\, \mu}_{\rho} \to -6 i W^{+\, \nu}_{\mu}W^{-\, \rho}_{\nu}
  \left(\cos\theta_W Z^{\ \mu}_{\rho} +\sin\theta_W A^{\ \mu}_{\rho}
\right)\,.
\end{equation}
\item$\mathcal {O}_{8HDHB}=i\phi^\dagger \phi\, (D_{\mu}\phi)^\dagger D_{\nu}\phi\, B^{\mu\nu}$
  
We first select the contributions that give rise to triple-gauge vertices
\begin{equation}
	i\phi^\dagger \phi\, (D_{\mu}\phi)^\dagger D_{\nu}\phi\, B^{\mu\nu}\to i\phi^\dagger \phi\, \left(-i\frac{g}{2}\sigma_I W^I_{\mu}\phi\right)^\dagger \left(-i\frac{g}{2}\sigma_J W^J_{\nu}\phi\right)\, B^{\mu\nu}\,.
\end{equation}
After EW symmmetry breaking, we obtain
\begin{equation}
  \label{eq:phiphiDDB}
  i\phi^\dagger \phi\, \left(-i\frac{g}{2}\sigma_I W^I_{\mu}\phi\right)^\dagger \left(-i\frac{g}{2}\sigma_J W^J_{\nu}\phi\right)\, B^{\mu\nu}\to
  i g^2 \frac{v^4}{8} W^-_{\mu} W^+_{\nu} B^{\mu\nu} = i g^2 \frac{v^4}{8} W^-_{\mu} W^+_{\nu} B^{\mu\nu}\,.  
\end{equation}

We then generate the following couplings:
\begin{equation}
		i \phi^\dagger \phi\, (D_{\mu}\phi)^\dagger D_{\nu}\phi\, B^{\mu\nu} \to 
ig^2\frac{v^4}{8}W^+_{\mu} W^-_{\nu}\left(\cos\theta_W A^{\mu\nu}-\sin\theta_W Z^{\mu\nu}\right)\,.
\end{equation}

\item$\mathcal {O}_{8HDHW}=i\phi^\dagger \phi\, (D^{\mu}\phi)^\dagger \sigma_I D^{\nu}\phi\, W^I_{\mu\nu}$

  Considering only the contributions to triple-gauge vertices, we obtain
  \begin{multline}
    \label{eq:ppDsD}
  i \phi^\dagger \phi\, (D^{\mu}\phi)^\dagger \sigma_I D^{\nu}\phi\,
  W^I_{\mu\nu}\to i\phi^\dagger \phi\,
  \left(\left(-i\frac{g}{2}\sigma_I W^{I \mu} -
      i\frac{g\prime}{2}B^{\mu}\right)\phi\right)^\dagger
  \left(\sigma_J W^J_{\mu\nu}\right)\times \\ \times \left(\left(-i\frac{g}{2}\sigma_I
      W^{I\nu} - i\frac{g\prime}{2}B^{\nu}\right)\phi\right)\, .
\end{multline}
After EW symmetry breaking, we obtain
\begin{equation}
  \begin{split}
   \left(\left(-i\frac{g}{2}\sigma_I W^{I \mu}\!\!\!\right.\right.&\left.\left. -
      i\frac{g\prime}{2}B^{\mu}\right)\phi\right)^\dagger
  \left(\sigma_J W^J_{\mu\nu}\right)\left(\left(-i\frac{g}{2}\sigma_I
      W^{I\nu} - i\frac{g\prime}{2}B^{\nu}\right)\phi\right) \to\\  & \to g^2 \frac{v^2}{4} \begin{pmatrix}W^{-\, \mu}\,, & -\frac{Z^{\mu}}{\sqrt 2\cos\theta_W}\end{pmatrix}
  \begin{pmatrix}W^{3}_{\mu\nu} & \sqrt{2}\,W^{+}_{\mu\nu}\\ \sqrt{2}\,W^{-}_{\mu\nu} & -W^{3}_{\mu\nu}\end{pmatrix}
  \begin{pmatrix}W^{+\, \nu} \\ -\frac{Z^{\nu}}{\sqrt 2\cos\theta_W}\end{pmatrix}\,.
\end{split}
\end{equation}
Substituting the above expression into eq.~(\ref{eq:ppDsD}), and
ignoring the term containing a triple neutral gauge coupling, we
obtain
\begin{equation}
  \label{eq:ppDsD-simplified}
  \begin{split}
   i & \phi^\dagger \phi\, (D^{\mu}\phi)^\dagger \sigma_I D^{\nu}\phi\,
   W^I_{\mu\nu}\to i g^2 \frac{v^4}{8}\left(W^{-\, \mu}W^{+\, \nu}W^{3}_{\mu\nu}-\frac{1}{\cos\theta_W}\left(W^{-\, \mu}W^{+}_{\mu\nu}Z^{\nu}+Z^{\mu}W^{-}_{\mu\nu}W^{+\, \nu}\right)
   \right)\\
   & =i g^2 \frac{v^4}{8} \left(\left(W^{-}_{\mu\nu}W^{+\, \mu}-W^{+}_{\mu\nu}W^{-\, \mu}\right)\frac{Z^{\nu}}{\cos\theta_W}+\cos\theta_WW^{-\, \mu}W^{+\, \nu}Z_{\mu\nu}+\sin\theta_WW^{-\, \mu}W^{+\, \nu}A_{\mu\nu}\right)\,.
  \end{split}
\end{equation}
\item$\mathcal {O}_{8HDHW2}=i\, \epsilon_{IJK} \phi^\dagger \sigma^I \phi\, (D^{\mu}\phi)^\dagger \sigma^J D^{\nu}\phi\, W^K_{\mu\nu}$


After EW symmetry breaking, we obtain, the only non-zero contribution from $\phi^\dagger \sigma^I \phi$ corresponds to $I=3$: 
%
%
\begin{equation}
  \begin{split}
 	i\phi^\dagger \sigma^3 \phi\, &(D^{\mu}\phi)^\dagger \epsilon_{3JK}\sigma^JW^K_{\mu\nu}  D^{\nu}\phi\to \\
%
%
        & \to 
        -ig^2 \frac{v^4}{8}\,
        \begin{pmatrix}W^{-\, \mu}\,, & -\frac{Z^{\mu}}{\sqrt 2\cos\theta_W}\end{pmatrix}
        \begin{pmatrix}0 & i\sqrt{2}W^{+}_{\mu\nu}\\ -i\sqrt{2}W^{-}_{\mu\nu} & 0\end{pmatrix}  \begin{pmatrix}W^{+\, \nu} \\ -\frac{Z^{\nu}}{\sqrt 2\cos{\theta_W}}\end{pmatrix}\\ & =
        -\frac{g^2}{\cos\theta_W} \frac{v^4}{8} \left(W^{+}_{\mu\nu}W^{-\, \mu} + W^{-}_{\mu\nu}W^{+\, \mu}\right)Z^{\nu}\,.
  \end{split}
\end{equation}
%
%
%
%
%
This is a new Lorentz structure which does not match to anomalous
triple gauge couplings and is not implemented in MCFM-RE. 

\item$\mathcal {O}_{8W1}=\epsilon_{IJK} \phi^\dagger \phi\, W^{I\,\nu}_{\mu} W^{J\,\rho}_{\nu} W^{K\,\mu}_{\rho}$

  This has the same Lorentz structure as that of $\mathcal {O}_{WWW}$ for this channel, except with an additional factor $v^2/2$.


\item$\mathcal {O}_{8W2}=\epsilon_{IJK}\phi^\dagger \sigma^I \phi\, W^{J\,\nu}_{\mu} W^{K\,\rho}_{\nu} B^{\ \mu}_{\rho}$

  After EW symmetry breaking, only the contribution corresponding to
  $I=3$ survives, giving
\begin{equation}
  \begin{split}
    \epsilon_{3JK}\phi^\dagger \sigma^3 \phi\, W^{J\,\nu}_{\mu} W^{K\,\rho}_{\nu} B^{\ \mu}_{\rho} & \to -\frac{v^2}{2}\, \left(W^{1\,\nu}_{\mu} W^{2\,\rho}_{\nu} -W^{2\,\nu}_{\mu} W^{1\,\rho}_{\nu} \right)B^{\ \mu}_{\rho} \\ & =
    i \frac{v^2}{2}  \left(W^{+\,\nu}_{\mu} W^{-\,\rho}_{\nu} -W^{-\,\nu}_{\mu} W^{+\,\rho}_{\nu} \right)B^{\ \mu}_{\rho}\\
    & =i \frac{v^2}{2}  \cos\theta_W \left(W^{+\,\nu}_{\mu} W^{-\,\rho}_{\nu} -W^{-\,\nu}_{\mu} W^{+\,\rho}_{\nu} \right)A^{\ \mu}_{\rho}\\
    & -i \frac{v^2}{2}  \sin\theta_W \left(W^{+\,\nu}_{\mu} W^{-\,\rho}_{\nu} -W^{-\,\nu}_{\mu} W^{+\,\rho}_{\nu} \right)Z^{\ \mu}_{\rho}\\
    & =i\,v^2 W^{+\,\nu}_{\mu} W^{-\,\rho}_{\nu} \left( \cos\theta_W  A^{\ \mu}_{\rho}-\sin\theta_W Z^{\ \mu}_{\rho}\right)\,.
  \end{split}
\end{equation}
%
%
%
%
%
%
%
%

\item$\mathcal {O}_{8HW}=(\phi^\dagger \phi)^2\, W^{I,\mu\nu} W^{I}_{\mu\nu}$

  This has the same Lorentz structure as that of $\mathcal {O}_{HW}$ for this channel, except with an additional factor $v^2/2$.

\item$\mathcal {O}_{8HWB}=(\phi^\dagger \phi)\phi^\dagger  \sigma^I\phi\, W^I_{\mu\nu}B^{\mu\nu}$

  This has the same Lorentz structure as that of $\mathcal {O}_{HWB}$ for this channel, except with an additional factor $v^2/2$.

\end{itemize}

\newpage

\section{Sources of Error at the HL-LHC}
\label{sec:ERRs}

In this section, we give some details on how the errors on EW corrections and projected systematics have been calculated.

\subsection{SM $q\bar q$ + EW Predictions}
\label{sec:SMEW}

\begin{figure}[htbp]
  \includegraphics[width=.5\textwidth]{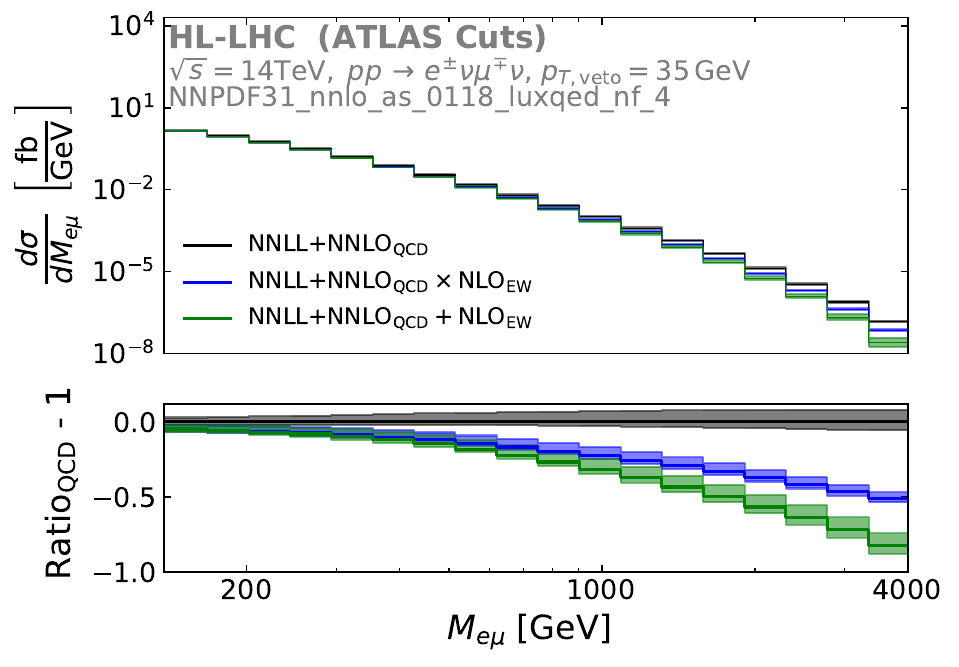} 
    \includegraphics[width=.5\textwidth]{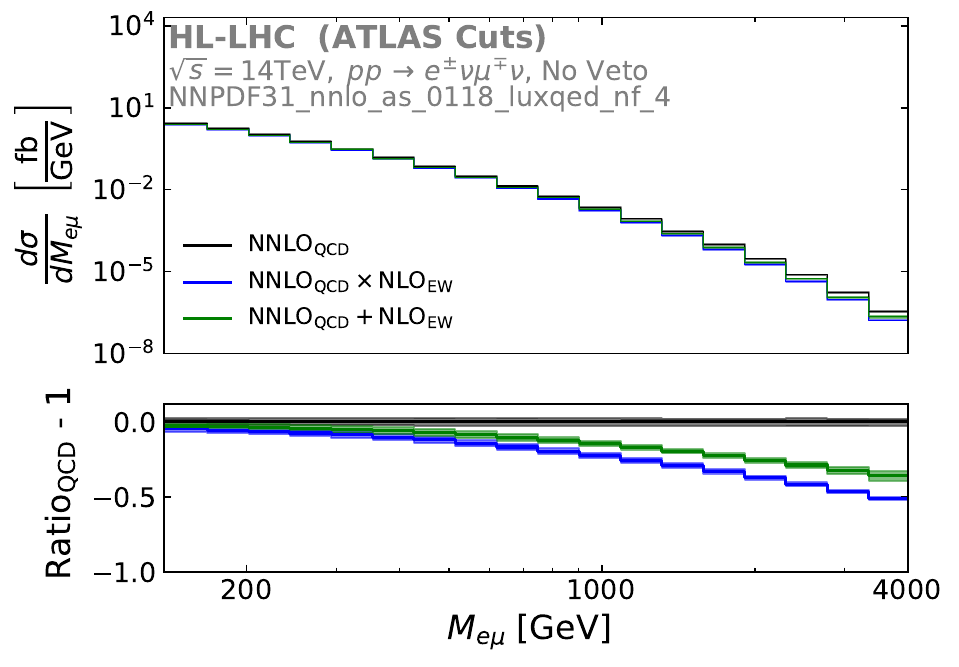} 
  \caption{Comparison of dilepton mass distribution with and without (black) electroweak
    corrections. Both multiplicative (blue) and additive (green) schemes are compared, with and without jet-veto.}
  \label{fig:EW_Effects}
\end{figure}

To understand the size of mixed QCD-EW corrections, we compare EW corrections obtained from 
additive and multiplicative combination schemes taken from~\cite{Grazzini:2019jkl}. 
These are defined as:
\begin{equation}
	\label{eq:multiplicative}
	\mathrm{d}\sigma_{\mathrm{NNLL+NNLO\,\,QCD}\times \mathrm{EW}_{q\bar{q}}}=\mathrm{d}\sigma^{q\bar{q}}_{\mathrm{NNLL+NNLO\,\,QCD}}\big(1+\delta^{q\bar{q}}_{\mathrm{EW}}\big)+\mathrm{d}\sigma^{\gamma\gamma}_{\mathrm{NLO}}+\mathrm{d}\sigma^{gg}_{\mathrm{NLL}}\,,
\end{equation}
and
\begin{equation}
	\label{eq:additive}
	\mathrm{d}\sigma_{\mathrm{NNLL+NNLO\,\,QCD}+\mathrm{EW}_{q\bar{q}}}=\mathrm{d}\sigma^{q\bar{q}}_{\mathrm{LO}}\big(1+\delta^{q\bar q}_{\mathrm{NNLL+NNLO\,\,QCD}}+\delta^{q\bar{q}}_{\mathrm{EW}}\big)+\mathrm{d}\sigma^{\gamma\gamma}_{\mathrm{NLO}}+\mathrm{d}\sigma^{gg}_{\mathrm{NLL}},
\end{equation}
where $\delta^{q\bar{q}}_{\mathrm{EW}}$ are the NLO EW corrections to the LO quark induced process and $\delta^{q\bar q}_{\mathrm{NNLL+NNLO\,\,QCD}}$ are similarly the QCD corrections. 
The multiplicative scheme adds an estimate for the size of the missing $\delta_{\mathrm{NNLL+NNLO\,\,QCD}}\,\delta_{\mathrm{EW}}$ cross 
terms (which are not present in the additive scheme). The two different schemes are shown in figure~\ref{fig:EW_Effects}.

It is found that the effect of the EW corrections grows rapidly with
energy, as expected.  The size of the missing QCD-EW cross terms
(given by the difference between additive and multiplicative schemes
in figure~\ref{fig:EW_Effects}) also grows with energy.
In the case of the jet-veto, the two large negative Sudakov
suppressions (one from the veto on QCD radiation and the other from
the veto on EW radiation) combine to give a large positive
contribution for the estimated mixed corrections relative to the
additive scheme.
In the case without a jet-veto, there is a large K-factor for the NNLO
prediction relative to the LO prediction but still a negative Sudakov
suppression for the NLO EW corrections. These combine to give a large
extra negative suppression relative to the additive scheme.

In both cases, the size of these corrections is large enough to
warrant adding the size of these corrections as a theoretical
uncertainty. This profiles the effect of not including the combined
QCD-EW corrections (which should formally be included at this
order). This prescription also has the advantage of signalling the
regions of phase space where the EW logarithms become so large that
they should not be considered without an EW
resummation~\cite{Denner:2024yut}. In these regions, the theoretical
error from the cross terms will greatly reduce the constraining power
of those bins on new physics fits. 

\begin{figure}[htbp]
\centering
  \includegraphics[width=.49\textwidth]{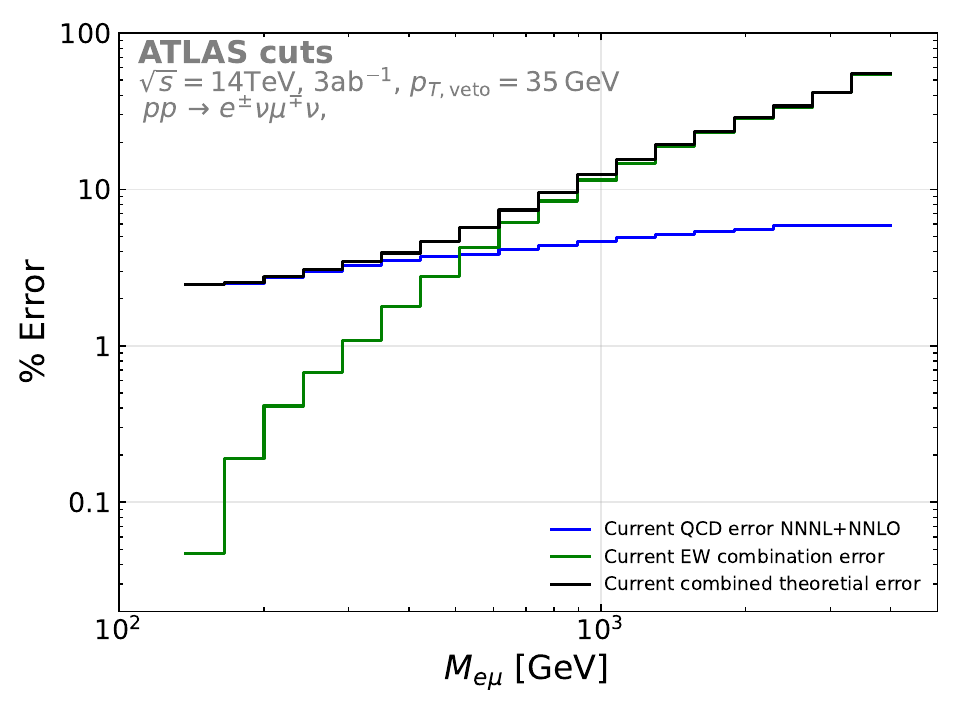}
    \includegraphics[width=.49\textwidth]{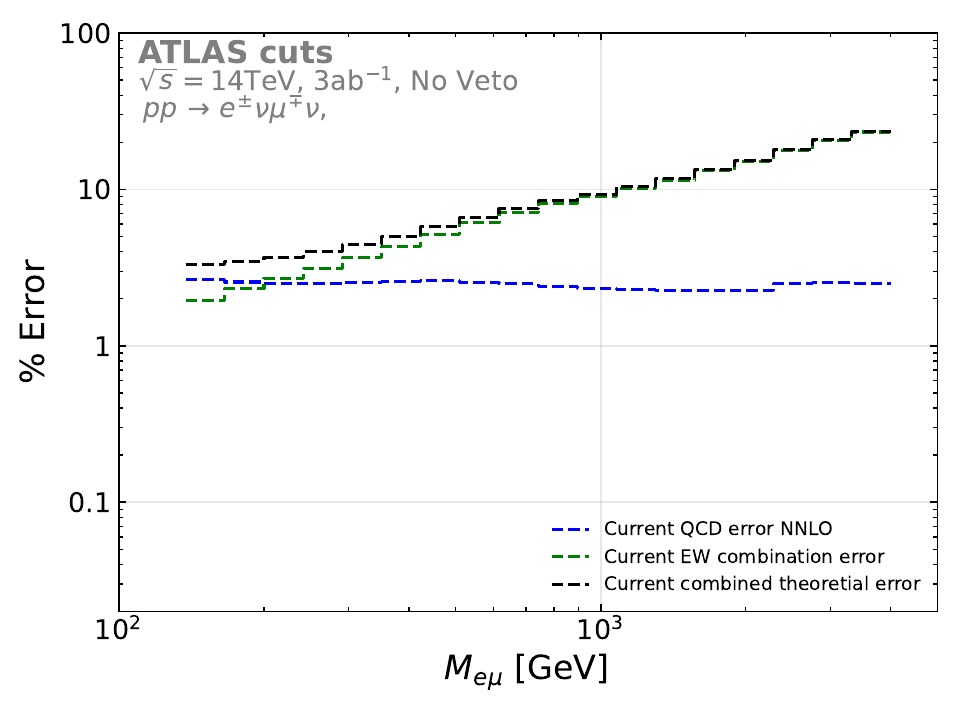}
  \caption{Comparison of theoretical uncertainties arising from QCD scale variation (blue), 
  and from the missing combined QCD-EW corrections (green) this uncertainty is profiled
  by taking the difference between the multiplicative and additive combination schemes given
  in equations~\eqref{eq:multiplicative} and~\eqref{eq:additive} respectively. The combined 
  theoretical uncertainty is also shown (black).}
  \label{fig:Eweffecterrors}
\end{figure}

In Figure~\ref{fig:Eweffecterrors}, we show the size of these
uncertainties relative to the QCD errors. In the case without a
jet-veto, at all energies there is a large NNLO K-factor which leads
to the QCD-EW error starting at $2\%$ at low $M_{e\mu}$ and rising to
$20\%$ at $M_{e\mu}=4\,$TeV. However, QCD errors remain small, about
$2.5\%$, at all energies.  This leads to a combined error starting at
$3.5\%$ and rising to $20\%$. With a jet-veto, the lack of enhancement
or suppression relative to LO at low energies causes missing QCD-EW
terms to be negligible.  However, at larger energies, the large
Sudakov suppression from the jet-veto leads to large missing terms
rising to $60\%$. The QCD errors also do not grow until the high
energy limit, implying that, at low energies, the combined error is
smaller than that of the no-veto case, starting at $2.5\%$, but
eventually rising much higher to $60\%$.

\subsection{Estimates of HL-LHC Errors}
\label{sec:hllhcerr}
\begin{figure}[htbp]
\centering
  \includegraphics[width=.49\textwidth]{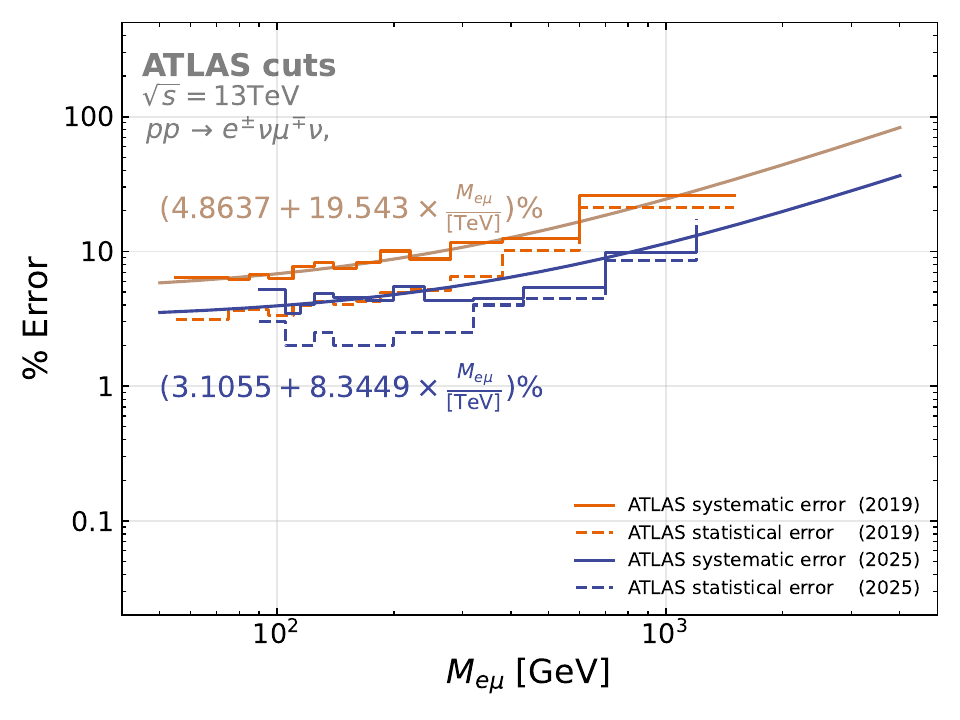}
    \includegraphics[width=.49\textwidth]{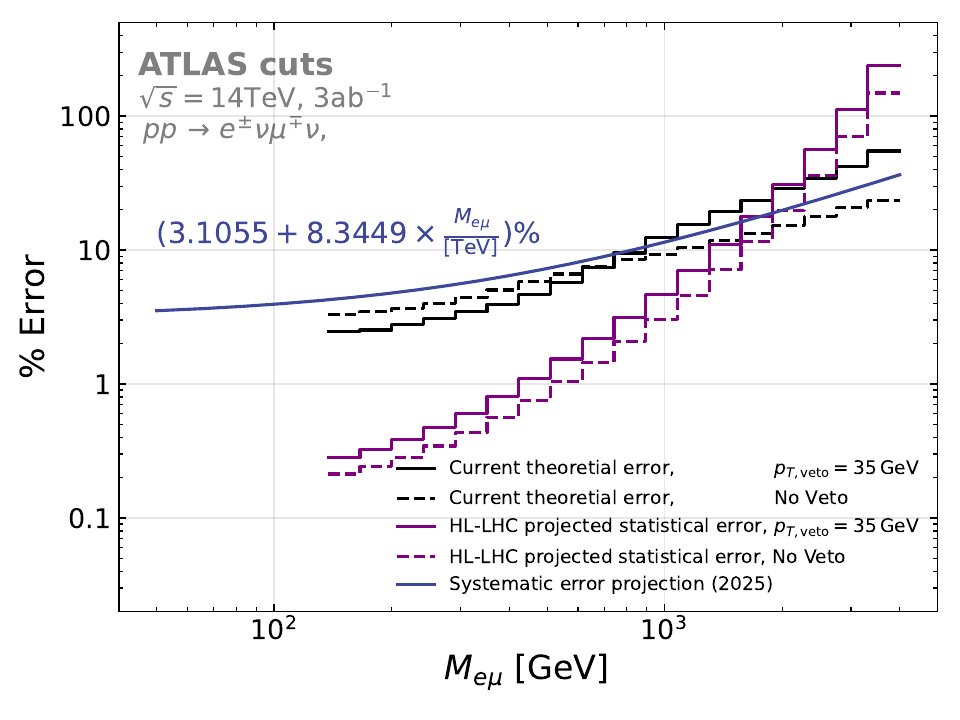}
    \caption{Comparison of statistical and systematic uncertainties
      between the ATLAS analyses of 2019~\cite{ATLAS:2019rob} and
      2025~\cite{ATLAS:2025dhf} for the dilepton invariant mass
      distribution in $WW$ production (left).  Alongside, we show
      projections for the sources of error at the HL-LHC ($14\,$TeV,
      $3\,$ab$^{-1}$) (right).}
  \label{fig:HLLHC_error_predictions}
\end{figure}

As well as theoretical and statistical uncertainties, systematic
uncertainties will form a large part of the overall error in fits at
the HL-LHC. ATLAS has recently updated their $W^+W^-$ prediction from
run 2 of the LHC~\cite{ATLAS:2025dhf} at $140\,$fb$^{-1}$.  Whilst we
have not yet performed the analysis for this new data set, we have
observed that systematic errors have been reduced. The size of this
reduction and the comparison of theoretical, EW and other errors is
presented in figure~\ref{fig:HLLHC_error_predictions}. Note that the
new ATLAS analysis employs $b$-quark tagging in order to separate the
$t\bar t$ background instead of using a jet-veto. This allows for the
use of fixed-order results for both signal and background. However,
the efficiency of the $b$-tagging must be taken into account, as well
as cancellations which could be spoiled in the case that an initial
state gluon emission decaying into a $b\bar b$ pair is vetoed.  We do
not consider these effects here and leave a more detailed study based
on the new ATLAS analysis to future work.

In left plot of figure~\ref{fig:HLLHC_error_predictions}, it can be seen that there has been a significant improvement
in the systematic errors between 2019~\cite{ATLAS:2019rob} and 2025~\cite{ATLAS:2025dhf}. For both analyses, we have included
a line of best-fit to extrapolate the errors to higher values of $M_{e\mu}$. Although this fit performs more poorly
for the most recent systematic errors, we still use it as a benchmark for a worst-case scenario for future
HL-LHC analyses. In the right plot of figure~\ref{fig:HLLHC_error_predictions}, we find that, for low energies, systematic
errors could continue to dominate the uncertainties. At higher energies, both statistical and EW uncertainties begin
to dominate, with the three sources of uncertainty contributing approximately equally. This motivates the
inclusion of resummed EW predictions for both signal and background as soon as possible, as well as more work on the inclusion
of mixed QCD-EW corrections. A similar improvement in systematic uncertainties between now and the HL-LHC could lead to them being around the
same size as current QCD theoretical uncertainties.
\begin{figure}[htbp]
\centering
  \includegraphics[width=.7\textwidth]{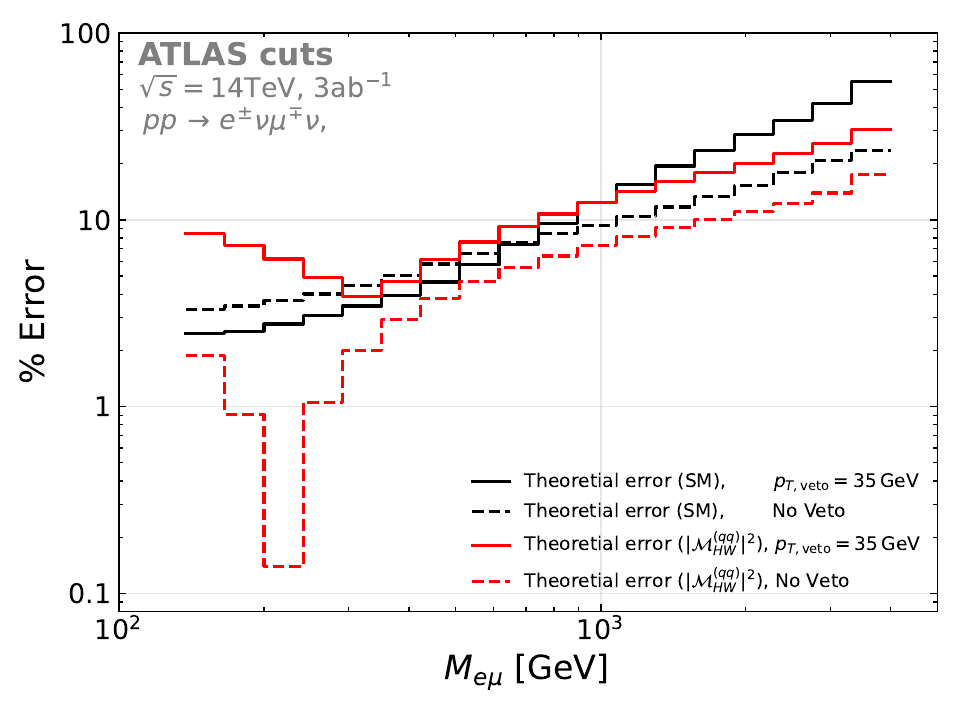}
  \caption{Comparison of Standard Model (black) and BSM (red) theoretical uncertainties.
  The squared contribution of $\mathcal{O}_{HW}$ is chosen to demonstrate the uncertainties
  although this percentage error is broadly similar between all the dimension-6 $q\bar q$ squared
  contributions. This is shown with a jet-veto at NLL and $p_{T\,,\mathrm{veto}} = 35\,$GeV and without
  a jet-veto at LO.}
 \label{fig:EFT_error_prediction}
\end{figure}

Since the EFT contribution is a fraction of the SM contribution, its
uncertainties will be less important than those on the SM
contribution. However, they can still be a relevant source of
uncertainty for fits and are a target for more accurate predictions,
especially if they are an order of magnitude larger than the SM
theoretical errors. Theoretical uncertainties are shown in
Figure~\ref{fig:EFT_error_prediction}.  It can be seen that, in the
low-energy regime, the jet-veto gives better uncertainties for the SM,
but worse uncertainties for the BSM contributions. In the high energy
regime, the uncertainties are always worse in the jet-veto case. It
should be noted that removing the jet-veto should have almost no
effect on the signal to background ratio as both are dominated by
radiation from the initial-state quark-antiquark pair.

\newpage
\section{Results at $140\,\mathrm{fb}^{-1}$ without EFT validity requirement}
\label{sec:Appendix140}

We present here the $1/\Lambda^4$ constraints for the ATLAS 2019
measurement of $WW$ production with a jet-veto~\cite{ATLAS:2019rob},
but at $140\,\mathrm{fb}^{-1}$. These predictions were obtained by
adjusting the luminosity and statistial errors accordingly. The aim of this exercise is
to compare with constraints obtained from a newer ATLAS measurement
for which data is not available yet~\cite{ATLAS:2025dhf}. Aside
from the fact that the ATLAS 2019 data has been modified, no EFT
validity check has been performed on these results, which means they should not be used as
constraints for any theory and are merely available for comparison.
\begin{figure}[!htbp]
	\centering
  \includegraphics[width=.49\textwidth]{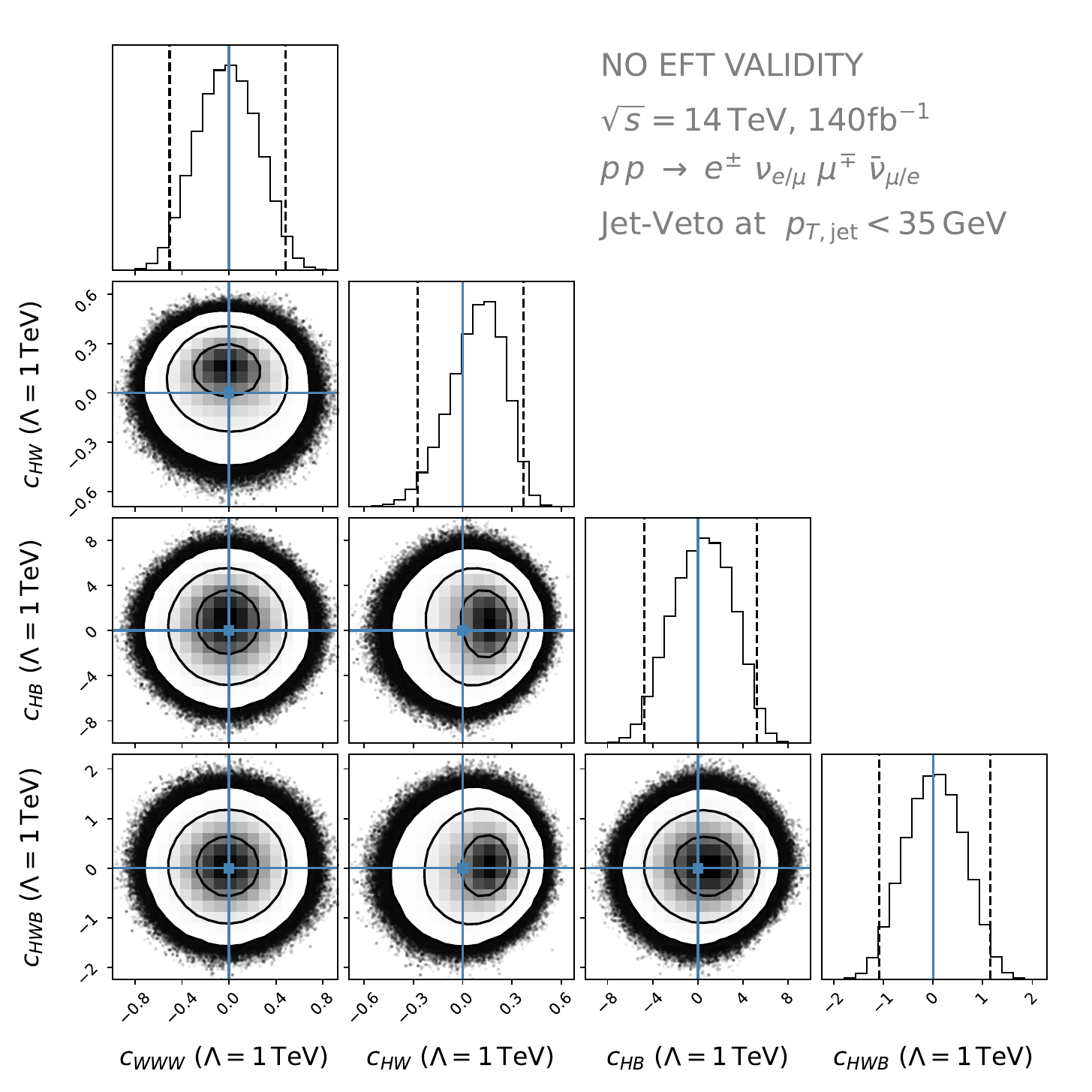}
  \caption{ Sensitivity plots for
    $\{c_{HW}, c_{WWW}, c_{HWB}, c_{HB}\}$ using 2019 ATLAS data
    ($13\,$TeV, $36.1\,$fb$^{-1}$) using $WW$ production with a
    jet-veto ($p_{T, \mathrm{veto}}=35\,$GeV)~\cite{ATLAS:2019rob},
    extrapolated to $140\,\mathrm{fb}^{-1}$, as explained in the text.
    The exclusion contours are placed at $1$, $2$, and $3\sigma$ and
    the dashed lines correspond to univariate constraints at
    $2\sigma$. The value of $\Lambda$ is set at $1\,$TeV. }
  \label{fig:vetonoveto_comparison}
\end{figure}

\bibliographystyle{JHEP}

\bibliography{dim8atgcWW.bib}

\end{document}